\documentclass[structabstract]{aa}  
\usepackage{txfonts}
\usepackage{ifpdf}
\ifpdf
\usepackage[pdftex]{epsfig}
\else
\usepackage[dvips]{epsfig}
\fi
\usepackage{graphicx}
\usepackage{natbib}
\usepackage{lscape}
\newcommand{\solar}{L$_{\odot}$\ }
\newcommand{\solm}{M$_{\odot}$}
\newcommand{\rf}{\par\noindent\hangindent 15pt {}}
\begin{document}

 \title{Near-infrared proper motions and spectroscopy 
 of infrared excess sources at the Galactic Center}



\author{
A. Eckart$^{1,2}$, 
K. Mu\v{z}i\'{c}$^{3}$, 
S. Yazici$^{1}$, 
N. Sabha$^{1,2}$,
B. Shahzamanian$^{1,2}$, 
G. Witzel$^{1}$,
L. Moser$^{1}$,
\\
M. Garcia-Marin$^{1}$,
M. Valencia-S.$^{1,2}$,
B. Jalali$^{1}$,
M. Bremer$^{1}$,
C. Straubmeier$^{1}$,
C. Rauch$^{1,2}$,
R. Buchholz$^{1}$,
\\
D. Kunneriath$^{4}$,
J. Moultaka$^{5,6}$
} 
\institute{1) I. Physikalisches Institut, Universit\"at zu K\"oln,
           Z\"ulpicher Str. 77,
           50937 K\"oln, Germany\\
           2) Max-Planck-Institut f\"ur Radioastronomie, 
           Auf dem H\"ugel 69, 
	   53121 Bonn, Germany\\
           3) European Southern Observatory, Alonso de Cordova 3107, Vitacura, Casilla 19, Santiago, 19001, Chile \\
           4) Astronomical Institute, Academy of Sciences, Bocni II 1401, CZ-141 31 Prague, Czech Republic\\
           5) Universit\'e de Toulouse; UPS-OMP; IRAP; Toulouse, France\\
           6) CNRS; IRAP; 14, avenue Edouard Belin, F-31400 Toulouse, France\\
           \email{eckart@ph1.uni-koeln.de} }

\date{Received  / Accepted }


\abstract
{There are a number of faint compact infrared excess sources in the central
stellar cluster of the Milky Way. Their nature and origin is unclear. 
In addition to several isolated objects of this kind 
there is a small but dense cluster of comoving sources (IRS13N) located 
$\sim$3'' west of SgrA* just 0.5'' north of the bright IRS13E cluster of 
Wolf-Rayet and O-type stars.
Based on the analysis of their color and brightness, there are two main possibilities: 
(1) they may be dust-embedded stars older than a few
Myr, or (2) very young, dusty stars with ages younger than 1~Myr.}
{We present a first K$_s$-band identification and proper motions of the 
IRS13N members, the high-velocity dusty S-cluster object 
(DSO, also referred to as G2), and other 
infrared excess sources in the central field.
Goal is to constrain the nature of these source.
}
{The $L'$- (3.8$\mu$m) 
K$_s$- (2.2$\mu$m) and H-band (1.65$\mu$m)
observations were  carried out using the 
NACO adaptive optics system at the ESO VLT.
Proper motions were obtained by linear fitting of the stellar positions 
extracted by StarFinder as a function of time, weighted by positional 
uncertainties, and by Gaussian fitting from high-pass filtered and deconvolved images.
We also present results of near-infrared (NIR) H- and K$_s$-band ESO-SINFONI 
integral field spectroscopy of the Galactic Center cluster ISR13N.}
{
We show that within the uncertainties, the 
positions 
and proper motions of the IRS13N sources 
in  K$_s$- and $L'$-band are identical.
The HK$-s$L' colors then indicate that 
the bright $L'$-band IRS13N sources are indeed 
dust-enshrouded stars rather than core-less dust clouds.
The proper motions also show that the IRS13N sources are 
not strongly gravitationally bound to each other.
Combined with their NIR colors, this implies that they have been 
formed recently.
For the DSO we obtain 
proper motions and a K$_s$-$L'$-color.
}
{Most of the compact $L'$-band excess emission sources have a compact H- or K$_s$-band 
counterpart and therefore are likely stars with dust shells or disks. 
Our new results and orbital analysis from our previous work favor 
the hypothesis that the infrared excess IRS13N members
and other dusty sources close to SgrA* are young dusty stars and that star formation 
at the Galactic Center (GC) is a continuously ongoing process. For the DSO the color 
information indicates that it may be a dust cloud or a dust-embedded star.}


\authorrunning{A. Eckart et al.} 
\titlerunning{Infrared excess sources at the Galactic Center}

   \maketitle
%

\section{Introduction}
\label{intro}

The formation of young and massive stars is an important 
process at the center of the Milky Way
(see e.$\,$g. Paumard et al. 2006, Ghez et al. 2005).
In the central half parsec of the Milky Way these objects are 
organized in at least one disk-like structure
of clockwise rotating stars (CWS; Genzel et al. 2003,
Levin \& Beloborodov 2003, Paumard et al. 2006) and 
the existence of a second, less populated 
disk of counter-clockwise rotating stars (CCWS)
has been proposed by Paumard et al. (2006).
It is unclear how these young stars have been formed in 
the strong tidal field of the super-massive black hole (SMBH) 
at the position of SgrA*. 
There are two prominent scenarios: One includes star formation
in-situ 
(in an accretion disk; Levin \& Beloborodov 2003, Nayakshin et al. 2006), 
the other requires the in-spiral of a massive 
stellar cluster that formed at a safe distance of 5-30 pc 
from the Galactic Center 
(Gerhard et al. 2001, McMillan \& Portegies Zwart 2003, Kim et al. 2004, Portegies Zwart et al. 2006). 
Currently the in-situ scenario appears to be favored by a number of authors 
(Nayakshin \& Sunyaev 2005, Nayakshin et al. 2006, Paumard et al. 2006).
Also, results by Stolte et al. (2007, 2010) seem to rule out
the possibility that known compact clusters close to the 
GC (such as the Arches cluster) 
could migrate inwards and fuel the young stellar population at the very center.

A prime object for testing these scenarios both observationally and theoretically 
is the IRS13 group of sources. 
IRS13E (located $\sim$$\,$3'' west and $\sim$$\,$1.5'' south of SgrA*)
is the densest stellar association after 
the immediate vicinity of SgrA* and contains several 
massive Wolf-Rayet (WR) and O-type stars 
(Maillard et al. 2004, Moultaka et al. 2005, Paumard et al. 2006, Fritz et al. 2010).
It is generally considered to be associated with the mini-spiral 
(Moultaka et al. 2005, Paumard et al. 2004).
For IRS13E four out of seven 
identified stars show highly correlated velocities 
(Maillard et al. 2004, Sch\"odel et al 2005),indicating that is 
probably bound.   
It is unclear what the origin of the IRS13E group of stars is.
It is conceivable that such a cluster could have been formed in an accretion disk
as described by  Milosavljevic \& Loeb (2004) and Nayakshin et al. (2005).
However, numerical simulations have shown that it is difficult to explain the 
formation of such a dominant feature in a star-forming disk (Nayakshin et al. 2007).
To support the cluster in-fall scenario,
an intermediate-mass black hole (IMBH) was proposed to reside in
the center of the cluster (Maillard et al. 2004).
The existence of an IMBH results in a higher 
dynamical friction and a higher stability of the in-falling cluster
such that this process in general is more efficient. It would allow compact
clusters to reach the central parsec of the Galaxy 
(with $>$10$^{6}$\solm)  within their lifetimes 
(Hansen \& Milosavljevic 2003, Berukoff et al. 2006, Portegies Zwart et al. 2006,
but see Kim et al. 2004 for a characterization of the problems 
with this hypothesis).
However, the presence
of the IMBH that would be required to stabilize IRS13E is highly disputed. 
Sch\"odel et al. (2005) analyzed the
velocity dispersion of cluster stars. The authors found that the mass
of such an object should be $\ge$7000$\,$\solm.
However, both the X-ray (Baganoff et al. 2003) and
the radio (Zhao \& Goss 1998) source at the position of IRS13E can be
explained by colliding winds of high-mass-losing stars 
(Coker et al. 2002, Zhao \& Goss 1998).
These findings make the presence of an unusual and massive and dark object
in IRS13E unnecessary.

In addition to the IRS13E association there are a number of 
faint infrared excess sources that may be essential for the
discussion of star formation in the central stellar cluster
(see Eckart et al. 2006a and Perger et al. 2008).
In some cases there is spectroscopic evidence 
for them being associated with young stars (Perger et al. 2008).
Here we present new proper motion and new spectroscopy data on
these objects 
\footnote{Based on observations collected at the European Southern Observatory, Chile}.
These comprise first K$_s$-band proper motion measurements of sources
that include the IRS13N association, and several sources 
at a projected distance of only a few arcseconds from SgrA*.

Approximately 0.5'' north of IRS13E, a small cluster of unusually
red compact sources has been reported (IRS13N; Eckart et al. 2004).
Muzic et al. (2008) gave a detailed analysis of this cluster and showed that
an orbital analysis supports that the cluster is a comoving group of young stars.
A strong infrared excess is due to the emission of warm 
dust (T $\sim$$\,1000$$\,$K;
Moultaka et al. 2005).
The authors proposed two possible explanations for the nature of IRS13N: 
(1) objects older than a few Myr and similar to bow-shock sources 
reported by Tanner et al. (2005), or (2) very young stars (0.1 - 1 Myr old).
The latter scenario implies more recent star formation than has been
assumed so far for the GC environment. 

Recently, Gillessen et al. (2012a) reported a fast-moving 
strong infrared excess source, which they interpreted as a core-less gas- and dust cloud
approaching SgrA*. Owing to its location and because it is apparently on
an elliptical orbit around SgrA*, we refer to it in the following as the
dusty S-cluster object (DSO). The community has started to call the newly found
fast-moving object G2 (e.g.  Burkert et al. 2012), reminiscent of the gas cloud G1 found by 
Clenet et al. (2005). 
As we show in the present paper, it cannot be excluded that the newly found object is 
instead a star with a gas/dust shell or disk around it, and not a core-less gas cloud 
as claimed by Gillessen et al. (2012a).
We therefore propose to call it DSO rather than G2 
since the acronym DSO (Dusty S-cluster object) reflects the true nature of the source in a more realistic way.
While essential properties of this source have already been described 
by Gillessen et al. (2012a), no continuum 
identification of it shortward of 3.5$\mu$m has been
published to date.
In other cases of near-infrared excess sources 
(e.g. see Fig.~14 in Eckart et al. 2006a) 
a detailed investigation of their nature had not yet been carried out.
We also present first K$_s$-band proper motion studies 
of the DSO.

After the introduction we describe the observations and 
data reduction in section~\ref{reduction}. 
In the results section~\ref{results} we 
derive the proper motions 
and then outline the physical properties of the sources
in section \ref{discussion}.
In section~\ref{summary} we give a summary and the 
appendix contains additional maps and graphs.

\begin{figure*}[ht!]
\centering
\includegraphics[width=19cm,angle=-00]{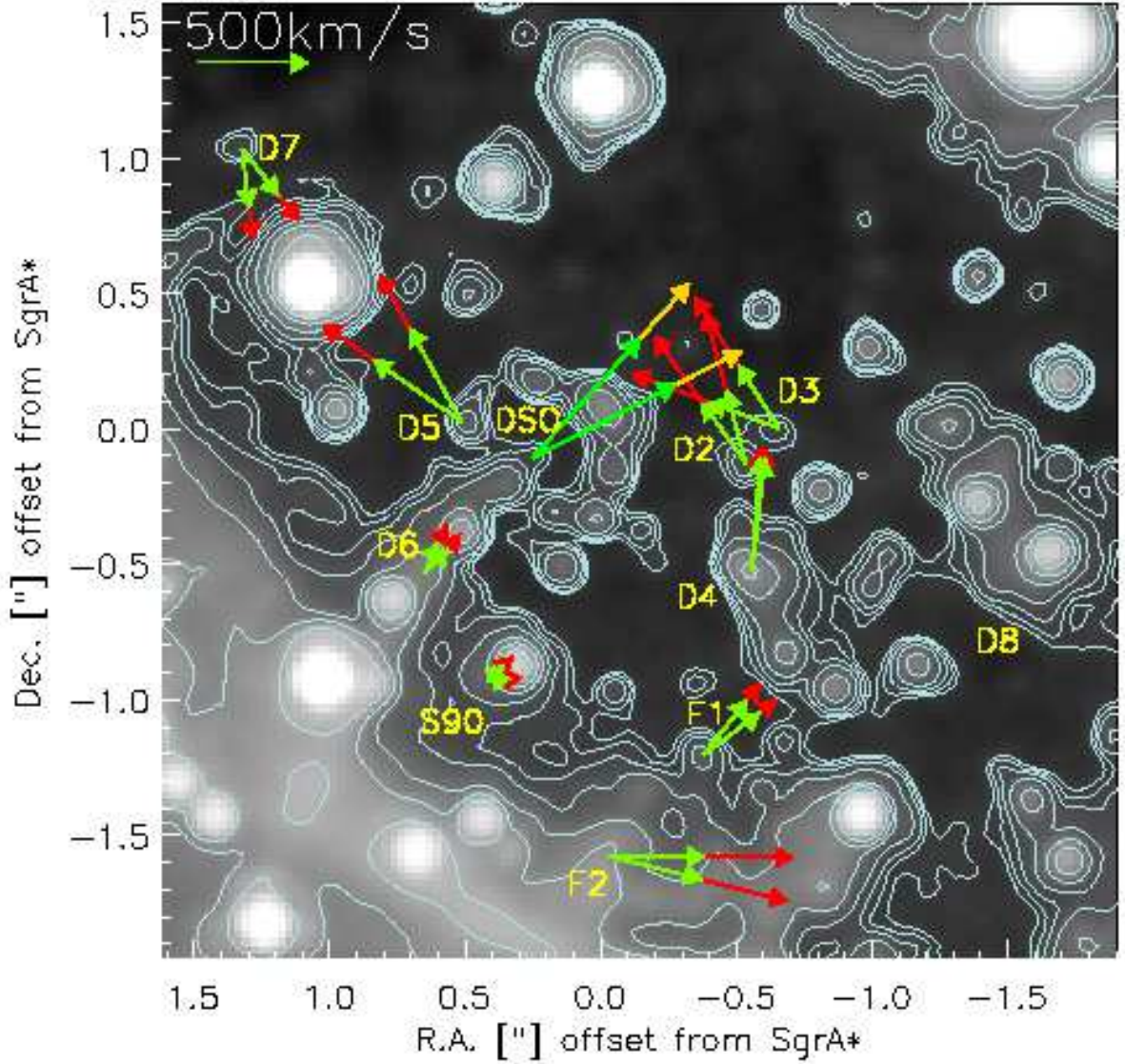}
\caption{\small 
Proper motion vectors of dusty objects in the central 
stellar cluster of the Milky Way (see Tab.~\ref{Tab:ExcessPM}).
We show the May 10, 2003, adaptive optics $L'$-band image of the central $4"\times4"$.
The arrows indicate the uncertainties in amount 
(red or yellow portion of the otherwise green arrow) 
and angle of the measured proper motion derived from data between 2002 and 2008
in the H, K$_s$, and L'-bands (see Tab.~\ref{Tab:useddata}).
}
\label{eckartfig-01}
\end{figure*}

\section{Observations and data reduction}
\label{reduction}                 

\subsection{NACO near-infrared imaging}
\label{NACOreduction}                 
The imaging data in the H-, K$_s$-, and $L'$-bands were obtained with the 
NAOS/CONICA adaptive optics (AO) assisted imager/spectrometer 
(Lenzen et al. 1998, Rousset et al. 1998, Brandner 2002) 
at the UT4 (Yepun) at the ESO Very Large Telescope (VLT). IRS7, a supergiant with 
K$_s$$\approx 6.5-7.0\,$mag and located $\sim$5.5'' north from SgrA*,
was used to lock the AO loop using the NIR wavefront sensor.
We used data whose observing dates and pixel scales for the sources 
in the central few arcseconds,
the IRS13N sources and the DSO are given in  Tab.~\ref{Tab:useddata}. 
Data reduction was standard, with sky subtraction, bad-pixel 
correction, and flat-fielding. Randomly dithered exposures were 
median-averaged to obtain final mosaics.
The sky background in the H- and K$_s$-band was measured on
a dark cloud about 400'' north and 713'' east of the GC. In the 
$L'$-band, the sky background was extracted from 
a median of the dithered science exposures. 
These are images in which the profiles of single stars are sharply 
peaked and have no broad wings. In the K$_s$-band one can see a diffraction-limited
stellar image (in some cases with an indication of a diffraction ring) 
on top of a more extended foot that stems from the 
residual of the AO seeing correction. From these images we selected data sets
in which the NIR PSF close to SgrA* has a FWHM of less that 0.25'' (estimated
through the second-order moment; optical guide star seeing typically $<$0.5'') and an 
estimated FWHM of the diffraction limited part of the PSF of better 
than 1.5 times the diffraction limit (by visual inspection of the PSF profiles).
The diffraction-limited FWHM values in the $H-$, K$_s$-, and L'-band 
are about  42~mas,  56~mas, and 98~mas (NaCo instrument manual).
Infrared PSF width estimates for all NACO H-band data until 2008 are given 
with the light curves by Bremer et al. (2011).
Active optics guide star PSF width estimates obtained in the optical for 
all NACO K$_s$-band data until 2010 are given in Fig.20 by Witzel et al. (2012).
For the years 2002 and 2009-2012 we did not find high-quality $H$-band imaging VLT data.
The work of Bremer et al. (2011) and Witzel et al. (2012) as well as the 
comparison to published data suggests that the flux densities and colors 
of the sources we investigated are constant to within 40\% and 80\%.

\begin{figure}
\centering
\includegraphics[width=8cm,angle=-00]{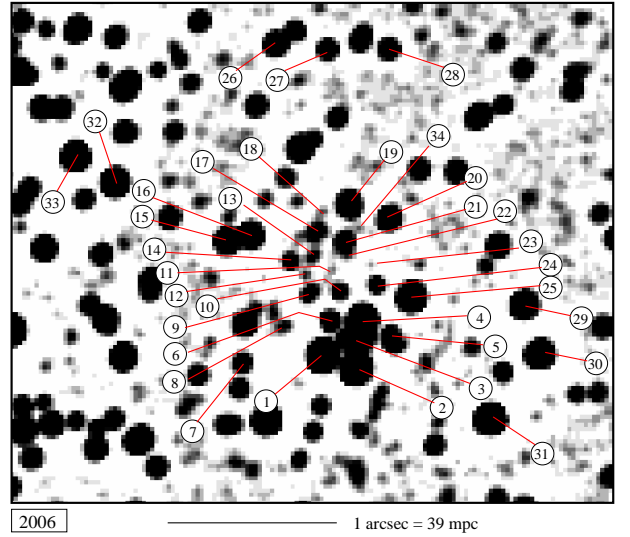}
\caption{\small
High-pass-filtered 
image of the IRS13E/IRS13N field in the K$_s$-band.
The labels correspond to those in Tab.~\ref{Tab:13N}.
Source 1 to 6 are sources E1 to E6 in IRS13E.
The angular resolution is close to the diffraction limit of 60mas.
}
\label{eckartfig-02}
\end{figure}

The sources in the wider field of the central star cluster are 
shown in Fig.~\ref{eckartfig-01}. 
The IRS13E and IRS13N clusters 
are shown in Figs.~\ref{eckartfig-02} and \ref{eckartfig-03}. 
The relative and absolute coordinates of the lower left (southeastern)
corner of the images are listed in Tab.~\ref{Tab:coordinates}.
In the IRS13N region eight sources were resolved and labeled 
$\alpha$ through $\kappa$, as shown in 
Figs.~\ref{eckartfig-04} - \ref{eckartfig-06}.
The nomenclature is taken from Eckart et al. (2004).

\begin{table}[ht!]
\centering
\begin{tabular}{c|c|c|c}
\hline
band & $H$ & $K_s$ & $L'$ \\
\hline
$\frac{arcsec}{pixel}$ & 0.0132 & 0.0132 \& & 0.0270 \\
                       &        & 0.0270 &        \\
\hline
field  & -       & 2002.67 & 2002.75 \\
       & 2003.54 & 2003.55 & 2003.36 \\
       & 2004.41 & 2004.82 & 2004.41 \\
       & -       & 2005.66 & 2005.45 \\
       & 2006.41 & -       &       - \\
       & 2006.65 & 2006.82 & 2006.50 \\
       & 2007.35 & 2007.31 & 2007.34 \\
       & -       & -       & 2008.49 \\
\hline
IRS13N &- & 2002.42 & 2002.66 \\
       &- & 2003.54 & 2003.36\\
       &- & 2004.66 & 2004.32 \\
       &- & 2005.83 & 2005.36 \\
       &- & 2006.55 & 2006.41 \\
       &- & 2007.33 & 2007.39\\
       &- & 2008.49 &-\\
       &- & 2009.47 &-\\
       &- & 2010.53 &-\\
       &- & 2011.47 &-\\
\hline
DSO    &2003.54 & 2002.42 & 2002.66 \\
       &2004.66 & 2003.54 & 2003.36 \\
       &2005.46 & 2004.66 & 2004.32 \\
       &2006.55 & 2005.83 & 2005.36  \\
       &2007.33 & 2006.55 & 2006.41  \\
       &2008.56 & 2007.33 & 2007.39 \\
       &      - & 2008.49 & 2008.49 \\
       &      - & 2008.19 &         \\
       &      - & 2008.70 &         \\
       &      - & 2009.47 & 2009.47 \\
       &      - & 2010.53 & - \\
       &      - & 2011.55 & 2011.55 \\
       &      - & 2012.54 & 2012.54 \\
\hline
\end{tabular}
\caption{
Observational data of the GC used in $H$-, $K_s$-, and $L'$-band 
for the sources show in Fig.~\ref{eckartfig-01}.
}
\label{Tab:useddata}
\end{table}

\subsection{SINFONI imaging spectroscopy}
\label{reductionSinfoni}                 

Spectroscopy of the region around IRS13 was carried out 
in June 2009 using SINFONI, 
the AO-assisted integral field spectrometer mounted at Yepun, 
Unit Telescope 4 of the ESO VLT in Chile 
(Eisenhauer et al. 2003). 
The AO guiding was carried out in the optical using an R$\sim$13.9 star 
located $\sim$19'' from SgrA*. We used the K$_s$-band grating, which provides 
spectral resolution of 4000; the spatial setting of 0.1'' per image 
slice results in a field-of-view of 3''$\times$3''.  
The mean Strehl ratio of the cube (measured from Br$\gamma$) is about 7\%. 
The sky was sampled on the same dark cloud as in the case of NACO
observations, and the total on-source time was 14$\times$300~s. 
Because of the non-quadratic shape of the SINFONI pixels 
(50~mas $\times$ 100~mas), 
the observations were split into two blocks observed at position 
angles of 0 and 90 degrees. The two cubes were combined into a single
data cube as a final step in the reduction procedure.  
To combine the two data cubes, we first de-rotated the second cube by 90 degrees to match the
orientation of the first one. Both data cubes were re-binned to a unique pixel scale 
of 50~mas $\times$ 50~mas using the IDL task {\it 'congrid'} and were median-collapsed.
The resulting images were then cross-correlated to determine the offsets that can be 
supplied to the pipeline recipe  
{\it 'sinfo\_utl\_cube\_combine'}. 
The recipe shifts the two cubes and co-adds them to produce the final data cube.

The 3D cubes were reduced and reconstructed with 
the ESO-supported SINFONI reduction 
software package. Bad pixel, cosmic rays, and flat-field corrections 
were applied to the 2D raw frames. The 3D cubes were 
reconstructed using calibration frames for the slit-let distances 
and the light dispersion. 
Intermediate standard-star observations of a nearby G2V 
star were used to correct for strong atmospheric (telluric) 
absorptions. The standard-stars were observed near in time 
and airmass to the target exposures. The G2V 
spectral characteristics were removed by dividing the standard-star 
spectrum by the well-known, high signal-to-noise solar 
spectrum (NSO/Kitt Peak spectra, as provided at the ISAAC/VLT web page).

The flux was calibrated with the known fluxes of 
the IRS13N cluster sources (Eckart et al. 2004; Muzic et al. 2008), 
since we are primarily interested in measuring line ratios. 
The total flux within the SINFONI 
FOV was scaled according to these values. 
The positional registration of the data cube was performed via the know
positions of the bright IRS13E sources and its uncertainty is about 0.05''.
\\
\\
\noindent

\noindent
\begin{figure}
\centering
\includegraphics[width=8cm,angle=-00]{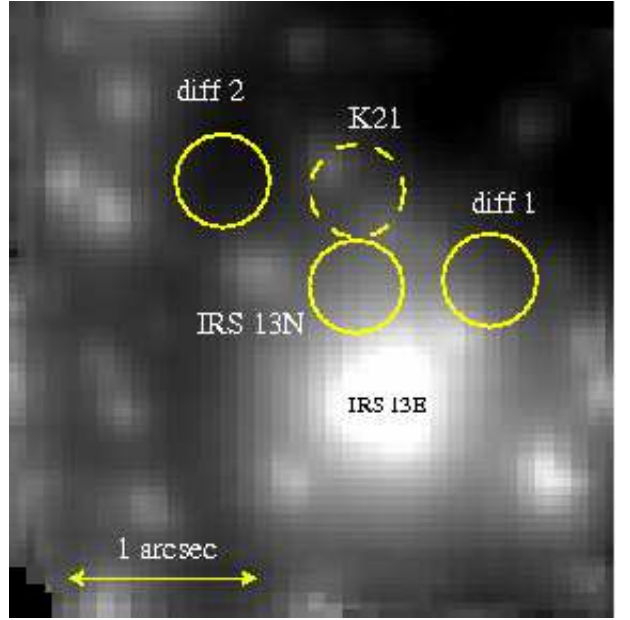}
\caption{\small
Wavelength-integrated 2$\mu$m SINFONI data cube. Yellow circles mark
the regions in which the spectra were obtained.
The angular resolution is about 0.11 arcseconds.
}
\label{eckartfig-03}
\end{figure}

From the final reconstructed 3D cube we extracted spectra
and emission line maps.
In Figs.~\ref{eckartfig-07} and \ref{eckartfig-08} we show the
K- (intermediate continuum band IB 2.15$\mu$m) and $L'$-band (3.8$\mu$m) 
images with the Br$\gamma$ and [FeIII]-emission lines and the 
CO absorption line maps in contours overplotted.
The contours in
Figs.~\ref{eckartfig-07}, \ref{eckartfig-08},  and  \ref{eckartfig-09}   
are labeled with percentages of the peak flux.
The emission line maps presented in Fig.~\ref{eckartfig-07}  were 
created by summing the flux of the emission line 
and then subtracting the average flux density of the 
continuum to the left and right of the line. 
The wavelength ranges we used for the maps are 
Br$\gamma$: 2.16 - 2.17 $\mu$m;
[FeIII] (2.242 $\mu$m): 2.2395 - 2.2438 $\mu$m;
[FeIII] (2.3479 $\mu$m): 2.3460 - 2.3515 $\mu$m;
[FeIII] (2.2178 $\mu$m): 2.2154 - 2.2197 $\mu$m;
CO-absorption: 2.2932 - 2.3015 $\mu$m.
These wavelength ranges are several times larger than a typical radial velocity offset. This is true both
for the stellar sources and the extended emission in the observed region, because the absolute 
values of the radial velocity do not exceed 150 kms$^{-1}$ (Paumard et al. 2004, 2006; Zhao et al. 2009).
The absorption line maps presented in Fig.~\ref{eckartfig-08} were 
created by extrapolating the continuum shortward of 
2.29$\mu$m and subtracting it from the measured spectrum.
The maps were then produced by plotting the negative value of 
the line integral over the strongest 
CO-bandhead between 2.290$\mu$m and 2.300$\mu$m.

In Fig.~\ref{eckartfig-09} we show the radio image from Zhao et al. (2009) 
with line maps in contours overplotted.
The radio data and our imaging field spectroscopy data
cubes were essentially taken at the same epoch.
In addition, we extracted spectra
for specific regions as shown in Fig.~\ref{eckartfig-03}.
The radius of the extraction regions is
5 pixels (0.25'').
In Fig.~\ref{eckartfig-10} we show the
integrated K$_s$-band spectrum of IRS13N 
and for comparison the spectra of the nearby regions east and west
of IRS13N that are dominated by diffuse emission.
the integrated K$_s$-band spectrum of 
the [FeIII]-emission line blob north of IRS13~N
is presented in Fig.~\ref{eckartfig-11}.
As indicated by the shape of the [FeIII] line at 2.3479$\mu$m, this line 
might be a blend with the weaker HeII line that is close in wavelength.

\section{Results}
\label{results}                 

\subsection{Deriving the proper motions}
\label{pmderivation}

At a distance of 8~kpc the GC is well-suited to derive 
proper motions of stars 
(Eckart \& Genzel 1996, Ghez et al. 2000, Eckart et al. 2002)
and dusty filaments (Muzic et al. 2007).
In the framework of the GC, proper motions have mostly been 
used to derive a mass estimate of the potential the 
stars are moving in
(e.g. Sch\"odel, Merritt \& Eckart 2009, Sch\"odel et al. 2005, Ghez et al. 2005)
or to explore the origin and dynamics of young and old 
stellar populations in the central cluster
(e.g. Genzel et al. 2003, 
Levin \& Beloborodov 2003, 
Ghez et al. 2005,
Paumard et al. 2006,
Sch\"odel et al. 2005,
Muzic et al. 2008,
Bartko et al. 2010,
Sch\"odel, Merritt \& Eckart 2009,
Lu et al. 2008).
However, comparing the proper motions is 
also a unique tool to identify sources across different wavelength bands.
In this paper we use this technique to identify stellar counterparts 
of dusty sources in the GC stellar cluster.
No correction for image distortions was applied. However, the comparison to
published positions and proper motions shows that the linear and quadratic distorions across
the small ($<<$5'') fields that we considered are less than 5$\times$10$^{-4}$ and 5$\times$10$^{-5}$.
The small corrections for image rotation that we applied are less then 3$\times$10$^{-2}$ degrees.
Details are given in the following subsections.

\subsubsection{Proper motions of IRS13N and IRS13E}
\label{pmderivationIRS13N}

To derive the K$_s$-band proper motions of the IRS13N cluster sources 
it is necessary to work on images with a very high point source sensitivity
for each epoch.  
In each epoch the faint K$_s$-band counterparts of the
significantly brighter $L'$-band sources 
had to be identified. 
A first identification for epoch 2002 was given by 
Eckart et al. (2004).
The K$_s$-band sources are faint compared to the IRS13E sources and 
deconvolution or point spread function (PSF) fitting is not sufficient to 
provide source identifications for all epochs.
Therefore, we produced high-pass filtered images of the highest Strehl
K$_s$-band images. 
In these images the peak of the diffraction-limited core of the AO
images is highlighted and the surrounding emission
is strongly suppressed.

The high-pass filtered image can be produced by calculating a
smooth-subtracted image $\Sigma$ as

\begin{equation}
\Sigma = I*G - I = (\Delta + E) * P * G - (\Delta + E) * P~.
\end{equation}

The symbol $*$ denotes the convolution operator,
$\Delta$ denotes the distribution of stellar sources,
and $E$ represents the distribution of extended emission.
Here $P$ is the PSF
and $I$ the image that results from the convolution of 
$P$ and the object distribution $O$ = $\Delta + E$.
We used a narrow Gaussian $G$ normalized to an integral power of unity. 
It is our experience that it is best to choose the
Gaussian $G$ to have a width that 
corresponds to about 30\% to 50\% of the full width at half maximum of 
a diffraction-limited core of a stellar AO image.
The smooth-subtracted image can then be rewritten as

\begin{equation}
\Sigma =  \Delta * (P * G - P) +  (E * P * G - E * P)~.
\end{equation}

Since the width of the extended emission is  
much larger than the width of the PSF
(for the one-dimensional (1D) case see discussion on narrow dust features 
by Muzic et al. 2007),
we can then infer that $E * P * G \sim  E * P$
and the high-pass filtered differential image is

\begin{equation}
\Sigma \sim  \Delta * (P * G - P)~
\end{equation}

(see also Sabha et al. 2010).
If the image is clipped at zero flux, density the remainder of
$P * G - P $ represents the top section of the diffraction-limited
PSF core of the AO image
and $\Sigma$ is an approximation of an image of the point 
sources in the field.
To give a higher weight to signals close to and below the spatial frequency of the
PSF we convolved the high-pass-filtered image again with the Gaussian $G$.

This high-pass filtering method has the advantage that no estimate of the
PSF is required. The PSF is typically variable across the imaging field
and has a finite size. The corresponding apodization radius and residuals from
stars that have to be removed from the PSF wings (via a median procedure or explicit 
subtraction) are usually the cause for additional artifacts that spoil the image quality if
high sensitivity is required.
In addition, a deconvolution may suffer from small number division problems at high
spatial frequencies, or it runs with a limited number of iterations, or a loop gain
and a stopping or convergence criterion. 
A limited number of sources needs to be identified for PSF fitting.
All these restrictions and their potential drawbacks do not have to be 
considered for the high-pass filtering method.
The noise property of the high-pass filtering is intimately linked to the original noise 
in the image. All noise contributions with spatial frequencies significantly lower than
the characteristic cutoff frequency of the filter (i.e. given by the FWHM of the Gaussian kernel)
are strongly suppressed by subtracting the flux-conserved image from the original.
Only the (already existing) image power around and above the characteristic cutoff frequency 
of the filter will remain.  This includes power at the diffraction limit.
Hence, there is no mechanism over which additional and spurious noise peaks can be 
introduced by the filtering process. 
High-pass filtering has, however, the disadvantage that PSF imperfections at
high spatial frequencies, which may be introduced by the combined imaging optics 
(residual astigmatism, trefoil, etc.) are not corrected for.

\begin{table}[ht!]
\centering
\begin{tabular}{lrrll}
\hline
Source                     & $\Delta$$\alpha$& $\Delta$$\delta$& R.A.(J2000)& Dec.(J2000)\\
Figure                     &        &        &                       &             \\ \hline
SgrA*                      &  0.000 &  0.000 & 17$^h$45$^m$40.050$^s$& -29$^o$ 00' 28.120"\\
Fig.~\ref{eckartfig-02}    & -0.931 & -2.603 & 17$^h$45$^m$40.121$^s$& -29$^o$ 00' 30.723"\\
Fig.~\ref{eckartfig-03}    & -1.065 & -2.842 & 17$^h$45$^m$40.131$^s$& -29$^o$ 00' 30.962"\\
Fig.~\ref{eckartfig-04}    & -1.965 & -1.875 & 17$^h$45$^m$40.200$^s$& -29$^o$ 00' 29.995"\\
Fig.~\ref{eckartfig-05-1} & -1.965 & -1.875 & 17$^h$45$^m$40.200$^s$& -29$^o$ 00' 29.995"\\
Fig.~\ref{eckartfig-05}    & -2.780 & -1.171 & 17$^h$45$^m$40.262$^s$& -29$^o$ 00' 29.291"\\
\hline
\end{tabular}
\caption{Relative and absolute coordinates of the lower left (southeastern) 
corner of the image sections shown in the corresponding figure.
The relative coordinates are given with respect to the SgrA* radio position in the first line.
}
\label{Tab:coordinates}
\end{table}

We derived the positions of individual sources in the high-pass filtered images
from (1) Gauss-fits to the
resulting source profiles and (2) by determining the peak position 
on a sub-pixel level.
Within the uncertainties (see below) both methods resulted in 
the same proper motion values.
To provide an accurate relative positional reference system
the low proper motion stars 26-33 identified in Fig.~\ref{eckartfig-02}
and the May 2005 positions of stars E1 through E6 in the IRS13E cluster 
(Fritz et al. 2010) were used to establish a calibrated reference 
frame that allowed us to compare 
the positions and proper motions to values in the literature:
in particular Fritz et al. (2010)
mainly for the IRS13E sources in K$_s$-band,
Muzic et al. (2008) for the IRS13E and IRS13N sources in $L'$-band, and 
Sch\"odel et al. (2005) for the IRS13E sources and stars in the 
wider 3'' diameter field around the cluster obtained in the K$_s$-band.

\begin{figure}
\centering
\includegraphics[width=10cm,angle=-00]{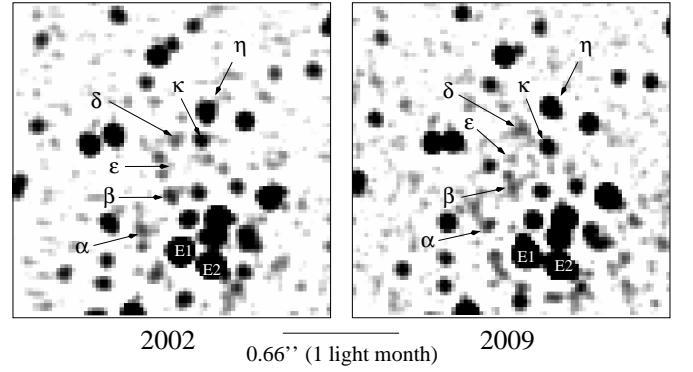}
\caption{\small
K$_s$-band identification of individual IRS13E and IRS13N sources
at two different epochs
and visualizations of the positions of the IRS13N  sources 
$\alpha$, $\beta$, $\delta$, $\epsilon$, $\eta$, and $\kappa$ at two different epochs.
The source labels in the high-pass filtered images correspond to those in Tab.~\ref{Tab:13N}.
}
\label{eckartfig-04}
\end{figure}

\begin{figure}
\centering
\includegraphics[width=08cm,angle=-00]{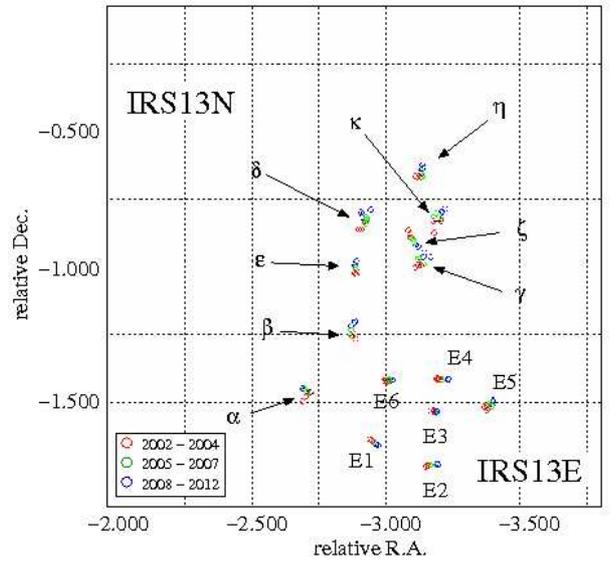}
\caption{\small
Proper motion of IRS13E and IRS13N.
For source labels see Figs.\ref{eckartfig-04} and \ref{eckartfig-05}.
}
\label{eckartfig-05-1}
\end{figure}

\begin{figure*}
\centering
\includegraphics[width=16cm,angle=-00]{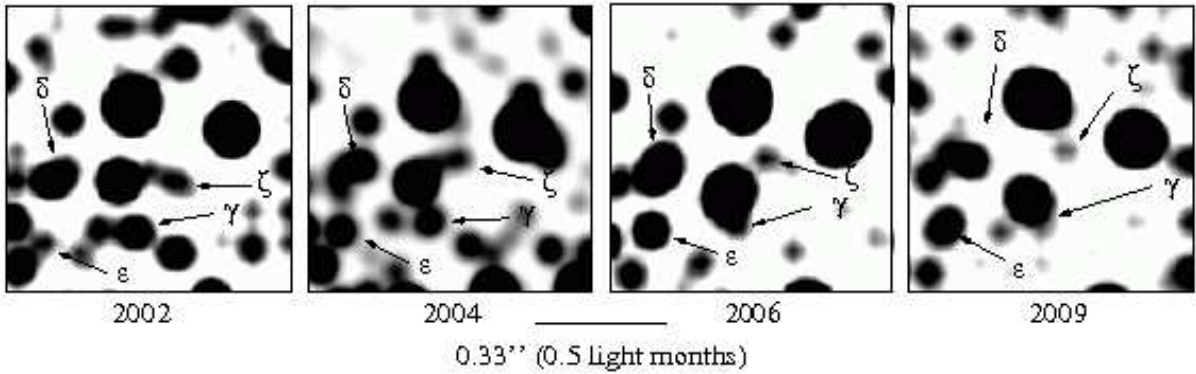}
\caption{\small
Visualization of the K$_s$-band identifications of the IRS13N  sources 
$\gamma$, $\delta$, $\epsilon$, and $\zeta$ at four different epochs
in high-pass filtered images..
The source labels correspond to those in Tab.~\ref{Tab:13N}.
}
\label{eckartfig-05}
\end{figure*}

The proper motions of sources $\alpha$ through $\kappa$ can be
inspected in Figs.~\ref{eckartfig-04}, \ref{eckartfig-05-1} and \ref{eckartfig-05}.
For demonstration purposes the pixel scale of 27~mas has been decreased
(by interpolation) by a factor of 8.
Gray scales have been adapted to the quality of the frame.
Sources may in addition be influenced by blends with nearby background sources.
In Tab.~\ref{Tab:13N} we present the source name (column 1), a reference (2), 
a source identification (3),
coordinates relative to SgrA* and the corresponding uncertainties (4-7),
and proper motion values and the uncertainties (8-11).
The uncertainties for the positions and proper motion values were derived from the 
linear regression fit to the positions at all epochs.

\begin{figure}
\centering
\includegraphics[width=8cm,angle=-00]{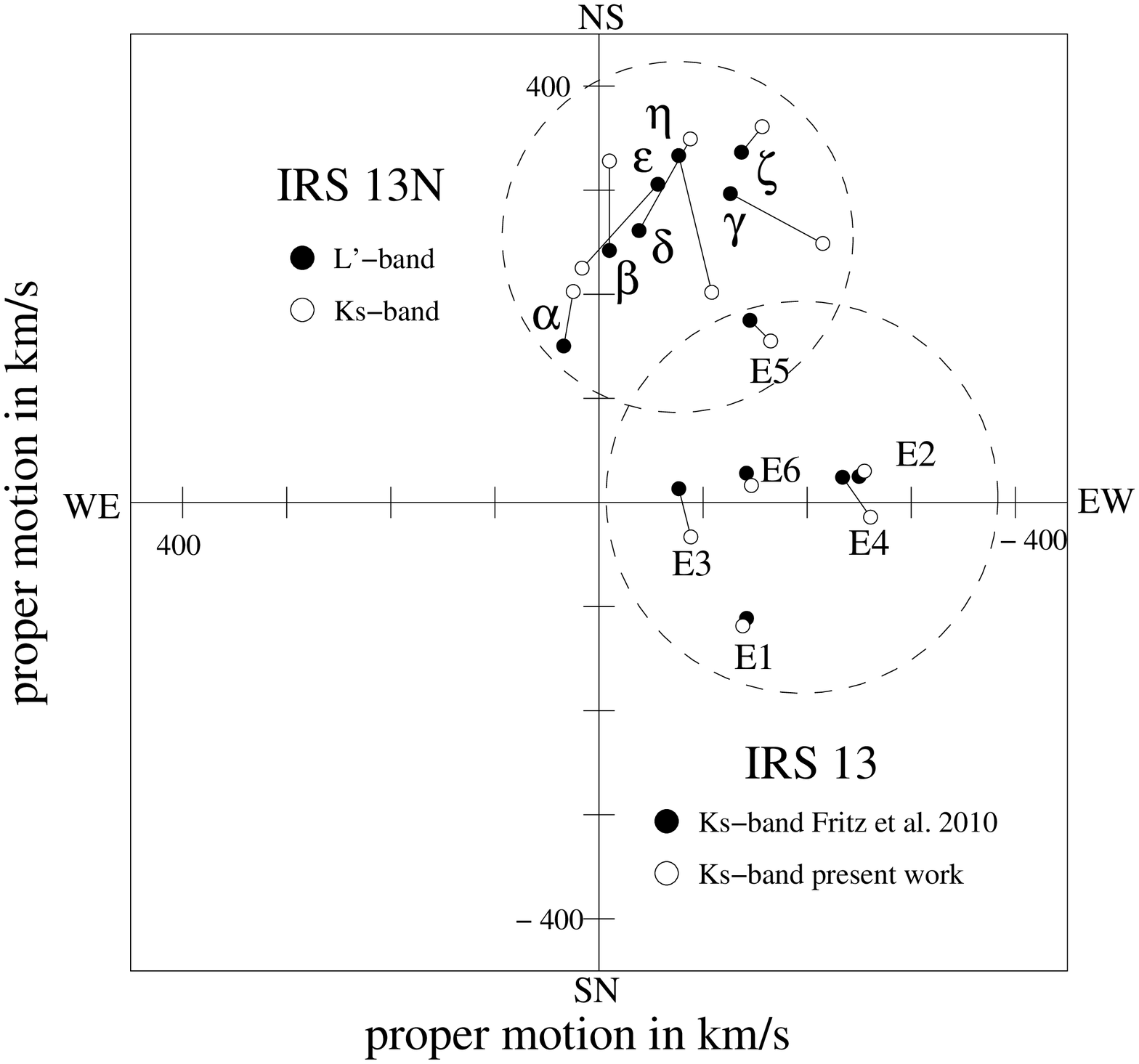}
\caption{\small
Comparison of K$_s$- and $L'$-band proper motions
in right ascension and declination for the GC 
IRS13E and the IRS13N cluster (indicated by circles in dashed lines).
The direction of the motion, i.e., S to N, or E to W, is labeled.
The source labels correspond to those in Tab.~\ref{Tab:13N}.
The K$_s$-band motions derived here are labeled with open dots.
For IRS13 and for IRS13N the filled dots indicate the proper motions published by
Fritz et al. (2010) and Muzic et al. (2008).
The open circles depict the $K_s$-band proper motions 
described here.
}
\label{eckartfig-06}
\end{figure}

In Fig.~\ref{eckartfig-06} we compare the proper motion values of IRS13E and IRS13N
sources obtained in $L'$-band (Muzic et al. 2008) to those obtained here
in K$_s$-band.
We compared our proper motion results with published values and quote 
the root mean square deviations in right ascension $\sigma_{R.A.}$ and declination $\sigma_{Dec.}$ .

These values have to be compared to the expected 1D proper motion 
velocity dispersion value of about 200~km/s at the 
location of the IRS13E and IRS13N clusters (Sch\"odel, Merritt \& Eckart 2009).

The IRS13E sources E1 to E6 we agree well with 
the results of Fritz et al. (2010) 
$\sigma_{R.A.}$ = 11 km/s and
$\sigma_{Dec.}$ = 22 km/s and with
in comparison to the $L'$-band proper motion by  Muzic et al. (2008) with
$\sigma_{R.A.}$ = 38 km/s and
$\sigma_{Dec.}$ = 44 km/s.
We regard the difference to Fritz et al. (2010) as 
a measure of our measurement uncertainty for bright sources in the K$_s$-band
and the differences to Muzic et al. (2008) as a measure of the combination of 
our measurement uncertainties in the K$_s$-band and the corresponding 
$L'$-band uncertainties.
The deviations from Sch\"odel et al. (2005) are larger
with
$\sigma_{R.A.}$ = 96 km/s and
$\sigma_{Dec.}$ = 77 km/s, but here the baseline in time used to derive the
motions was only three years.
For the IRS13N sources $\beta$ to $\eta$ we 
compared our K$_s$-band proper motions to 
the $L'$-band values by Muzic et al. (2008) and
found
$\sigma_{R.A.}$ = 65 km/s and
$\sigma_{Dec.}$ = 95 km/s.
Correcting for the uncertainties derived for bright sources 
from the comparison to the data by Muzic et al. (2008), we obtain a
residual uncertainty value between the K$_s$-band and L'-band proper motions of
$\sigma_{IRS13N, R.A.}$ = 52 km/s and
$\sigma_{IRS13N, Dec.}$ = 84 km/s
that can be solely attributed to the faint co-moving IRS13N sources $\beta$ to $\eta$
(see section \ref{discussion}).

\noindent

\noindent

\noindent

\subsubsection{Proper motions of the DSO}
\label{pmderivationScluster}
In the central arcsecond the source density is highest in the overall cluster.
In this region we need the highest angular resolution in addition to high 
point source sensitivity 
to separate the faint and bright stars from each other and to search for 
a stellar counterpart of the dusty high-velocity S-star cluster source
reported by Gillessen et al. (2012a).
In addition to the high pass-filtering we therefore employed Lucy- and linear Wiener filtering.
A detailed comparison of these with two algorithms with the CLEAN algorithm,
which is sensitive to structures that are favored by high-pass filtering,
is given in Ott, Eckart \&  Genzel (1999).
In the K$_s$-band we used the same images as for IRS13N.
In the $H$-band we used the images employed for the HK-spectral index
analysis of SgrA* presented by Bremer et al. (2011)
that covers the years 2003 till 2008.
In both bands we investigated the images at the derived L'-band positions of the dusty S-cluster object 
relative to the corresponding positions of the star S2 and SgrA*.

In addition, we compared the data with detailed simulations of the field including important, neighboring
sources such as S23, S54, and S63. For these sources the proper motions, and in the case of S23 the 
curvatures, are listed in Tab.9 by Gillessen et al. (2009). From S63 we derived the proper motions from 
our H- and K$_s$-band data, since the values listed by Gillessen et al. (2009) indicated a motion almost 
perpendicular to what we found. 
Our simulations clearly show that the DSO and star S63 blend each other from 2002 until 2007.
In 2007 the DSO starts to emerge from the S63 PSF and is clearly distinguishable from the neighboring stars.

This results in K$_s$-band identifications for all epochs since 2007.
We can use the 2008 H-band data to derive a lower limit for the DSO magnitude.
Taking the background from the neighboring stars and the 3$\sigma$ variation across the diffraction limited
beam into account, we find a limit of m$_H$$>$21.2.
See Figs.~\ref{eckartfig-14-1}, 
\ref{eckartfig-14-2}, 
\ref{eckartfig-14-3}, 
\ref{eckartfig-34},
\ref{eckartfig-36},
and \ref{eckartfig-21}.
In Fig.~\ref{eckartfig-13}  we show 
the proper motion of the DSO projected onto the sky 
compared to the orbital 
solution given by Gillessen et al. (2012a).

\subsubsection{Proper motions in the wider field of the central star cluster}
\label{pmderivationField}
The central parsec of the GC reveals different 
structures when observed at different wavelengths.
In Fig.~14 shown by Eckart et al. (2006)
these sources are denoted with D followed by a number. 
In Fig.~\ref{eckartfig-01} 
(taken on July 6, 2004 in the $L'$-band)
we show the proper motions of 
some of these sources in the central few arcseconds
(see also Figs.~\ref{eckartfig-18} and \ref{eckartfig-19} in the appendix). 
They are typically brighter than 14.6 in $L'$-band.
Some of them appear to be circular symmetric on the sky (e.g. D2 and D5), 
others appear to be elongated (D3 and D4)
or even filamentary, such as the source D8 at the eastern edge of
Fig.~\ref{eckartfig-01} about -1.5'' east and -0.5'' south of SgrA*.
The extended source D8 was excluded from our investigation since 
it is a very complex region in projection on the sky for which 
proper motions are difficult to obtain. 
Comparing Fig.~1 shown by Gillessen et al. (2009) 
to our Fig.~\ref{eckartfig-01},
we can associate the position 
of the S-star S43 with the position of D2
(see also Tab.~\ref{Tab:ExcessPM}).
Within the uncertainties both objects also compare well 
in their $H-$, $K_s-$, and $L'-$band proper motions.
For sources D3, D5, and D7 we could only determine L'-band proper motions.
For D4 the data are given by Muzic et al. (2010).
For D2, D6, S90, F1, and F2 the K$_s$- and L'-band proper motions
agree within the uncertainties given in Tab.~\ref{Tab:ExcessPM}.

The mosaics in $L'$-band were put into an astrometric frame
(see Muzic et al. 2008).
The $H$- and $K_s$-band data were transformed into 
the coordinate system of the $K_s$-band mosaic 
obtained on July 31, 2002, using the 
coordinate transformation procedure described 
by Muzic et al. (2008).
The positions of the 30 reference stars are evenly distributed over the field 
and were determined using the IDL code STARFINDER (Diolati et al. 2000)
for the coordinate transformation.
The known proper motions were calculated from the reference 
stars positions taken from Sch\"odel et al. (2009).

The peak positions of the objects were determined 
using two dimensional (2D) Gaussian fits 
in each mosaic and band. 
The uncertainties in the graphs are the $1\sigma$ uncertainty 
in the peak position and the uncertainties in velocities 
are the $1\sigma$-uncertainty of the linear fits. 
Figs.~\ref{eckartfig-18} and \ref{eckartfig-19} show detailed maps of the
immediate environment of the infrared excess sources close to SgrA* in the 
central stellar cluster.
Details of the proper motions are shown in  
Figs.~\ref{eckartfig-29} and \ref{eckartfig-30}
and are listed in 
Tabs.  \ref{Tab:Field1}, \ref{Tab:Field2}, and \ref{Tab:Field3}.
As an overview the results are summarized in Tab.~\ref{Tab:ExcessPM}
and in Fig.~\ref{eckartfig-01}. 
The figure shows that the motions in the different bands 
for each source can be related to each other, 
since the vectors point in the same direction within the error bars, 

\noindent

\begin{table*}[htb]                      
\begin{center}                          
\begin{tabular}{rrrrrrrrrrrrrc}         
 \hline                                 
name&ref&ID&
$\Delta\alpha$&$\sigma_{\Delta\alpha}$&
$\Delta\delta$&$\sigma_{\Delta\delta}$&
$v_{\alpha}$&$\sigma_{v_{\alpha}}$&
$v_{\delta}$&$\sigma_{v_{\delta}}$\\
\hline \\
E1         &smf & 1& -2.955&  0.001& -1.645&  0.001& -143&    8& -115&    5&\\
E2         &smf & 2& -3.167&  0.001& -1.730&  0.001& -245&    8&   34&    5&\\
E3         &smf & 3& -3.180&  0.001& -1.532&  0.001&  -97&   18&  -14&   15&\\
E4         &sf  & 4& -3.206&  0.001& -1.413&  0.001& -235&   12&  -11&    4&\\
E5         &mf  & 5& -3.378&  0.002& -1.502&  0.002& -150&   32&  155&   28&\\
E6         &sf  & 6& -3.005&  0.001& -1.420&  0.001& -146&    9&   16&   14&\\
$\alpha$   &    & 8& -2.707&  0.004& -1.459&  0.003&   27&   63&  200&   41&\\
$\beta$    &    & 9& -2.876&  0.002& -1.255&  0.004&   -8&   31&  325&   58&\\
$\gamma$   &    &22& -3.140&  0.004& -0.990&  0.003& -208&   60&  262&   44&\\
$\delta$   &    &17& -2.929&  0.004& -0.825&  0.003& -101&   63&  352&   41&\\
$\epsilon$ &    &13& -2.889&  0.001& -1.010&  0.001&   22&   14&  222&   20&\\
$\zeta$    &    &34& -3.183&  0.004& -0.812&  0.003& -162&   65&  362&   53&\\
$\eta$     &s   &19& -3.134&  0.002& -0.660&  0.003& -114&   24&  194&   46&\\
$\kappa$   &s   &21& -3.094&  0.001& -0.891&  0.002& -172&   18& -283&   28&\\
           &    & 7& -2.430&  0.001& -1.697&  0.002&   31&   15&  126&   33&\\
           &s   &10& -3.071&  0.001& -1.235&  0.001&  -10&   18&  114&   22&\\
           &    &11& -2.982&  0.004& -1.162&  0.003& -312&   56&  358&   54&\\
           &s   &12& -2.836&  0.001& -1.116&  0.001& -180&   11&  -30&   22&\\
           &s   &14& -2.757&  0.001& -1.017&  0.003&  274&   17& -261&   40&\\
           &s   &15& -2.334&  0.001& -0.898&  0.001& -119&   19&   62&   19&\\
           &s   &16& -2.486&  0.001& -0.851&  0.002&   65&   19& -249&   37&\\
           &    &18& -2.955&  0.002& -0.726&  0.003&  111&   27& -269&   42&\\
           &s   &20& -3.392&  0.001& -0.746&  0.001& -102&    9& -130&   22&\\
           &    &23& -3.253&  0.003& -1.063&  0.001& -155&   46&  105&   20&\\
           &s   &24& -3.312&  0.001& -1.188&  0.001&   22&   15&   -1&   20&\\
           &s   &25& -3.544&  0.001& -1.268&  0.001&  133&   15&   81&   20&\\
           &    &26& -2.651&  0.004&  0.398&  0.002& -436&   61& -215&   30&\\
           &    &27& -2.995&  0.002&  0.362&  0.002&   38&   24& -201&   25&\\
           &    &28& -3.392&  0.003&  0.362&  0.002& -142&   41& -115&   27&\\
           &    &29& -4.271&  0.002& -1.314&  0.001&   81&   28& -226&   21&\\
           &    &30& -4.370&  0.001& -1.638&  0.001& -139&   20&  -71&   20&\\
           &    &31& -4.049&  0.001& -2.064&  0.001&   93&   22& -145&   17&\\
           &    &32& -1.607&  0.002& -0.514&  0.001&  196&   24&  -57&   17&\\
           &    &33& -1.346&  0.002& -0.342&  0.001&   10&   25& -270&   18&\\
 \hline                                 
\end{tabular}                           
\end{center}                            
\caption{
K$_s$-band positions in arcseconds relative to SgrA* and proper motions in km/s
of IRS13E and IRS13N sources at epoch 2005.83.
The abbreviations s,m,f stand for
Sch\"odel et al. (2005), Muzic et al. (2008) and,
Fritz et al. (2010).
The identifications correspond to those in Fig.~\ref{eckartfig-02}.}        
\label{Tab:13N}
\end{table*}

\begin{table*}[htb]                      
\begin{center}
\begin{tabular}{lrrrrrrrrrrrrrrr}
\hline 
name & 
$\Delta \alpha$ & 
$\pm$&
$\Delta \delta$ & 
$\pm$&
v$_{\alpha}$ &
$\pm$&
v$_{\delta}$ &
$\pm$&
H&
K$_s$&
$L'$&
H-K$_s$&
K$_s$-$L'$
 \\ \hline 
D2, S43    &-0.515 &0.010 &  -0.114 &0.010& +226 & 44 &+480 & 203 &18.7  & 16.5 & 13.5&	2.2& 3.0\\
D6, S79    &+0.699 &0.010 &  -0.520 &0.010& -112 & 20 &+172 &  50 &17.6  & 15.4 & 13.5&	2.2& 1.9\\
S90        &+0.585 &0.010 &  -0.959 &0.010& -127 & 20 & +44 &  60 &17.4  & 15.3 & 14.6&	2.1& 0.7\\
F1         &-0.321 &0.025 &  -1.238 &0.025& -265 & 43 &+260 &  50 &18.3  & 16.0 & 13.7&	2.3& 2.3\\
F2         &-0.048 &0.025 &  -1.546 &0.025& -635 &200 & -77 &  81 &18.0  & 15.8 & 12.9&	2.2& 2.9\\
D3         &-0.623 &0.020 &  -0.045 &0.020& +410 &200 &+331 & 134 &21.3	 & 19.8	& 13.8&	1.5& 5.7\\
D5         &+0.547 &0.020 &  +0.030 &0.020& +419 &150 &+468 & 131 &19.8  & 17.3 & 13.4&	2.5& 3.9\\
D4, S50, X7&-0.540 &0.010 &  -0.520 &0.010&  -52 & 12 &+546 &  15 &19.3  & 17.1 & 12.6&	2.2& 4.5\\
D7         &+1.338 &0.020 &  +1.003 &0.020& -132 & 40 &-318 &  40 &$>$23.4& 18.5& 14.1&	4.9& 4.4\\
DSO        &+0.187 &0.010 &  +0.065 &0.010& -705 & 170&+560 & 160 &$>$21.2 & 18.9 & 14.4&$>$2.3& 4.5\\
\hline
\end{tabular}
\end{center}
\caption{Names, positional offsets from SgrA* in arcseconds,
proper motions in km/s of 
compact infrared excess sources in the central 2'' field close to SgrA*.
The data are given for epoch 2008.}
\label{Tab:ExcessPM}
\end{table*}

\noindent
\begin{figure}
\centering
\includegraphics[width=9.25cm,angle=-00]{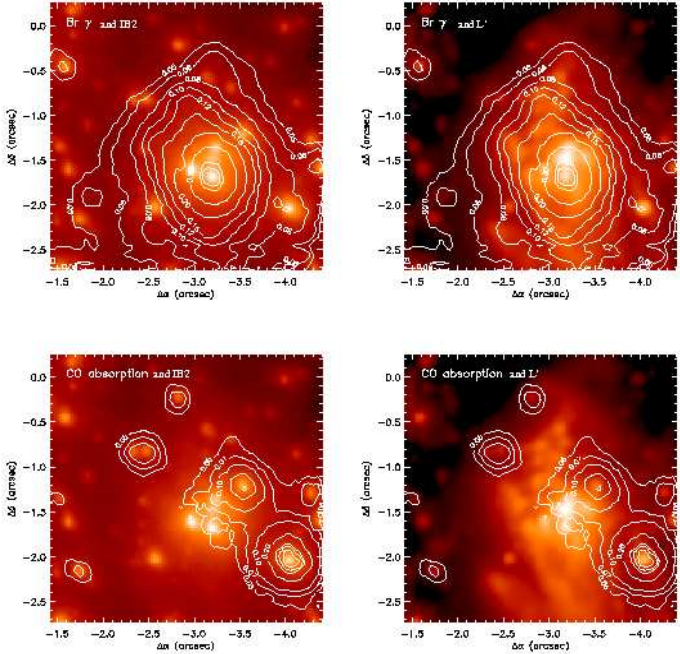}
\caption{\small
2$\mu$m-band image (left; IB 2.15$\mu$m) and 
$L'$-band image (right; 3.8$\mu$m) 
with the Br$\gamma$ emission (top) and 
$CO$ absorption (bottom) line maps in contours overplotted.
}
\label{eckartfig-07}
\end{figure}

\noindent
\begin{figure}
\centering
\includegraphics[width=9.25cm,angle=-00]{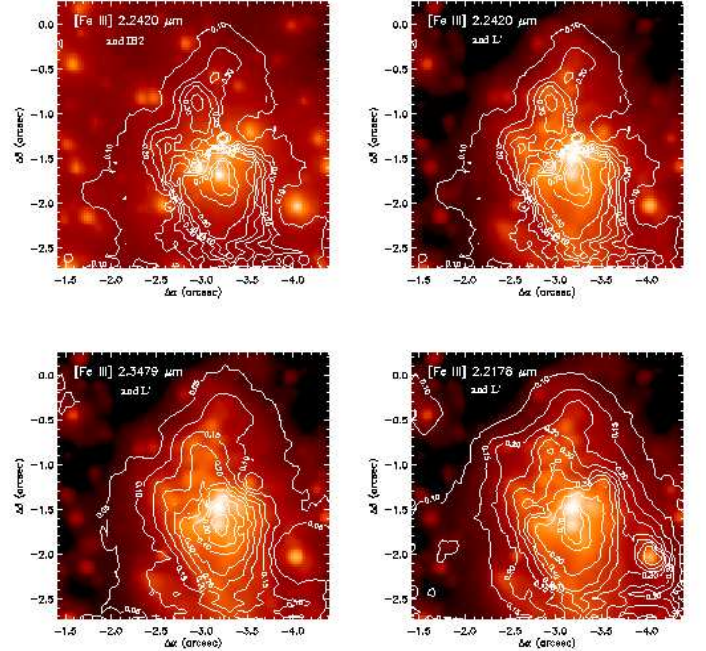}
\caption{\small
2$\mu$m-band image (left; IB 2.15 micron) and 
$L'$-band image (right; 3.8$\mu$m) 
with [FeIII]-emission line maps in contours overplotted
}
\label{eckartfig-08}
\end{figure}

\noindent
\begin{figure}
\centering
\includegraphics[width=9.00cm,angle=-00]{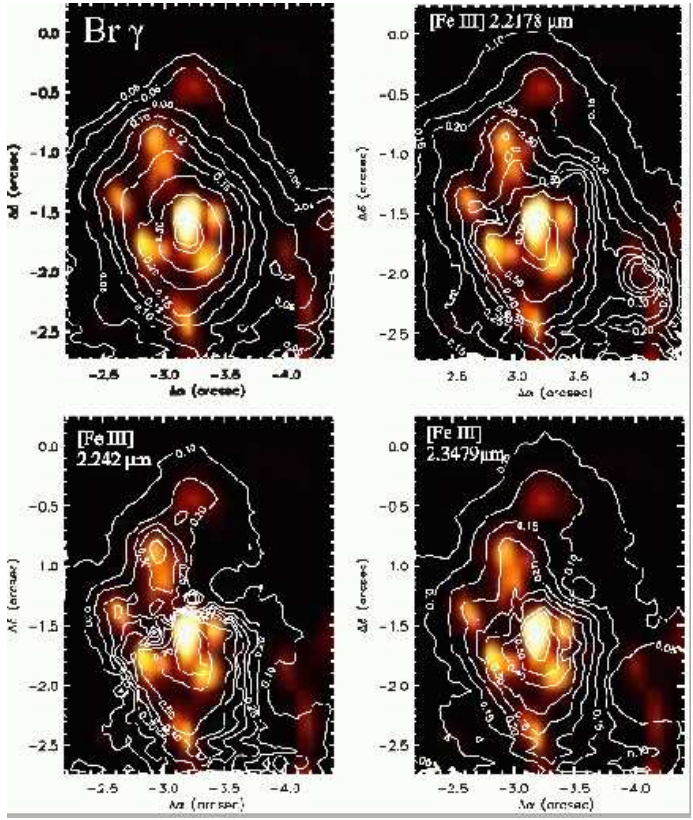}
\caption{\small
13mm VLA radio image from Zhao et al. (2009) with our 
Br$\gamma$- and [FeIII]-emission line maps in contours
overplotted.
}
\label{eckartfig-09}
\end{figure}

\noindent
\begin{figure}
\centering
\includegraphics[width=9cm,angle=-00]{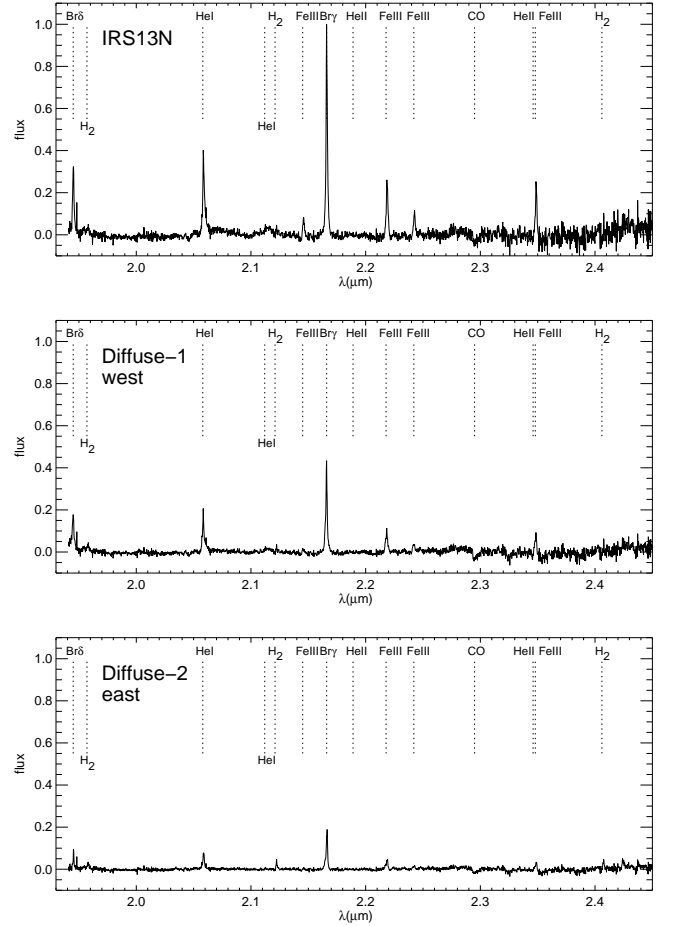}
\caption{\small
Integrated K$_s$-band spectrum of IRS13N 
(top panel), spectrum of the nearby region Diffuse-1 west of IRS13N
(middle panel) and Diffuse-2 east of IRS13N (bottom).
The three spectra were normalized to the peak value of the 
Br$\gamma$ line in the IRS13N spectrum.
The spectra are continuum subtracted.
}
\label{eckartfig-10}
\end{figure}

\noindent
\begin{figure}
\centering
\includegraphics[width=9cm,angle=-00]{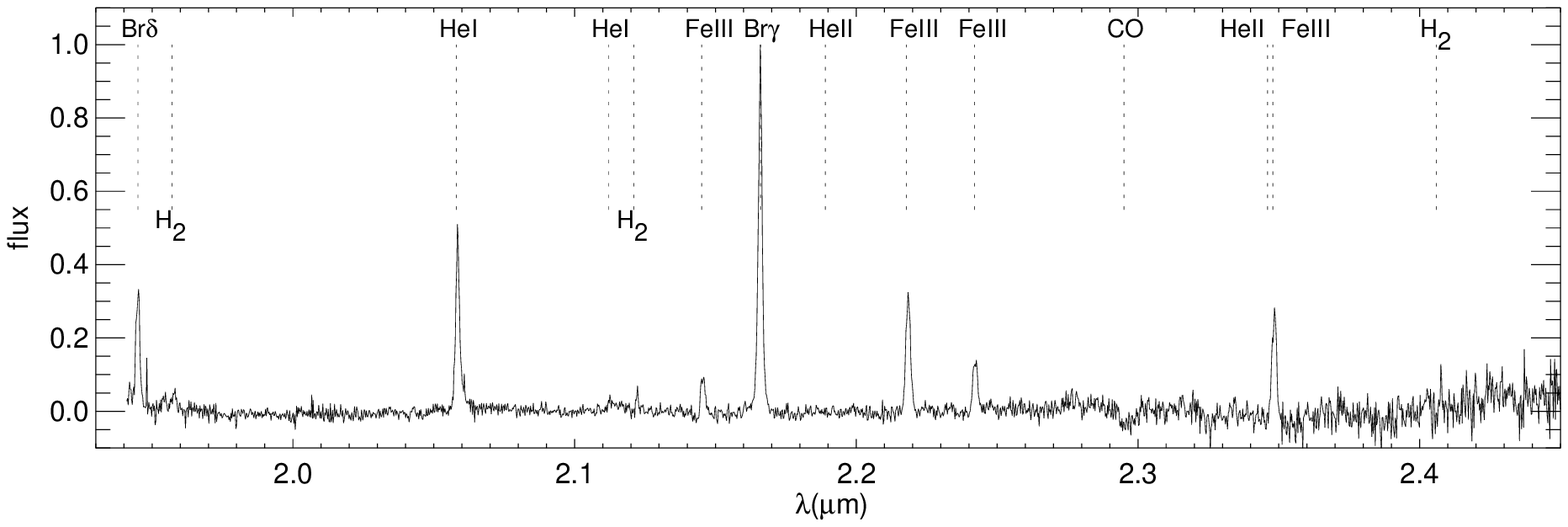}
\caption{\small
Integrated K$_s$-band spectrum of the [FeIII]-emission line source 
at the position of source K21 (Zhao et al. 2009) north of IRS13 N.
This spectrum is continuum subtracted normalized to the peak of the Br$\gamma$ line.
}
\label{eckartfig-11}
\end{figure}

\section{Discussion}
\label{discussion}

\subsection{Identification and proper motions of IRS13N sources}
The derived stellar 1D velocity dispersion at that location
is three times higher than the root mean square difference between the 
proper motions of the sources identified in the
K$_s$ and $L'$-band.
In addition, the difference between our 2005.83 1D 
IRS13N source positions in the K$_s$-band and the $L'$-band positions
by Muzic et al. (2008) at a mean epoch closer to 2005 is
about 20~mas, i.e., two thirds of a K$_s$-band pixel of 27~mas size.
This is on the same order (or below) of what is expected as the 
maximum change in position due to proper motions of the sources 
during the time difference between these mean epochs.
We conclude that over the past 10 epochs we used to analyze the
motion of the IRS13N sources the K$_s$-band sources can indeed be identified
as being the co-moving counterparts of the $L'$-band sources.
This also means that the $L'$-band sources are not unrelated 
(core-less) dust clouds as speculated by Fritz et al. (2010).

It is likely that not all sources $\alpha$ to $\kappa$ that
had been discussed in Eckart et al. (2004) belong to the same
association.
We pointed out already in Eckart et al. (2004) that $\kappa$ has 
significantly bluer colors than the other IRS13N objects.
Infdeed, it has not yet been identified with an $L'$-band source.
Our K$_s$-band proper motion analysis also shows that it is
moving south and west rather than predominantly north as all the
other IRS13N members. From this we conclude that $\kappa$ does not
belong to the small IRS13N cluster of co-moving infrared excess 
sources.

Source $\alpha$ is farthest south of the Eckart et al. (2004) sources
and has also a velocity offset from the other IRS13N sources.
Here we give the $L'$-band velocities derived by Muzic et al. (2008) 
a higher weight, since in K$_s$-band the source is apparently located 
in a crowded region as well as in the wings of the bright IRS13E member E1.
We conclude that $\alpha$ is also probably not a member of the 
small IRS13N cluster of co-moving sources.
We are therefore left with the sources $\beta$ to $\eta$ 
that have been identified as co-moving and whose 
dynamics were analyzed in detail by Muzic et al. (2008).

Source $\eta$ is significantly brighter in $H$-band than all
other IRS13N sources.
It also has a the highest
velocity discrepancy between K$_s$ and L' with a difference of $\sim$150 km/s.
If we assume that the IRS13N sources are embedded in their own dust
disk or envelopes and that they all have a random orientation 
towards the observer, this implies 
an average opening angle of 180$^o$/6=30$^o$ under which the
source can be seen unobscured. From this one may conclude that the 
circum-stellar material almost entirely surrounds the objects.
If the sources have disks, it means either that not all 
material has settled into the disk, or that these disks are 
as thick as they are long in diameter, or that they are significantly warped.
All of this could be expected if the sources indeed belong to a 
dynamically young compact cluster of co-moving young stellar objects.

The velocity dispersions of sources $\beta$ to $\eta$
corrected for their measurements uncertainties is about
45 km/s for the L'-band data discussed by Muzic et al. (2008).
For the less well determined motions derived from our K$_s$-band 
identifications this value is about a factor of 1.5 higher.
The implications for the possibly associated stellar mass of the 
IRS13N association has been discussed in detail by Muzic et al. (2008).
The result is that if the IRS13N cluster is gravitationally bound, 
these velocity dispersions imply a stellar mass that 
should be detectable. Hence, it is likely to be unstable, implying 
that the association and its members are (dynamically) young.

\noindent
\begin{figure}
\centering
\includegraphics[width=8cm,angle=-00]{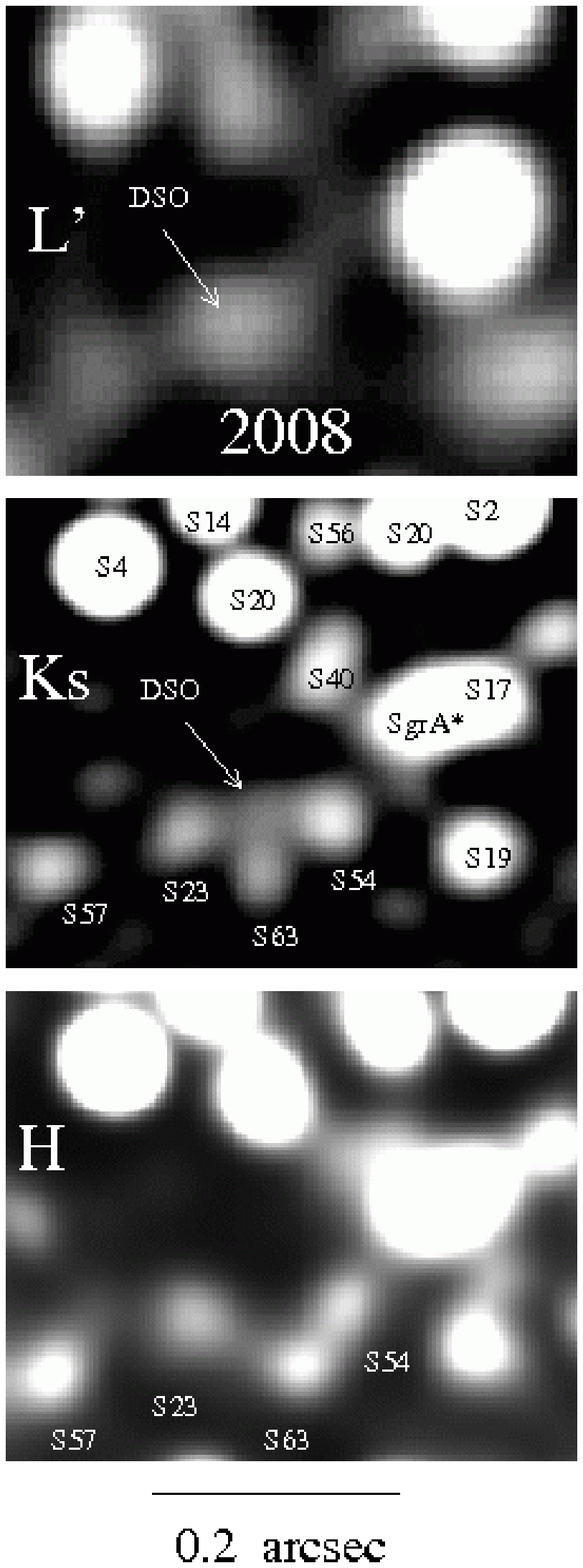}
\caption{\small Distinguishing the DSO against stars 
S57, S23, S54, and S63 in 2008 in L'- and K$_s$-band. 
Only an upper limit can be given in the H-band.
}
\label{eckartfig-35}
\end{figure}

\noindent
\begin{figure}[ht!]
\centering
\includegraphics[width=9cm,angle=-00]{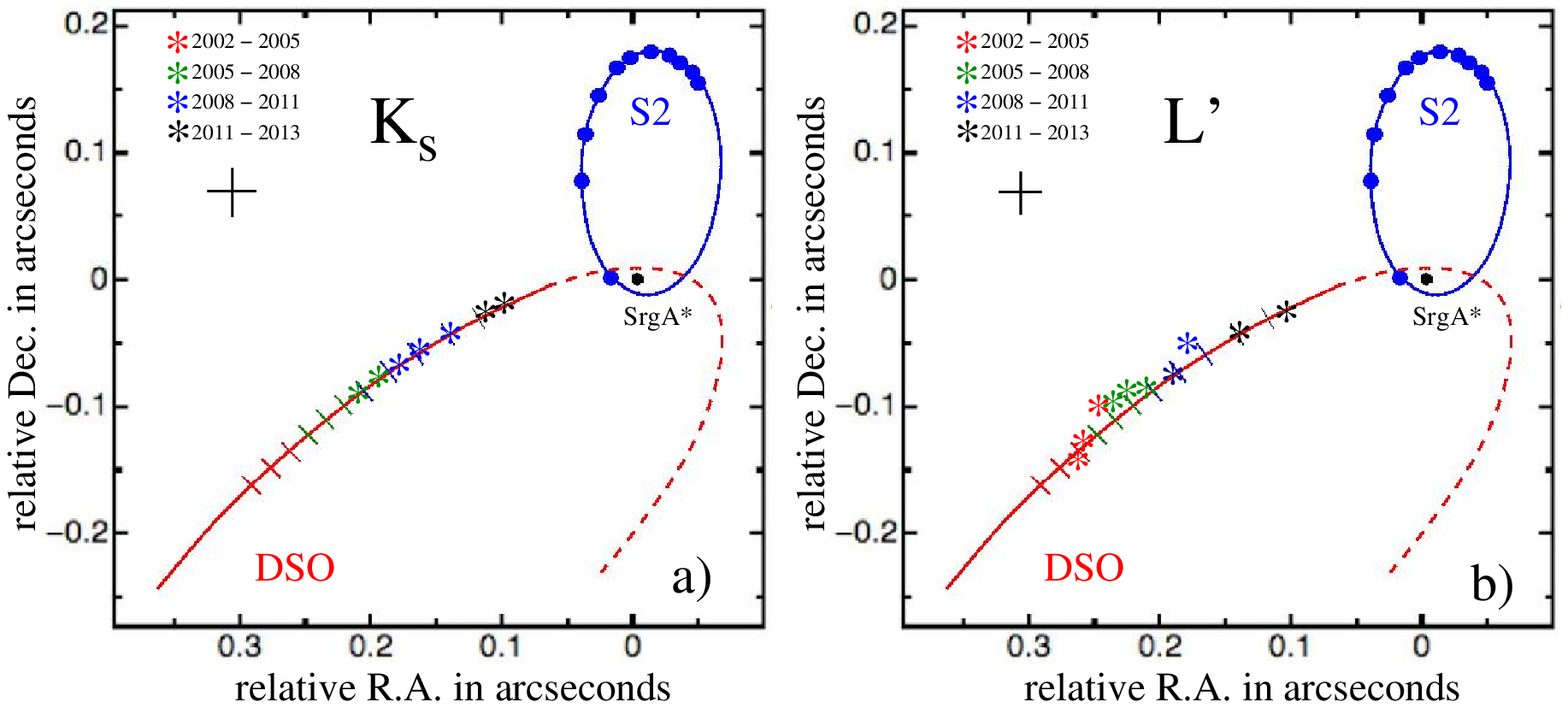}
\caption{\small 
Positions of the DSO relative to S2 and SgrA* 
plotted in comparison to the orbital tracks provided by Gillessen et al. (2012a) 
(red for the DSO, blue for S2, SgrA* is at the position of the black dot at the origin). 
The results for the K$_s$- and L'-band 
are shown in panels a) and b) (see data in Tab.~\ref{Tab:DSO}). 
Thin lines indicate the approximate expected positions of the DSO on its orbit 
from 2002.0 to 2012.0 (based on our L'-band reductions and Gillessen et al. 2012a).
The epochs are color-coded and the measurement uncertainty is shown as a black cross
(see \ref{eckartfig-24} in the appendix).
}
\label{eckartfig-13}
\end{figure}
\subsection{Nature of the IRS13N sources}
\label{resultsIRS13N}                 

In this paper we used the SINFONI line and continuum
images for a qualitative interpretation of the 
nature of the extended and stellar sources only.
A quantitative analysis of the line fluxes is beyond 
the scope of the paper and will be published elsewhere.
The comparison between the Br$\gamma$ line emission and the
near-infrared continuum emission of the stars in the IRS13E/IRS13N field 
in Fig.~\ref{eckartfig-07} shows that the line emission peaks on one of the 
brightest components (E2) of IRS13E, covers the entire IRS13E cluster
and extends to the IRS13N complex.
The strength of the emission is correlated with the presence of
stars in the field and is clearly not only associated with
the stars, but comprises a diffuse component as well.
The maps of the [FeIII]-emission depicted in Fig.~\ref{eckartfig-08} show that
the line flux is peaked 
both on IRS13E (E2) and IRS13N.
The [FeIII] flux also shows contributions from a diffuse component.
In IRS13N the [FeIII] integrated line flux density peaks to within $\le$0.1''
close to sources $\epsilon$ and $\delta$. 
Zhao et al. (2009) found a radio source labeled K23 that lies between 
$\epsilon$ and $\delta$ and has to within 3-4$\sigma$ the same 
velocity as the mean velocity of the NIR counterparts of 
$\epsilon$ and $\delta$.

The integrated [FeIII] line emission is also peaked 
on a source about 0.15'' north of $\eta$ with
offsets from SgrA* of $-3.25``$$\pm$0.03`` 
in right ascension and $-0.45``$$\pm$0.03`` in and declination.
This position agrees with that of 
source K21 identified by Zhao et al. (2009)
to within about 0.1''.
The authors found a proper motion velocity of 
$v_{\alpha}$=-224$\pm$16~km~s$^{-1}$ and
$v_{\delta}$=-421$\pm$20~km~s$^{-1}$ toward the NNW.
In the $L'$-band we do not find a clearly defined point source 
at the position of K21. The continuum  flux density 
extension from IRS13N at that position is consistent with a source
that is about a magnitude fainter (i.e. $L'$$\sim$12) than source $\eta$.
At K$_s$-band no clear association to a source brighter than 
K$\sim$16.3 can be established.
This indicates that K21 is either associated with a stellar source even
more reddened than the IRS13N sources or that the source K21 is a
core-less dust source bright in [FeIII]-emission.
The integrated line emission of Br$\gamma$ is also extended and
covers the position of K21.
A spectrum of the infrared flux density at the position of K21 is shown in 
Fig.~\ref{eckartfig-11}.

The comparison between the CO absorption and the 
near-infrared continuum emission 
in Fig.~\ref{eckartfig-07} shows that the IRS13N complex is clearly located 
at a minimum of the CO-absorption in the IRS13E/IRS13N field.
The absorption is concentrated on a few stars in the field 
to objects about 0.5'' east, west, and north of IRS13N.
The minimum  in the absorption strength centered on IRS13N
comes close to the background value due to the overall central
stellar cluster.
Given the contribution through PSF wings of these stars at the location of
IRS13N, this is fully consistent with the statement that most of the
IRS13N sources (about 0.5'' north of IRS13E) do not show
any significant CO line absorption.

The comparison of the radio data by Zhao et al. (2009) (Fig.~\ref{eckartfig-09})
shows that the radio continuum emission peaks to within $\le$0.10''
at the peaks or contour line excursions of the [FeIII] line emission.
The radio continuum is more peaked on the IRS13E source E1.
The correspondence to the [FeIII]-emission (in particular the 2.242$\mu$m line
emission) is best. The Br$\gamma$ emission is clearly more extended
than the bright radio continuum emission, highlighting that the
radio emission is more dominated by the compact stellar sources than 
the more diffuse emission traced by Br$\gamma$.

The combined spectra on IRS13N and the two off-positions east and west
(Figs.~\ref{eckartfig-03} and \ref{eckartfig-10}) demonstrate that the emission
in the Br$\gamma$, H$_2$ and in particular in the [FeIII] lines is
brighter on IRS13E than in the off-positions.
The spectra also demonstrate that the integrated CO-band head absorption
toward IRS13E is as strong as toward the off-positions. This indicates 
that the CO-absorption at this position can be fully explained 
by the CO-absorption in the PSF wings of the bright nearby late-type stars 
and the overall stellar cluster background.

\subsection{NIR colors and spectra of the IRS13N cluster}
\label{COhead}

The analysis of the HK$_s$L' (i.e. a significant 
section of the SED) colors presented in Fig.16
shows that they - if dereddened -
would not come to lie even close to the positions
expected for colors of pure dust-emitting sources but rather lie at the locations
expected for objects that show a mixture of stellar and dust emission
(see also Eckart et al. 2004).

Compared to the immediate surroundings (except for the IRS13E cluster),
the sources in the IRS13N region are clearly correlated with a minimum in CO absorption 
and an excess in Fe- and H-recombination line emission.
While the excess in emission may in part be due to a local concentration
of gas associated with the 13N association or possibly with the mini-spiral,
the lack of CO absorption can be attributed to the predominantly early 
spectral type, as suggested from the NIR colors of the sources (Eckart et al. 2004). 

While NIR CO absorption and emission is often present,
it is not a necessary feature to be observed toward young massive stars.
Out of 201 young massive stars in M17, Hoffmeister et al. (2006)
found 30\% to have featureless spectra. Half of those show
an infrared excess that corresponds to basically the same percentage 
of the remaining 60\% of the sources that show CO in absorption or emission.
Almost a third of the featureless sources also tend to show 
X-ray emission and a fifth shows simultaneous infrared excess 
and X-ray emission.

Featureless sources have the tendency to be brighter in the NIR
compared to other young stellar objects.
From the distribution in Fig.~3 in Hoffmeister et al. (2006) we
can conclude that featureless objects are typically up to two magnitudes
more luminous compared to CO emission or absorption sources.
They are apparently later than spectral class B3.

Compared to previous CO bandhead studies in star-forming regions,
the investigation by Hoffmeister et al. (2006) is the most extensive 
and a fairly unbiased one. 
Casali \& Eiroa (1996) argued that Class\,II sources tend to show 
CO absorption, while Class\,I sources are featureless. 
The thermal continuum emission from dusty in-falling envelopes 
can exceed the combined stellar and disk photospheric emission by
almost an order of magnitude (Calvet et al. 1997, Hoffmeister et al. 2006).
Hence, this process can result in featureless spectra. 
In addition, the veiling of an envelope will also weaken the 
CO absorption lines.
The presence of envelopes with large-scale heights is consistent with
the presence of a single bright source $\eta$ in addition to five faint objects,
as elaborated above.
The presence of a disk, on the other hand, is likely to amplify the
CO feature. Therefore, Hoffmeister et al. (2006) concluded that 
the 60 featureless objects in their sample of 201 sources in M17 
are very likely Class\,I sources that have just started to build up
an infrared excess that is progressively moving toward the 
MIR and NIR domain.

From model calculations Hoffmeister et al. (2006) infered
that the observed CO absorption is most likely a sign of heavily 
accreting proto-stars. 
The mass accretion rates may be above $10^{-5}\,$M$_\odot$\,yr$^{-1}$.
High accretion can be provided through angular momentum loss within a 
disk.
We conclude that the absence of that feature
may indicate that these stars are currently not strongly accreting 
or have not yet reached that phase.
This would be the case if the luminosity of the in-falling envelope 
is strong against the emission from the disk.

We infer from the combination of the fact that the 
stellar sources are in a co-moving group (Muzic et al. 2008, 
Eckart et al. 2004) and the proper motions 
and the spectroscopy presented here (see below) 
that the IRS13N sources may be featureless 
young stellar objects, like Class\,I sources, in which a 
bright NIR disk has not yet been formed but in which the NIR/MIR
excess is still dominated from surrounding and in-falling material.

\subsection{Age of the IRS13N sources}
\label{IRS13Nages}

Possible stellar identifications of the IRS13N sources have already been
discussed extensively by Eckart et al. (2004) and Muzic et al. (2008).
The fact that our data supports the finding that the sources 
($\beta$ to $\eta$; see below) are indeed co-moving poses serious 
problems to source identifications that involve stars with ages $>$3~Myr.
Here we highlight a few more aspects concerning the identification 
of the sources with young stars:

Broad-band colors can give information on the nature of sources.
In Fig.~\ref{eckartfig-12}  (left) we compare the colors of dusty objects in the 
central few arcseconds of the Milky Way to the colors of young 
stellar objects as given by Hoffmeister et al. (2006).
The sources of IRS13N are more highly extincted than the remaining 
dusty objects in the field.
Their dereddened colors agree with those young 
dust enshrouded objects (details in Eckart et al. 2004).
The assumption (see section~\ref{COhead})
that the IRS13N cluster members are Class~I sources
implies that they are young.
In a recent study of stellar and circumstellar properties 
of six Class I proto-stars Prato et al. (2009) estimated their ages 
as $<$2~Myr.
However, even the ages of young Class~II/Class~III sources are 
estimated to range between 0.1 and 1~Myr
(Green \& Meyer 1995, Luhman \& Rieke 1999, see also
van Kempen et al. 2009).
This is consistent with the assumption of Eckart et al. (2004) that 
the IRS13N sources are good candidates for being YSOs or 
young Herbig Ae/Be stars with ages that may range between 0.1 
and about $\sim$3~Myr.
Molinari et al. (2008) found that the time scales of the formation of stars 
are about 1-4$\times$10$^5$ years, with
shorter times for higher masses (6 to 40 \solm)
(Hillenbrand et al. 1992, Fuente  et al. 2002, Ishii et al. 1998).
The orbital time scale for the combined motion of the entire 
IRS~13E/IRS13N complex and the mini-spiral gas of about 
10$^{4}$ years (see also Muzic et al 2008) is short compared to the 
plausible ages of the massive young stars.
We also assume that their associated envelopes and 
disks can survive the hostile GC environment for at 
least 10$^4$ years (see discussion in Scally \& Clarke 2001).

\begin{table*}[htb]
\begin{center}
\begin{tabular}{rrrrrrrrr}\hline 
epoch & DSO &  DSO & DSO & DSO & S2 & S2\\  
      & K$_s$ &  K$_s$& L' & L' & HK$_s$L' & HK$_s$L'\\ 
     &
$\Delta$$\alpha$ &
$\Delta$$\delta$ &
$\Delta$$\alpha$ &
$\Delta$$\delta$ &
$\Delta$$\alpha$ &
$\Delta$$\delta$  \\ \hline
   2002.42  &       &           & 0.255 &-0.130 &   0.020 &    0.000 \\
   2003.54  &       &           & 0.249 &-0.097 &   0.041 &    0.079 \\
   2004.66  &       &           & 0.255 &-0.121 &   0.034 &    0.125 \\
   2005.46  &       &           &       &       &   0.027 &    0.147 \\
   2005.83  &       &           & 0.232 &-0.095 &   0.024 &    0.150 \\
   2006.55  & 0.213 &   -0.094  & 0.225 &-0.082 &   0.011 &    0.168 \\
   2007.33  & 0.196 &   -0.074  & 0.208 &-0.083 &   0.000 &    0.176 \\
   2008.49  &       &           & 0.186 &-0.073 &  -0.015 &    0.179 \\
   2008.56  & 0.187 &   -0.065  &       &       &  -0.016 &    0.179 \\
   2009.47  & 0.165 &   -0.055  & 0.182 &-0.048 &  -0.026 &    0.177 \\
   2010.53  & 0.147 &   -0.033  &       &       &  -0.035 &    0.173 \\
   2011.55  & 0.120 &   -0.022  & 0.144 &-0.040 &  -0.044 &    0.167 \\
   2012.54  & 0.101 &   -0.017  & 0.108 &-0.023 &  -0.053 &    0.161 \\
\hline 
\end{tabular}
\end{center}
\caption{K$_s$-, and L'-band positions of the DSO 
and the star S2 relative to the SgrA* position. 
Typical uncertainties range between 13~mas and 25~mas for the DSO and 
(see Fig.~\ref{eckartfig-24})
less than 10~mas for the position of S2.
}
\label{Tab:DSO}
\end{table*}

\begin{table*}[htb]
\begin{center}
\begin{tabular}{lrrrrrrrr}\hline
name &
flux &
epoch &
$\Delta$$\alpha$ &
$\Delta$$\delta$ &
pm$_{\alpha}$~~~ &
pm$_{\delta}$~~~ &
acc$_{\alpha}$~~~ &
acc$_{\delta}$~~~ \\
     &
ratio &
year &
mas &
mas &
mas yr$^{-1}$ &
mas yr$^{-1}$ &
mas yr$^{-2}$ &
mas yr$^{-2}$ \\ \hline
S23 & 1.01 & 2005.47&  307.4&   -89.1&  14.81&   11.17&  -0.953&   0.525\\
S57 & 1.24 & 2007.46&  393.6&  -147.4&   9.99&    4.05&   0.000&   0.000\\
S63 & 0.83 & 2004.00&  245.0&  -150.0&  13.00&   -3.60&   0.000&   0.000\\
DSO & 0.30 & 2002.00&  295.0&  -160.0&  17.60&  -14.00&  -0.160&  -0.240\\
S54 & 1.36 & 2006.59&  115.4&   -60.1&   1.05&   26.90&   0.000&   0.000\\
\hline
\end{tabular}
\end{center}
\caption{Epochs, flux ratios, positions, proper motions, and accelerations used to simulate the DSO and
the neighboring stars with results shown in 
Figs.~\ref{eckartfig-14-1}, \ref{eckartfig-14-2}, and \ref{eckartfig-14-3}.
The flux ratios have been calculated with respect to the mean flux derived (m$_K$=17.6) from the K$_s$-band 
magnitudes of S23, S54, S57, and S63 as given by Gillessen et al. (1999). 
They correspond (with an estimated uncertainty of 30\%) to the
relative fluxes suggested by the modeling presented in the figures listed above.
From the quality of the fit in these figures we estimate that the uncertainties 
are 10 mas for the coordinates, 1 mas~yr$^{-1}$ for the proper motions,
and  0.1 mas~yr$^{-1}$ for the accelerations.
}
\label{Tab:DSOsim}
\end{table*}

\noindent
\begin{figure}
\centering
\includegraphics[width=9cm,angle=-00]{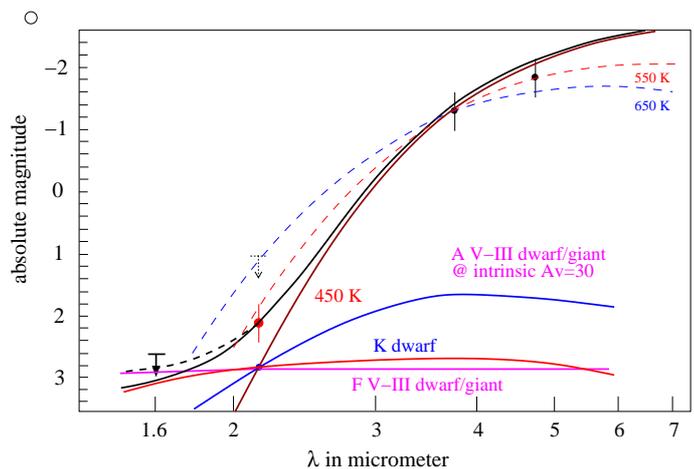}
\caption{\small 
Decomposition of the DSO spectrum including our K$_s$-band detection and
H-band limit. 
Here we demonstrate that a mixture of dust and stellar contribution is possible.
The points correspond to the L- and  K$_S$-band magnitudes, and H-band upper limit from this work, 
and the M-band measurement and K-band upper limit of Gillessen et al. (2002). 
Red and blue dashed curves also show their 550\,K and 650\,K warm dust fits. 
In solid blue, red, and magenta lines the emission from three different possible stellar 
types of the DSO core are plotted. Any of these stars embedded in 450\,K dust (solid brown line) 
can produce the black line that fits all the NIR DSO photometric measurements.
}
\label{eckartfig-33}
\end{figure}

\noindent
\begin{figure*}[ht!]
\centering
\includegraphics[width=19cm,angle=-00]{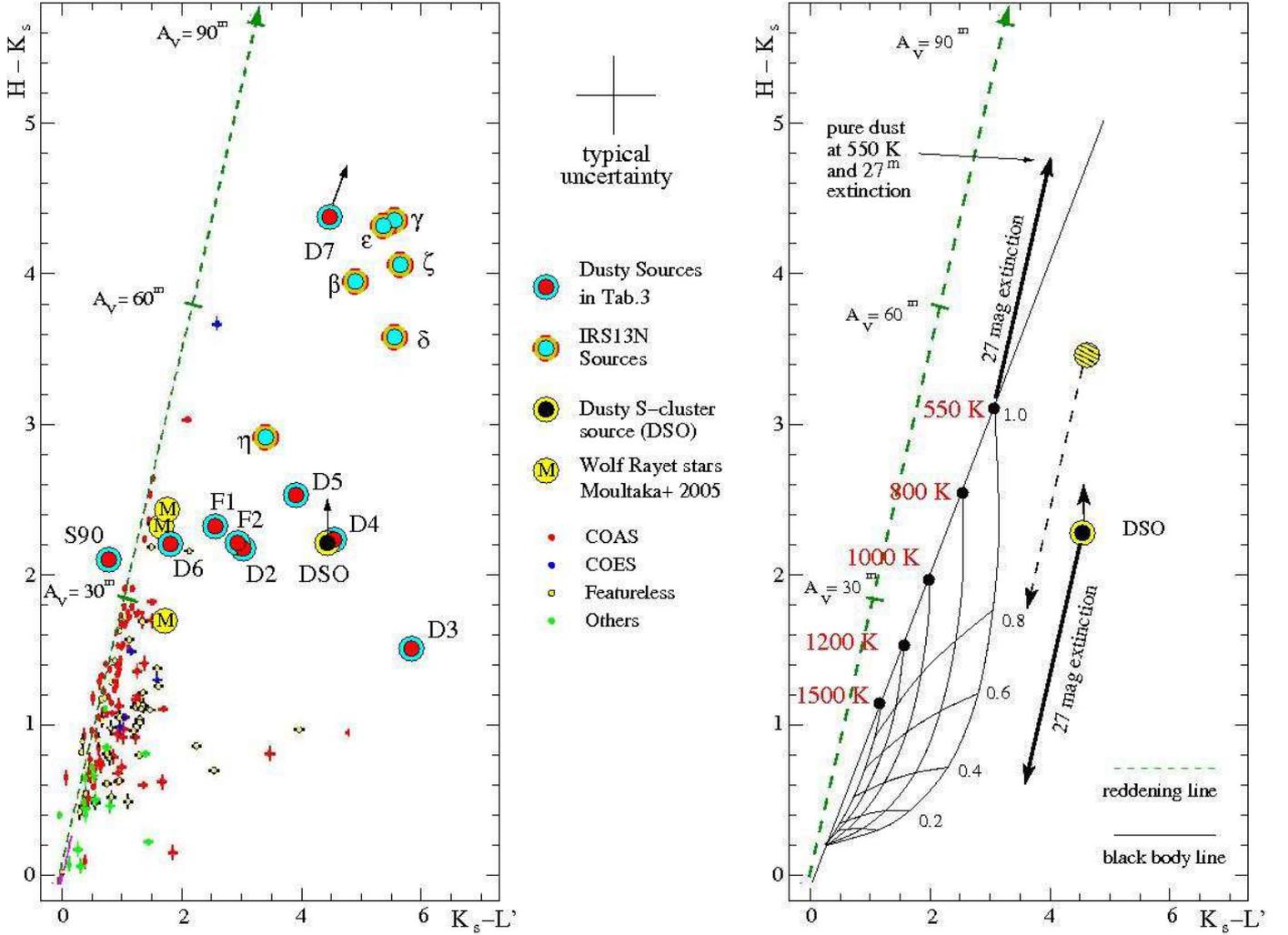}
\caption{\small 
Left: Colors of dusty objects in the central parsec of the Milky Way compared to the colors of
young stellar objects as given by Hoffmeister et al. (2006).
The labels COAS and COES indicate that the sources show CO absorption  or emission in their spectra.
Right: Colors of dusty objects in the central parsec of the Milky Way 
compared to colors of pure dust (temperatures given in red numbers) 
and mixtures of a stellar and a dust contribution (fractional numbers given in black;
see also Glass \&  Moorwood 1985).
Here we demonstrate that a mixture of dust and stellar contribution is possible.
Because the H-K$_S$ color of DSO is an upper limit, its NIR emission can be explained 
as pure dust at $550\,{\rm K}$ reddened by $\sim 27\,{\rm mag}$, or as a star embedded in 
warm dust, e.g., if its color were detected to be ${\rm  H-K_S\sim 3.5\, mag}$ (dashed circle), 
after dereddening by 27\,mag, its emission would be consistent with 20\%stellar and 80\% dust 
contributions.
}
\label{eckartfig-12}
\end{figure*}

\noindent
\begin{figure*}
\centering
\includegraphics[width=15cm,angle=-00]{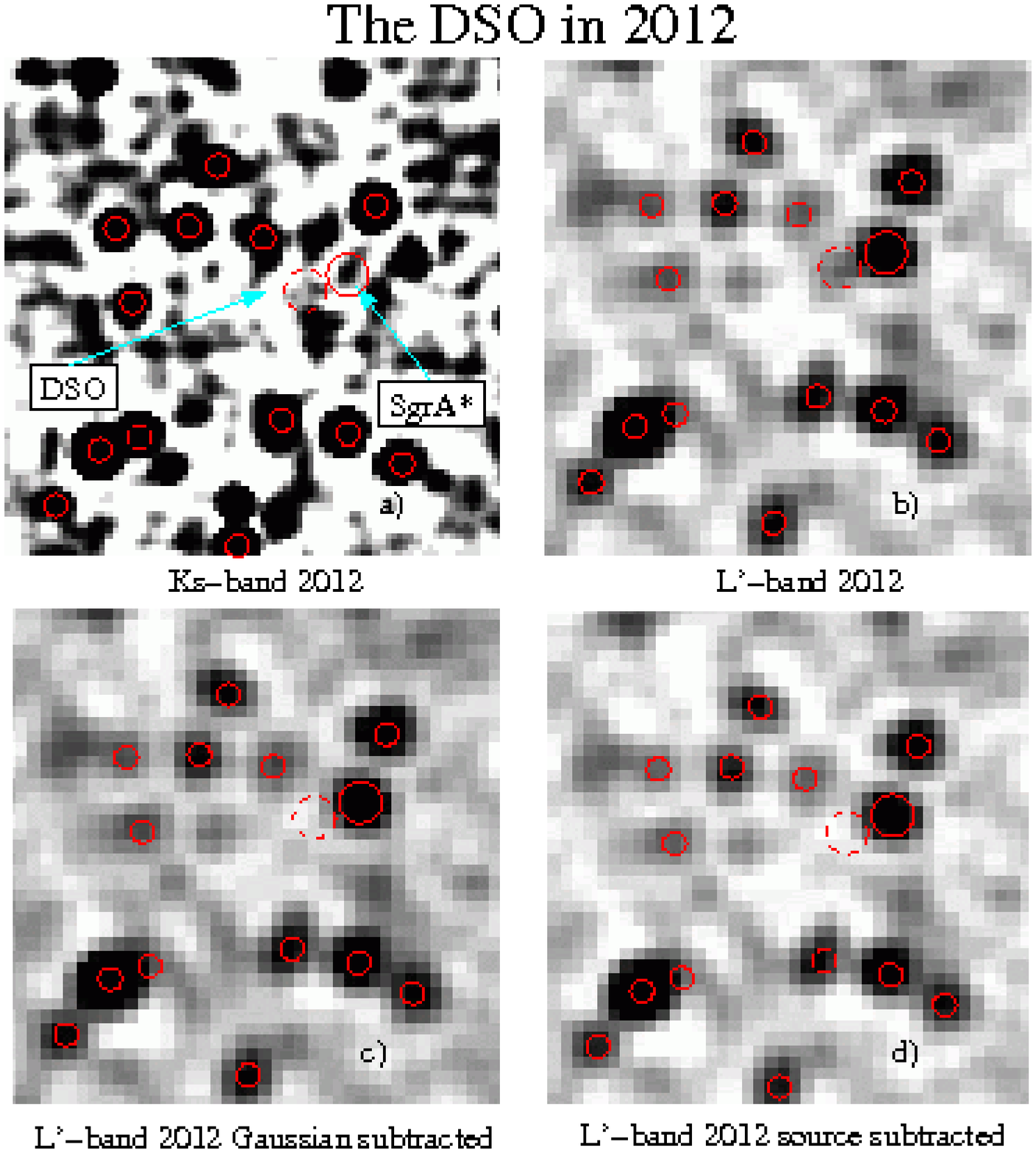}
\caption{\small Comparison of DSO images in the a) Ks- (13.2~mas pixel scale) 
and b) L'-band (27~mas pixel scale).
The image section size is 1''$\times$1''.
In c)  and d) a Gaussian of 98~mas FWHM and a PSF from the 
image have been scaled to the DSO peak brightness and subtracted at 
its position.
An astrometric grid of red circles 
marks some bright stars that can be used as positional references
to compare the images at different wavelengths.
}
\label{eckartfig-28}
\end{figure*}

\subsection{Broad-band spectrum of IRS13N}
Inspection of the 1.2'' angular resolution 1.3mm map of the 
mini-spiral presented by Kunneriath et al. (2011)
shows that the IRS13N complex is associated with 1.3mm continuum emission.
Just north of IRS13E it shows itself as a compact source component with a 
flux density of 20$\pm$3~mJy 
(see Fig.~\ref{eckartfig-16}). 
Arithmetically de-convolved with the beam, its size is less than 0.6~arcsec. 
In NE direction it is partially blended 
with IRS13E and a northern neighboring mm-source.
In the 13mm map presented by Zhao \& Goss (1998; Fig.3/PLATES therein) we can read off the 
map in their Fig.~3 an extended flux density of 11$\pm$2mJy with a compact 
source of about 3mJy.
These flux densities are consistent with a dominant nonthermal stellar wind contribution
with an approximate mass load of 10$^{-5}$ \solm~yr$^{-1}$ 
(e.g. Montes et al. 2009; Lang et al. 2005).
This a mass load is typical for young luminous stars
(Drake \& Linsky 1989, Kudritzki et al. 1999, Crowther 2001).

Eckart et al. (2004) quoted H, K$_s$ and $L'$ flux densities for individual IRS13N sources.
Viehmann et al. (2005, 2006) gave NIR/MIR flux density values 
obtained in an $\approx$1'' diameter circular aperture on IRS13N.
In the J- and $H$-band and probably even in the K$_s$-band 
these 1'' aperture flux densities are dominated
by confusing fore- and background sources of the central cluster
and the values by Viehmann et al. (2006) can probably only be taken as an upper limit.
Due to the very red colors of the sources the 1'' flux densities at longer
wavelengths are most likely characteristic for the IRS13N sources. 
However, at these wavelengths dust emission becomes dominant.
Our investigation of the K$_s$- and $L'$-band proper motions shows that the 
dust-emission-dominated $L'$-band flux densities can be fully associated with the
stellar sources found at shorter wavelength. Hence, we assume that a dominant portion
of the emission longward of the $K_s$-band can in fact be attributed to the IRS13N sources.

\subsection{Proper motions and nature of the central dusty sources}

Examination of the proper motions of the dusty sources in the central
few arcseconds shows that 5 of the 10 L'-band excess sources identified in 
Eckart et al. (2004, see Fig.~\ref{eckartfig-01}) have K$_s$-band counterparts.
The color-color diagram in Fig.~\ref{eckartfig-12} (left) shows that the colors 
of dusty objects in the 
central few arcseconds of the Milky Way are bluer than those of the 
IRS13N sources.
Correcting for the Galactic foreground extinction of approximately 
27 magnitudes in the visible
places them at the location of the young stars discussed by Hoffmeister et al. (2006).
The IRS13N sources are more highly extincted than the remaining dusty objects in the field.
Correcting for the Galactic foreground extinction places them at the
location of extincted young stellar objects in Fig~\ref{eckartfig-12}i (left).
Sources $\beta$ to $\eta$ in IRS13N are therefore excellent candidates 
for being young stars with an average extinction around $A_V \sim 27$.

\subsection{Pure dust versus photospheric emission}
\label{pure}
Here we investigate if the NIR emission of the infrared excess sources can be
caused by pure dust sources or if significant contributions from photospherice 
emission of stars need to be included.
If the objects were pure dust sources their observed colors must be 
consistent with reddened colors of pure dust sources. 
For the IRS13N this is apparently not the case.
The DSO is clearly detected in the K$_s$-band at m$_K$=18.9.
Using A$_K$=2.22 and approximating $L_K$[\solar] with
$1.1 \times 10^4 \times D[Mpc]^2 \times S_K [mJy]$ that results into
$S_K=0.13 mJy$ and $L_K$ [\solar] =92$\times$10$^{-3}$\solar.
This is 46 times more than the luminosity in the Br$\gamma$ line
reported for the head of the DSO by
Gillessen et al. (2012b).
This implies that the 2$\mu$m luminosity is dominated by the
continuum rather than the Br$\gamma$ line emission.
In the H-band we find an upper limit m$_H$$>$21.2 if we consider the 
background level from the neighboxuring stars S23, S54i, and S63 in 2008, and the
3$\sigma$ variation at the position of the DSO in the K$_s$- and L'-band.
With the DSO L'-band brightness of m$_{L'}$=14.4$\pm$0.3 obtained from the 
observing epochs listed in Tab.~\ref{Tab:useddata} we obtain
H-K$_s$=2.3 as a lower limit (Fig.~\ref{eckartfig-12} right).
This results in a measured color K$_s$-L'=4.5.
With A$_H$=4.21, A$_{K_s}$=2.22 and A$_{L'}$=1.09,
A$_{M}$$\sim$A$_{L'}$
(e.g. Viehmann et al. 2005, Lutz et al. 1996)
the absolute magnitudes can be found and  plotted in a spectrum
(Fig.~\ref{eckartfig-33}).
In Fig.~\ref{eckartfig-12} (right) we compare the data to the black-body line
and extincted colors of different mixtures between a stellar and a dust contribution.
(Glass \&  Moorwood 1985).
The pure 550~K dust colors (K$_s$-L'$\sim$3.0) reddened by A$_V$=27  
are consistent with the K$_s$-L' color of the DSO.
However
(with the current H-K$_s$ color limit
and with potential H-K$_s$  colors that range up to the pure 550~K color),
the dereddened positions of the DSO 
show that the DSO colors are also consistent with a mixture between a stellar and a 
dust contribution.
Withn its current (H-K$_s$ color limited) position in the diagram, 
mixtures with 60\% stellar emission and 40\% dust with temperatures less than 550~K
are possible.
If the DSO will be detected half way between its current H-K$_s$ and that of pure 
550~K dust (striped circle in Fig.~\ref{eckartfig-12}), 
20\% stellar and 80\% 550~K dust contribution are possible.
In Fig.~\ref{eckartfig-33} we show a possible spectral decomposition 
of the DSO using the M-band measurement by Gillessen et al. (2012a).
In addition to their 550~K and 650~K fit and their m$_{K_s}$=17.8 limit (dashed lines) 
we plot our K$_s$-band  DSO detection and the H-band limit.
The fit shows that the integrated flux can amount up to about 30~\solar.
We show a decomposition in which we assume that 50\% of the current K$_s$ band flux
is due to a late dwarf (blue line) or an AF giant or AGB star (magenta line).
We used dust at a temperature of 450~K plotted in red (see dereddening of the 
DSO in the color color diagram in Fig.~\ref{eckartfig-12} right).
The sum of the spectra is shown by the thick black line (dashed at short 
wavelengths for the AF giant/AGB case).

Similarly, after dereddening the IRS13N sources for A$_V$=27$^m$, they miss the 
pure black-body line by about two magnitudes (see Fig.~\ref{eckartfig-12} left).
For fully dereddened objects an offset of two magnitudes in K$_s$-L' calls for
temperatures of about 800~K or below with flux density contributions of a pure dust 
component of less than 50\%
(see also Fig.7 by Glass \&  Moorwood 1985).
This temperature estimate is consistent with the overall IRS13N temperature 
of around 650~K given by Fritz et al. (2010).
This clearly implies that the IRS13N sources cannot be pure dust emitters
but must have a stellar core. This is also fully consistent with the fact that the colors
of these objects are similar to those of early dust enshrouded stars with colors shown
the diagram in Fig.~\ref{eckartfig-12} (left).
Hence, for IRS13N sources the evidence for photospheric emission from
a star is strongly supported by the H-band detections (Eckart et al. 2004 Tab.2 -
here all sources except $\epsilon$ and $\gamma$ could be identified).

Inspection of Fig.~\ref{eckartfig-12} (left), the black-body line shown in 
Fig.~\ref{eckartfig-12} (right) and comparison to Fig.7 by 
Glass \&  Moorwood (1985) shows that within the uncertainties
the sources F1 and D7 are good candidates for being pure dust sources.
Sources D6 and S90 will have mostly stellar colors after dereddening with 
a possible contribution of dust $>$1000~K.
Sources F2, D2, D3, D4, and D5 are clear candidates for being objects that show a 
mixture from photospheric emission and emission from dust with 
temperatures well below 1000~K.

\subsection{Statistical robustness of the DSO identifications}
\label{robust}

Here we analyze the statistical robustness of the DSO identifications 
in the H- and K$_s$-bands.
In Sabha et al. (2012) and Eckart et al. (2012) we tabulated the probability for
finding a blend star above the confusion limit in a single K$_s$-band spatial
resolution element in the overall region of the S-star cluster and at the
position of SgrA*. 
With two independent methods we determined the central K$_s$-band confusion limit observationally.
Sabha et al. (2010, 2012) employed successive subtraction of stars in the S-star cluster.
Witzel et al. (2012) used the shape of the SgrA* flare amplitude histogram.
The otherwise straight power spectrum with which this histogram can be described
turns over at a flare flux density at which a mere flare detection becomes a 
flare flux density measurement (details in Witzel et al. 2012).
This indicates a confusion limit in the range of 0.5 to 1.5~mJy.

The result of the statistical investigation of the S-star cluster by
Sabha et al. (2012) is that for a range of observationally supported parameters 
that describe the structure and luminosity function of the cluster, 
the authors can give the probability of finding a blend star in a 
single spatial resolution element above the confusion limit.
Here we use a power law index of the projected spatial distribution of 
$\alpha$=0.30$\pm$0.05 and a K-band luminosity function exponent 
of KLF=0.18$\pm$0.07.
Projecting the S-star cluster onto the sky blend stars are created by line of sight 
overlaps of fainter stars within a finite angular resolution element. 
The probability of finding blend stars 
in the central 1.3$\times$1.3 arcsec$^2$ overall
can be as high as 1.0 and 
reaches a value of about 0.5 at the position of SgrA* itself. 
In the following we use these values although
they can be much lower if the immediate vicinity of 
SgrA* is emptied of low-mass stars through stellar dynamical processes.

The probability of finding a K$_s$-band source in an
L'-band resolution element (that corresponds to 2.8 K$_s$-band resolution elements)
at a specific predicted position within the S-cluster is about 170 times lower than unity, 
i.e., 6$\times$10$^{-3}$. As explained in Sabha et al. (2012), the live time of these blend stars
(also indicated through observations) is about 3 years during which they dissolve 
again into individual objects with flux densities below the confusion limit
due to their proper motions.
If one aims to explain the 2007-2012 K$_s$-band identifications of the DSO 
by blend stars, one requires about 2-3 of these transient apparent objects.
This happens with a probability of 3$\times$10$^{-7}$.
Since there is no guarantee that apparent proper motions during the dissolving phase 
of the blend stars follow the observed orbit and since we have chosen upper limits
to start with, these low-probability values can be regarded as a safe 
upper limits for a scenario in which the DSO can be explained as 
the result of blend stars.
Even if a blend star is assumed, one can think of at least 5-10 proper motion velocity values 
and 5-10 directions that would be expected given the accuracy of 13-25~mas and the involvement of 2 to 6 epochs.
This will lower the probabilities quoted above by a factor of 25-100.
This implies that the combined probability of being confused by a comoving blend star 
is less than 10$^{-4}$.

However, the number of bright, single stars in the central region is high.
We can therefore consider the likelihood of finding faint stars above the
confusion limit at the position of the DSO over a time span of several years.
From the KLF analysis by Sabha et al. (2010, 2012) we infer 
a maximum of about 40 stars fainter than 16 magnitudes (dereddened) within the central 1.3''x1.3''.
This implies 0.2 sources per resolution element and observing epoch. For six consecutive epochs 
this results in a probability of about a percent to find a random configuration of the
central cluster such that a faint star is found at consecutive positions given by the L'-band detections.
If we include the immediate neighborhood (i.e., 9 to 25 surrounding resolution elements), the 
probability can rise to 25\%.
With a limited accuracy of determining the direction and value of the proper motion,
this implies that the likelihood of finding a star that is close in position and velocity 
is in the percentage range.
This may explain the confusion of the DSO with S63 in the interval 2002 to 2007.

In summary, the contamination with a bright field star over the entire time interval 2002 to 2012
is likely (around a few percent)  but being confused by blend stars is very unlikely ($\le$10$^{-4}$).
For the IRS13N and all other infrared excess sources discussed here the projected stellar 
densities are even lower than those close to the center, such that serendipitous 
identifications are also not very likely.

\subsection{Proper motions and identification of the DSO}

The analysis presented here shows that the dusty S-cluster object 
reported by Gillessen et al. (2012a) does indeed have a K$_s$-band 
counterpart.  
Its H-K$_s$-color limits indicate that it cannot be excluded that the DSO is
an embedded star rather than a pure dust cloud.
Gillessen et al. (2012a) showed that both in proper 
motions and radial velocities the trajectory 
of the source perfectly fits a closed elliptical orbit. 
The authors interpreted this source as a dust cloud 
that is approaching SgrA* and will be disrupted during 
the peri-bothron passage.  
The combination of the fragility of the cloud and the 
fact that it is on a bound elliptical orbit implies that
the object must have been on that orbit for a
couple of revolutions. 
In the harsh environment of the 
central stellar cluster the evaporation time scale 
for a pure dust cloud is incompatible with this fact.
The dust cloud would disappear before it can establish 
a relaxed orbit.
If the object is on a bound orbit as a cloud and 
since it will be heavily stretched 
out by tidal forces during the peri-bothron passages
it should not be as compact as it currently appears.
Variations in the shape should (with lower angular 
resolution than in the H- and K$_s$-band) be traceable in the bright
L'-band continuum where it appears to be rather compact
(Gillessen et al. 2012a).
In the K$_s$-band the DSO is close to the confusion limit
(see discussion in subsection \ref{robust}).
Given the strong and extended Br$\gamma$ recombination line emission 
from the mini-spiral the situation may be similar for the DSO Br$\gamma$ line emission.
As in the case of the continuum identifications this may influence the
brightness, position, shape, and velocity field of the source as seen in
the light of the line emission (Gillessen et al. 2012a).

If the DSO is indeed a dusty star, then it 
may develop a bow-shock if it passes through an accretion wind 
from SgrA* quite similar to the sources X3 and X7 described by 
Muzic et al. (2010).  
Since the DSO as a dusty source is an obvious probe for
strong winds possibly associated with SgrA*. 
Its presence and the 
fact that it has not yet developed a bow-shock, although it is
already closer than X3 and X7, indicates that the wind from SgrA* is
highly un-isotropic and possibly directed toward the mini-cavity
(see Scenario IV below and Muzic et al. 2010 for a more detailed discussion).

Comparing the luminosity and the magnitude of the DSO (Tab.~\ref{Tab:ExcessPM})
we obtained in the H- and K$_s$-band
to the absolute magnitudes of OB and Be stars as presented by
Wegner (2006) and stars of luminosity class I-V, 
we found that the DSO could preferentially be a K-dwarf (V) or an
F-dwarf/subgiant (V-III) at no additional extinction intrinsic to the
surrounding dust cloud or an A-dwarf/subgiant (V-III) at about 30 magnitudes of
intrinsic visual extinction (see Fig.~\ref{eckartfig-33}).
The AGB or early evolutionary phase may be included 
here to support the dusty envelope.
These identifications point toward a slightly later type as compared to the 
earlier-type B0-2.5~V stars in the S-cluster (Martins et al. 2008). 
This would imply a stellar mass of a few solar masses. 
However, the combination of local extinction (including geometrical effects 
like shadowing and volume-filling factors of the extincting material) and scattering 
may alter that identification and it could be even earlier or more luminous as well.
This leads us to discuss four additional scenarios:
\\
\\
{\it Scenario I:}
Since the DSO is prominent in its hydrogen line emission 
(Gillessen et al. 2012a) with a Br$\gamma$ line with of 100~km/s in 2008
but rather faint in the H and K$_s$ continuum 
bands (see Tab.~\ref{Tab:ExcessPM}),
it may very well be a dusty, faint narrow-line Wolf-Rayet (WR) star. 
They are known to be less luminous than most other WR or OB giants,
and there are a few of these objects in the GC field
(see e.g. Moultaka et al. 2005, Paumard et al. 2006).
The WR stars 
IRS7W with m$_K$=13.1, IRS7SE with m$_K$=13.0 (Martins et al. 2007)
and WR2 with m$_K$=12.9 (Moultaka et al. 2005)
are amongst the faintest known so far in the Galactic Center cluster
(for faint WR stars in the Milky Way see also  Shara et al. 1999, 2012,
Mauerhan, van Dyk, Morris 2011 with survey limits around m$_K$=15.0).
Given these apparent magnitudes, the DSO would have to be an exceptionally faint WR star.
The earliest WR stars with dense winds and strong free-free excesses
may belong to the intrinsically faintest of the WRs
(Mauerhan, Van Dyk, Morris 2011).
As indicated by the increasing Br$\gamma$ (Gillessen et al. 2012a), the harsh
environment of the S-star cluster and the possibly repeated interaction with the
SgrA* black hole may have altered the stars mass loss or shell size and hence its
luminosity in the thermal infrared and the Br$\gamma$ line width of an object
that started out as a narrow-line WR star at larger separations from SgrA*.
\\
\\
{\it Scenario II:}
Dong, Wang \& Morris (2012) have carried out a 
multi-wavelength study of evolved massive stars in the GC.
In their Fig.~12 they show a Pa$\alpha$ equivalent line width 
(EW) vs. K$_s$-[3.6$\mu$m] plot of these objects.
From this plot it becomes obvious that 
for WC stars and the fainter WNL stars/OB super-giants the 
EW values are lower than 100$\AA$ 
for K$_s$-[3.6$\mu$m]$>$2 magnitudes and even drop significantly with further increasing
infrared excess, reaching values of up to K$_s$-[3.6$\mu$m]$\sim$5.
Dong, Wang \& Morris (2012)  suggested that the 
free-free emission from the strong winds of the WNL stars/OB
giants could potentially dominate the emission in the mid-infrared (Wright \& Barlow 1975).
This could explain the positive EW vs. K$_s$-[3.6$\mu$m] correlation they found. 
These properties fit with the DSO well if we assume that 
K$_s$-[3.6$\mu$m]$\approx$K$_s$-L' and that a small EW in the Pa$\alpha$ line 
is correlated with a small EW of the Br$\gamma$ line.
The Br$\gamma$ line width in 2008
indicated in Fig.2 by Gillessen et al. (2012a) is on 
the order of or smaller than 100 km/s in 2008 and 200 km/s in 2011.
Taking the Galactic foreground extinction 
into account, its position in the color-color diagram in Fig.~\ref{eckartfig-12} (left)
comes to lie in the WNL stars/OB supergiant domain with 
small EW and high infrared excess,
as described by Dong, Wang \& Morris (2012).
\\
\\
{\it Scenario III:}
Murray-Clay \& Loeb (2012) have
proposed that the dusty S-cluster object is a proto-planetary disk 
that has been brought in from a young stellar ring.
However, to change the ellipticity of the previously rather circular orbit in 
one single event to the high ellipticity orbit on which it is now requires a similarly 
violent interaction as the upcoming peri-bothron passage. If the object were
only a core-less dust cloud, this event would therefore
lead to an early destruction of the object at the time it enters the current orbit.
If the source is indeed a core-less dust cloud, the only alternative is that the
ellipticity of its orbit has been changed gradually. However, this implies many 
peri-bothron passages, imposes a conflict with evaporation timescales, and will most likely
lead to destruction as well.
The only alternative is indeed that the dusty S-cluster object has been formed on an 
orbit rather similar to the current one, or that it has been brought in 
gradually from 
outside the S-star cluster, as proposed earlier by Murray-Clay \& Loeb (2012).
This hypothesis is also supported by the analysis in the color-color diagram
shown in Fig.~\ref{eckartfig-12} (left).
The authors show that there are two objects, D3 and the DSO, which are 
dominated by their dust emission in the $L'$-band, which would be the 
case if they were relatively young stellar objects.
\\
\\
{\it Scenario IV:}
The DSO may be similar to the bow-shock sources X3 and X7 (Muzic et al. 2010)
but has not yet developed an obvious bow-shock structure.
For X3 and X7 Muzic et al. (2010) discussed possible stellar counterparts,
including the possibility of late-type and main-sequence sources.
However, the preference is clearly on late B-type main-sequence stars (B7-8V) 
and low-luminosity WR stars as well.
Other explanations have problems: The central star 
of a planetary nebulae will remain at the
required brightness for less than 1000 yr before it becomes
too faint, and a main-sequence star cannot be an appropriate
source of a dust-rich envelope, as is apparently observed for the 
DSO in the L -band.

\subsection{Will the DSO be disrupted?}
Gillessen et al. (2012a) have reported that the Br$\gamma$ emission of the DSO
is spread over 200~mas. Similar sizes are also predicted in the
models presented by Burkert et al. (2012) and Schartmann et al. 2012.
In Fig.\ref{eckartfig-28} we compare images of the DSO in the Ks- and L'-band.
We also show the L'-band images after removing the DSO using a Gaussian 
of 98~mas FWHM (i.e., the diffraction limit in the L'-band) and a star 
located 0.5'' west and 0.42'' north of SgrA* (i.e., a PSF obtained from 
an image section close to SgrA*; see Fig.~\ref{eckartfig-01}).
The analysis of the images shows that $>$90\% of the DSO emission at 
3.8$\mu$m wavelength is compact (FWHM$\le$20~mas). Only up to 10\% of the
flux density contained in the compact part of the DSO can be extended
on the scale size of the PSF.
This indicates that the warm ($\sim$550~K) dust emission is very compact in comparison 
to the hotter ($\sim$10$^4$~K) hydrogen recombination line emission.
Alternatively, the MIR continuum emission contains a significant 
compact free-free contribution (see below). In this case, however, one 
would expect that the recombination line emission is dominated by
contributions from this free-free component and should hence be
more compact than observed.

\begin{figure}
\centering
\includegraphics[width=8cm,angle=-00]{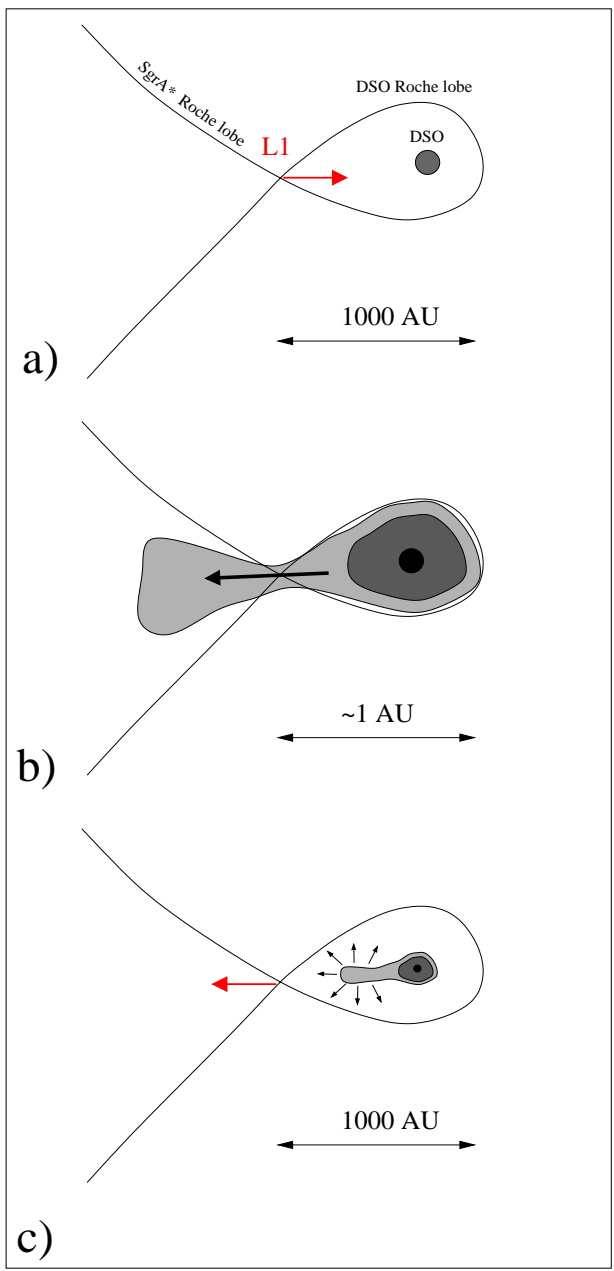}
\caption{\small Sketch of the relative position and motion of the Lagrange point L1 and the DSO.}
\label{eckartfig-17}
\end{figure}

Since the K$_s$-band identifications of the DSO suggest that it 
can also be associated with a star it is only  
its dusty envelope that can potentially be disrupted. 
To investigate that possibility we have to determine the location of
the Lagrange point $L1$ through which mass can be transferred and compare it to 
the size of the dust shell or disk.
In Fig.~\ref{eckartfig-17}
we show a sketch that demonstrates the location of the DSO and the motion of the
Lagrange point L1 with respect to it. 
In panel a) in Fig.~\ref{eckartfig-17} we show the situation about one year 
before the peri-bothron passage. The L1 point is rapidly approaching the DSO
through its orbital motion toward SgrA*. 
At peri-bothron, $L1$ will also be closest to the star (panel b) in Fig.~\ref{eckartfig-17}).
If the star has a mass of about 1\solm,  the separation of $L1$ from it will be
about 0.1~AU. For a Herbig Ae/Be stars with 2-8\solm ~that distance will be 
0.2 and 0.5~AU. For a typical S-cluster stellar mass of $\sim$20-30\solm ~the separation 
will be closer to one AU. 
Interferometrically determined typical inner ring sizes for young Herbig Ae/Be and
T~Tauri stars can be as small as 0.1-1~AU (Monnier \& Millan-Gabet 2002),
however, a disk or shell may be much larger than this.
Based on its MIR-luminosity, Gillessen et al. (2012a) determined a source
size of about 1~AU.
Hence, during peri-bothron passage a significant amount of the dusty circumstellar material
may pass beyond $L1$ and start to move into the Roche lobe associated with SgrA*. 

\noindent

However, this  peri-bothron passage will only last for a few months, after which the DSO will 
move away from SgrA* at a speed of around a 1000 km/s within the first year.
After one year the separation of the $L1$-point from the DSO is already about 1000 times larger
than during the peri-bothron passage (panel c) in Fig.~\ref{eckartfig-17}).
This implies that any cold dusty material
that passed the $L1$ toward the SgrA* will be overtaken again by the 
receding $L1$-point. 
However, the time will not be sufficient to completely disrupt the 
dust shell or disk of the DSO, especially if it is very compact.
The material that passed he $L1$-point during the peri-bothron passage will not stay 
in the SgrA* Roche lobe and will not be immediately accreted.
Only thin, hot material associated with a possible stellar wind of a 
hot and massive object may stay past the $L1$-point, but at the typical escape 
velocities of such volatile material of a few 100 km/s it would have left the star in any case.

Shcherbakov \& Baganoff (2010) have discussed the feeding rate of SgrA* as a function of radius.
Their modeling suggests that sources X3 and X7 (Muzic et al. 2010) are  
still in the regime in which most of the in-flowing mass is blown away again
(see their Fig.3).
This may partly explain the bow-shock structure of X3 and X7.
During its peri-bothron passage the star S2 has been well within the zone in which matter 
of its stellar wind could have been accreted. 
However, no effect on the variability of SgrA* has been reported.
The DSO peri-bothron will be at a larger radius than that of S2. This implies that if the
stellar wind from its central star is not stronger than that of S2, no enhanced accretion effect will
result from it. 
If the radius-dependent accretion flow is clumpy or un-isotropic, the DSO may still develop
a bow-shock like X3 and X7 and a major portion of the cold and dusty material that 
was detached from the DSO during its peri-bothron passage will be blown away.

\section{Summary}
\label{summary}

We have shown that most of the $L'$-band excess sources are indeed associated with
stellar objects that can be identified as H- or K$_s$-band sources with the same 
proper motions as the $L'$-band objects.
While this is true for many of the infrared excess sources we found in the inner
few arcseconds of the GC, it is also true for all objects that
constitute the small association of co-moving IRS13N sources 
(sources $\beta$ to $\eta$ in the nomenclature of Eckart et al. 2004).
The colors of the excess sources in the field (excepting D3 and the DSO) have brightnesses 
and colors that agree well with those of the narrow-line WR star candidates 
found in the central stellar cluster by Moultaka et al. (2005).

The IRS13N sources have visual extinctions that are higher than those of the 
excess sources in the field by up to 30 magnitudes.
The  integral field spectroscopy shows that the IRS13N sources are located at 
a minimum of the CO-band-head absorption in that region.
This makes it unlikely that they are dusty late-type stars older than a few Myr.
Given the colors and extinctions of the IRS13N sources ($\beta$ to $\eta$),
they are the best candidates for being young stellar objects 
in the overall GC stellar cluster with ages younger than 1~Myr.
They show no evidence for CO-bandhead absorption or emission and are then
likely to belong to the Class\,I sources
of spectrally featureless dusty young stars.

For the fast-moving DSO reported by Gillessen et al. (2012a)
we were able to provide a K$_s$-band identification as well.
While the H-K$_s$ color limit point at a significant contribution of dust, there is also the 
possibility that the DSO is a dust-enshrouded stellar source
and not a core-less cloud of gas and dust.
The DSO may be a stellar source and is close to the SMBH for a few months only.
This may have a significant influence on the possible disruption of its 
dusty envelope or disk. 
While the dusty material will certainly experience
the gravitational pull by the SMBH SgrA*, it may be rapidly enclosed again by the DSO's
Roche lobe a few months after the peri-bothron passage (Fig.~\ref{eckartfig-17}).
While this is a possible
scenario for the cold and dusty material, any fast wind possibly associated 
with the DSO may be accreted if it is a more massive, hot object.
However, the peri-bothron passage distance to SgrA* is more than twice as large as 
that of S2 which did not have any reported effect on the variability of SgrA*.
It may not be excluded, however, that if the DSO passes SgrA* by
it may develop a bow-shock similar to sources X3 and X7 (Muzic et al. 2010).  
Additional monitoring of the trajectory and size evolution of the DSO as well
as the variability of SgrA* are essential to achieve a complete picture of the
scenario.
It is probably currently not possible to reliably determine if and how much
gas will be accreted. It is unclear if the object is already quite compact
(as indicated by the L-band data) through the influence of earlier passages.
It is unclear what the interaction with a wind from SgrA* is and how much
of the material will be blown away. New and detailed simulations 
(beyond those given by Burkert et al. 2012 and Schartmann et al. 2012) 
will be needed that include the 
DSO identification as a stellar system and the nature of SgrA*.

If the IRS13N sources, in addition to the infrared excess sources including the DSO,
are young stellar objects, this implies 
that the conditions of star formation at the center of the Milky Way  
may support a significant constant star formation rate in the overall central stellar cluster
in addition to larger star formation events
(in an accretion disk; Levin \& Beloborodov 2003, Nayakshin \& Sunyaev 2005,
Nayakshin et al. 2006).

\begin{acknowledgements}
We thank the anonymous referee for the useful comments and
suggestions that helped to improve the paper. 
This work was supported in part by the Deutsche Forschungsgemeinschaft
(DFG) via the Cologne Bonn Graduate School (BCGS), 
the Max Planck Society through 
the International Max Planck Research School (IMPRS) for Astronomy and 
Astrophysics, as well as 
special funds through the University of Cologne.
B. Shahzamanian, N. Sabha and M. Valencia-S.,
are members of the IMPRS.
N. Sabha is member of the Bonn Cologne Graduate School for
Physics and Astronomy.
Part of this work was supported by the German Deutsche
Forschungsgemeinschaft, DFG, via grant SFB 956, and by fruitful discussions
with members of the European Union funded COST Action MP0905: Black
Holes in a violent Universe and PECS project No. 98040. 
This work was co-funded under the Marie Curie Actions of the European
Commission (FP7-COFUND).
Macarena Garc\'{\i}a-Mar\'{\i}n is supported by the German 
federal department for education and research (BMBF) under 
the project number 50OS1101.
NSO/Kitt Peak FTS data used here were produced by NSF/NOAO.
We are grateful to all members
of the NAOS/CONICA and the ESO PARANAL team.  
\end{acknowledgements}

\vspace*{0.5cm}

\rf{ Baganoff, F.K. et al., 2001, Nature 413, 45}

\rf{ Berukoff, Steven J.; Hansen, Bradley M. S., 2006, ApJ 650, 901}

\rf{ Bremer, M.; Witzel, G.; Eckart, A.; Zamaninasab, M.; Buchholz, R. M.; Sch\"odel, R.; Straubmeier, C.; Garcia-Marin, M.; Duschl, W., 2011, A\&A 532, 26}

\rf{ Buchholz, R. M.; Sch\"odel, R.; Eckart, A, 2009, A\&A 499, 483}

\rf{ Burkert, A.; Schartmann, M.; Alig, C.; Gillessen, S.; Genzel, R.; Fritz, T. K.; Eisenhauer, F.,	2012, ApJ 750, 58}

\rf{ Calvet, N.; Hartmann, L.; Strom, S.E., 1997, ApJ 481, 912	}

\rf{ Casali, M. M., \& Eiroa, C. 1996, A\&A, 306, 427}

\rf{ Coker, R. F.; Pittard, J. M.; Kastner, J. H.,, 2002, A\&A 383, 568}

\rf{ Clenet, Y., Roun, D., Gratadour, D., Marco, O., Lena, P., Ageorges, N. \& Gendron, E. 2005, A\&A, 439, 9}

\rf{ Crowther, Paul A.; in 'The influence of binaries on stellar population studies', Dordrecht: Kluwer Academic Publishers, 2001, xix, 582 p. Astrophysics and space science library (ASSL), Vol. 264. ISBN 0792371046, p.215}

\rf{ Dong, H.; Wang, Q. D.; Morris, M. R.; in press MNRAS; 2012arXiv1204.6298D}

\rf{ Drake, S. A.; Linsky, J. L., 1989, AJ 98, 1831}

\rf{ Eckart, A.; Moultaka, J.; Viehmann, T.; Straubmeier, C.; Mouawad, N., 2004, ApJ 602, 760}

\rf{ Eckart, A., Genzel, R., Ott, T. and Sch\"odel, R., 2002, MNRAS 331, 917-934}

\rf{ Eckart, A. \& Genzel, R., 1996, Nature 383, 415-417}

\rf{ Eckart, A.; Sabha, N.; Witzel, G.; Straubmeier, C.; Shahzamanian, B.; et al.
               2012, Conf. Proc. SPIE Astronomical Telescopes \& Instrumentation 1 - 6 July 2012. 
               Amsterdam No. 8445-4, in press, 2012arXiv1208.1129E}

\rf{ Fritz, T. K., Gillessen, S., Dodds-Eden, K., Martins, F., Bartko, H., Genzel, R., Paumard, T., Ott, T., Pfuhl, O., and 3 coauthors, 2010, ApJ 721, 395}

\rf{ Fuente, A., Martin-Pintado, J., Bachiller, R., Rodriguez-Franco, A., Palla, F., 2002, A\&A 387, 977}

\rf{ Ghez, A., Morris, M., Becklin, E.E., Tanner, A. \& Kremenek, T., 2000, Nature 407, 349}

\rf{ Glass, I.S.,\& Moorwood, A.F.M., 1985, MNRAS 214, 429.}

\rf{ Green, T.P., Meyer, M.R., 1995, ApJ 450, 233}

\rf{ Hansen, B.M.S.; Milosavljevic  M., 2003, ApJ 593, L77}

\rf{ Hillenbrand, L.A., Strom, S.E., Vrba, F.J., Keene, J., 1992, ApJ 397, 613}

\rf{ Hoffmeister V. H., Chini, R., Scheyda, C.D., N\"urnberger, D., Vogt, N., and Nielbock, M., 2006, A\&A 457, L29}

\rf{ Ishii, M., Nagata, T., Sato, S., Watanabe, M., Y., Yongqiang, J., Terry J., 1998, AJ 116, 868}

\rf{ Kim, S.S.; Figer, D.F.; Morris, M., 2004 ApJ 607, L123	}

\rf{ Kudritzki, R. P.; Puls, J.; Lennon, D. J.; Venn, K. A.; Reetz, J.; Najarro, F.; McCarthy, J. K.; Herrero, A.,1999, A\&A 350, 970	}

\rf{ Genzel, R.; Sch\"odel, R.; Ott, T.; Eisenhauer, F.; Hofmann, R.; Lehnert, M.; Eckart, A.; et al., 2003, ApJ 594, 812}

\rf{ Gerhard, O., 2001, ApJ 546, L39	}

\rf{ Ghez, A. M.; Salim, S.; Hornstein, S. D.; Tanner, A.; Lu, J. R.; et al.,  2005, ApJ 620, 744}

\rf{ Gillessen, S.; Eisenhauer, F.; Trippe, S.; Alexander, T.; Genzel, R.; Martins, F.; Ott, T., 2009, ApJ 692, 1075	}

\rf{ Gillessen, S.; Genzel, R.; Fritz, T. K.; Quataert, E.; Alig, C.; et al., 2012 Nature 481, 51}

\rf{ Kunneriath, D.; Eckart, A.; Vogel, S. N.; Teuben, P.; et al., 2012, A\&A 538, 127}

\rf{ Levin, Y.; Beloborodov, A.M., 2003, ApJ 590, L33	}

\rf{ Lang, C.C.; Johnson, K.E.; Goss, W. M.; Rodriguez, L.F, 2005, AJ 130, 2185	}

\rf{ Lu, J. R.; Ghez, A. M.; Hornstein, S. D.; Morris, M.; Becklin, E. E., 2005, ApJ 625, L51}

\rf{ Luhman, K.L., Rieke, G.H., 1999, ApJ 525, 440}

\rf{ Lutz, D., Feuchtgruber, H., Genzel, R., et al. 1996, A\&A, 315, L269}

\rf{ Maillard, J. P.; Paumard, T.; Stolovy, S. R.; Rigaut, F.; 2004, A\&A 423, 155}

\rf{ Martins, F.; Genzel, R.; Hillier, D. J.; Eisenhauer, F.; Paumard, T.; Gillessen, S.; Ott, T.; Trippe, S., 2007, A\&A 468, 233}

\rf{ Martins, F.; Gillessen, S.; Eisenhauer, F.; Genzel, R.; Ott, T.; Trippe, S., 2008, ApJ 672, L119}

\rf{ Mauerhan, J.C.; Van Dyk, S.D.; Morris, P.W., 2011, AJ 142, 40}

\rf{ McMillan, S.L.W.; Portegies Zwart, S.F., 2003, ApJ 596, 314	}

\rf{ Milosavljevic, Mi; Loeb, A., 2004, ApJ 604, L45}

\rf{ Molinari, S.; Pezzuto, S.; Cesaroni, R.; Brand, J.; Faustini, F.; Testi, L., 2008, A\&A 481, 345}

\rf{ Monnier, J. D.; Millan-Gabet, R.;  2002, ApJ 579, 694}

\rf{ Montes, G.; Perez-Torres, M.A.; Alberdi, A.; Gonzalez, R.F., 2009, ApJ 705, 899}

\rf{ Moultaka, J.; Eckart, A.; Sch\"odel, R.; Viehmann, T.; Najarro, F; 2005, A\&A 443, 163}

\rf{ Murray-Clay, R.A. \& Loeb, A., 2011arXiv1112.4822M}

\rf{ Muzic, K.; Eckart, A.; Sch\"odel, R.; Meyer, L.; Zensus, A., 2007, A\&A 469, 993}

\rf{ Muzic, K.; Sch\"odel, R.; Eckart, A.; Meyer, L.; Zensus, A, 2008, A\&A 482, 173	}

\rf{ Muzic, K.; Eckart, A.; Sch\"odel, R.; Buchholz, R.; Zamaninasab, M.; Witzel, G., 2010, A\&A 521, 13	}

\rf{ Nayakshin, S.; Cuadra, J.; Springel, V.; 2007, MNRAS 379, 21	}

\rf{ Nayakshin, S.; Sunyaev, R.,  2005, MNRAS 364, L23}

\rf{ Ott, T.; Eckart, A.; Genzel, R., 1999, ApJ 523, 248O	

\rf{ Paumard, T.; Genzel, R.; Martins, F.; Nayakshin, S.; Beloborodov, A. M.; Levin, Y.; et al., 2006, ApJ 643, 1011}

\rf{ Portegies Zwart, S.F.; Baumgardt, H.; McMillan, S.L.W.; Makino, J.; Hut, P.; Ebisuzaki, T., 2006, ApJ 641, 319}

\rf{ Pott, J.-U.; Eckart, A.; Glindemann, A.; Kraus, S.; Sch\"odel, R.; Ghez, A. M.; Woillez, J.; Weigelt, G., 2008a, A\&A 487, 413}

\rf{ Pott, J.-U.; Eckart, A.; Glindemann, A.; Schoedel, R.; Viehmann, T.; Robberto, M.,  2008a, A\&A 480, 115	}

\rf{ Prato, L.; Lockhart, K.E.; Johns-Krull, C.M.; Rayner, J.T., 2009, AJ 137, 3931}

\rf{ Sabha, N.; Witzel, G.; Eckart, A.; Buchholz, R. M.; Bremer, M.; et al., 2010, A\&A 512, 2}

\rf{ Sabha, N.; Eckart, A.; Merritt, D.; Zamaninasab, M.; Witzel, G.; et al., 2012, A\&A 545, 70}

\rf{ Scally, A., \& Clarke, C., 2001, MNRAS 325, 449}

\rf{ Shcherbakov, Roman V.; Baganoff, Frederick K., 2010, ApJ 716, 504}

\rf{ Sch\"odel, R.; Eckart, A.; Iserlohe, C.; Genzel, R.; Ott, T., 2005, ApJ 625, L111}

\rf{ Sch\"odel, R.; Merritt, D.; Eckart, A., 2009, A\&A 502, 91	}

\rf{ Sch\"odel, R.; Najarro, F.; Muzic, K.; Eckart, A., 2010, A\&A 511, 18	}

\rf{ Shara, M.M.; Moffat, A.F.J.; Smith, L.F.; Niemela, V.S.; Potter, M.; Lamontagne, R., 1999, AJ 118, 390}

\rf{ Shara, M.M.; Faherty, J.K.; Zurek, D.; Moffat, A.F.J.; Gerke, J.; Doyon, R.; Artigau, E.; Drissen, L.,  2012, AJ 143, 149	}

\rf{ Stolte, A.; Morris, M. R.; Ghez, A. M.; Do, T.; Lu, J. R.; Wright, S. A.; Ballard, C.; Mills, E.; Matthews, K., 2010, ApJ 718, 810}

\rf{ Stolte, A.; Ghez, A.; Morris, M.; Lu, J.; Brandner, W.; Matthews, K., 2009, Ap\&SS, 324, 137	}

\rf{ Tanner, A.; Ghez, A. M.; Morris, M. R.; Christou, J. C., 2005, ApJ 624, 742}

\rf{ van Kempen, T.A.; van Dishoeck, E.F.; Salter, D.M.; Hogerheijde, M.R.; Joergensen, J.K.; Boogert, A.C.A., 2009, A\&A 498, 167}

\rf{ Viehmann, T.; Eckart, A.; Sch\"odel, R.; Pott, J.-U.; Moultaka, J, 2006, ApJ 642, 861}

\rf{ Viehmann, T.; Eckart, A.; Sch\"odel, R.; Moultaka, J.; Straubmeier, C.; Pott, J.-U., 2005, A\&A 433, 117	}

\rf{ Wegner, W., 2006, MNRAS 371, 185}

\rf{ Witzel, G., Eckart, A:, Bremer, M., Zamaninasab, M., Shahzamanian, B., et al., 2012, submitted to ApJ, 2012arXiv1208.5836W.}

\rf{ Wright, A. E.; Barlow, M. J.,1975, MNRAS 170, 41}

\rf{ Zhao, J.-H.,\& Goss, W.M., 1998, ApJ 499, L163 }

\rf{ Zhao, J.-H.; Morris, M.R.; Goss, W.M.; An, T., 2009, ApJ 699, 186}


\newpage
.
\newpage

\newpage
.
\newpage

\Online
\begin{appendix}
\section{Appendix}

\noindent
\begin{figure*}[ht!]
\centering
\includegraphics[width=15cm,angle=-00]{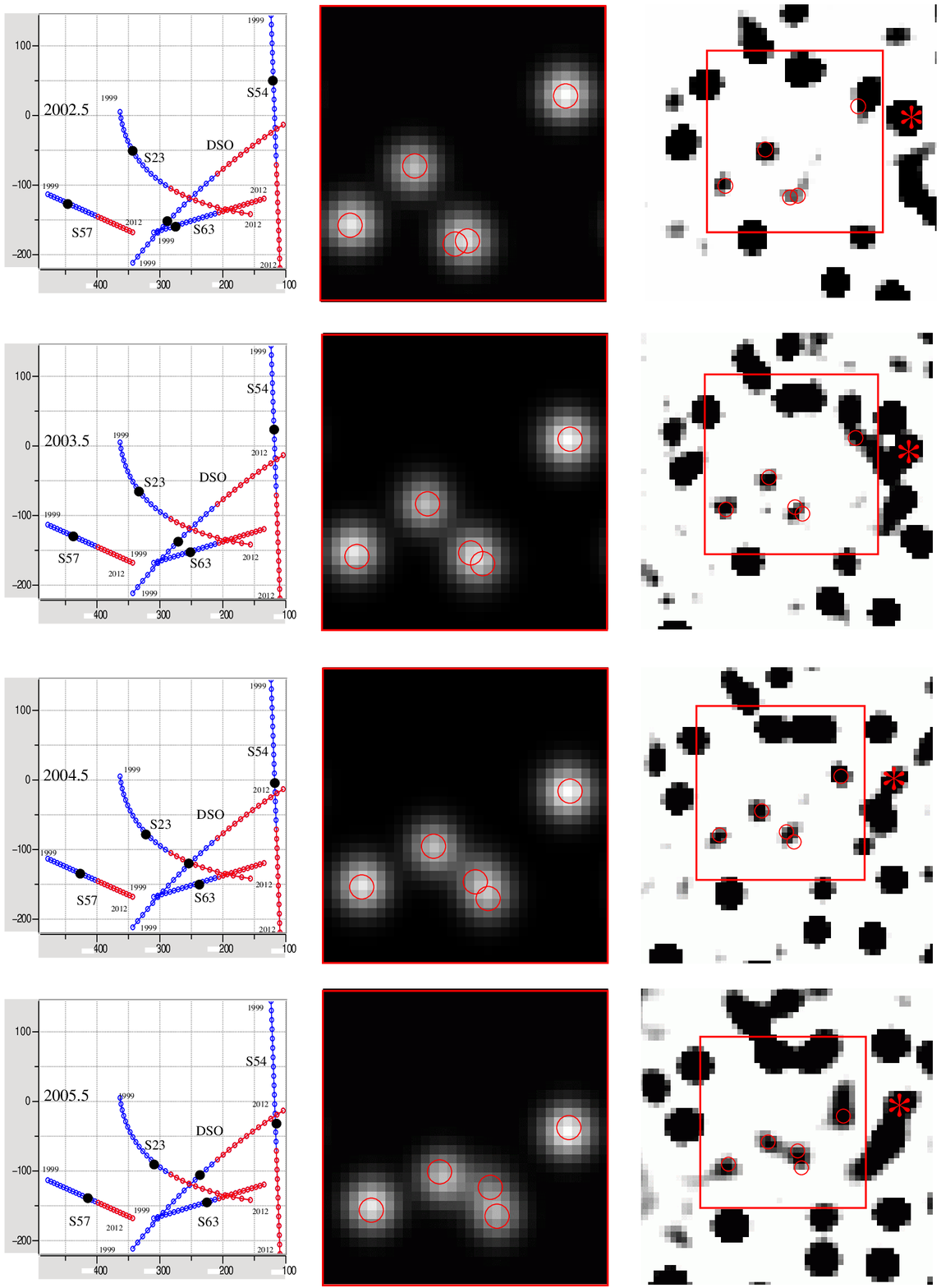}
\caption{\small
K$_s$-band identification of the DSO.
In Figs.~\ref{eckartfig-14-1}, \ref{eckartfig-14-2} and \ref{eckartfig-14-3} we show the relative positions 
of the stars S23, S57, S54, S63 and the DSO 
over the years 2002 - 2012
in comparison with K$_s$-band Lucy-deconvolved images.
The measuring uncertainties range between 13~mas and 25~mas.
We show the results of the model calculation (left) with the location of the sources on their sky-projected
track (dot interval 0.5 years; 1999-2006.5 in blue, 2007-2012 in red),
an image of the model with the sources indicated by red circles (middle; image size is 380mas$\times$380mas), and
a deconvolved K$_s$-band image with the modeled image section overlayed (right).
For the years 2006/7 to 2012 a source can clearly be 
identified at the $L'$-band position of the DSO.
The position of SgrA* is indicated by a red asterix.
In the years 2002 to 2005 the DSO is confused with S63.
In 2006-2008 it emerges from the
confusion and shows up between the stars S23, S63, and S54 just north of S63.
For 2009, 2011, 2012 the m$_K$=18.9 source cannot be clearly identified. This
is to a large part due to the stronger crowding at the center and to the
limited data quality.
}
\label{eckartfig-14-1}
\end{figure*}

\noindent
\begin{figure*}[ht!]
\centering
\includegraphics[width=15cm,angle=-00]{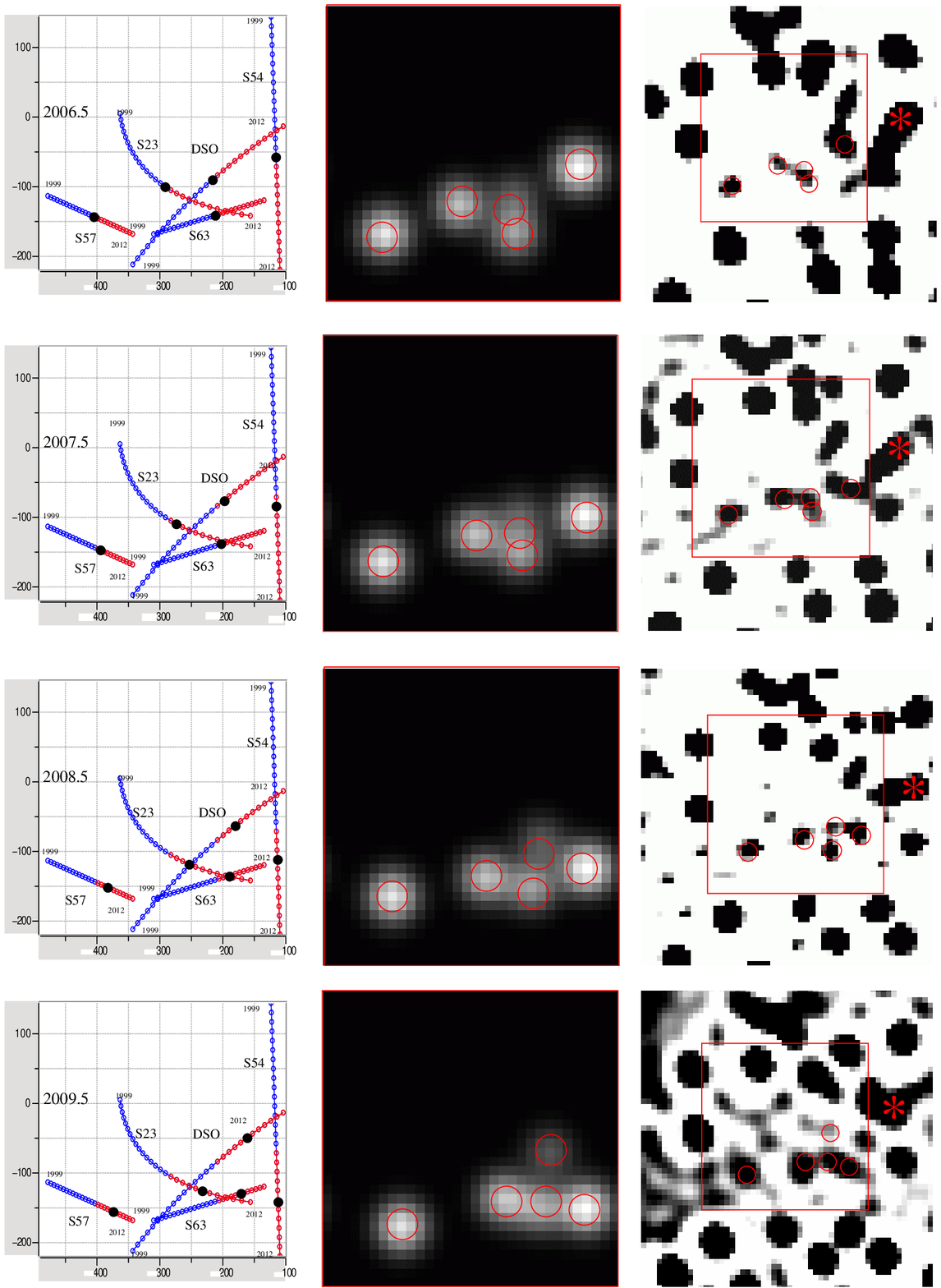}
\caption{\small
K$_s$-band identification of the DSO - continued.
}
\label{eckartfig-14-2}
\end{figure*}

\noindent
\begin{figure*}[ht!]
\centering
\includegraphics[width=15cm,angle=-00]{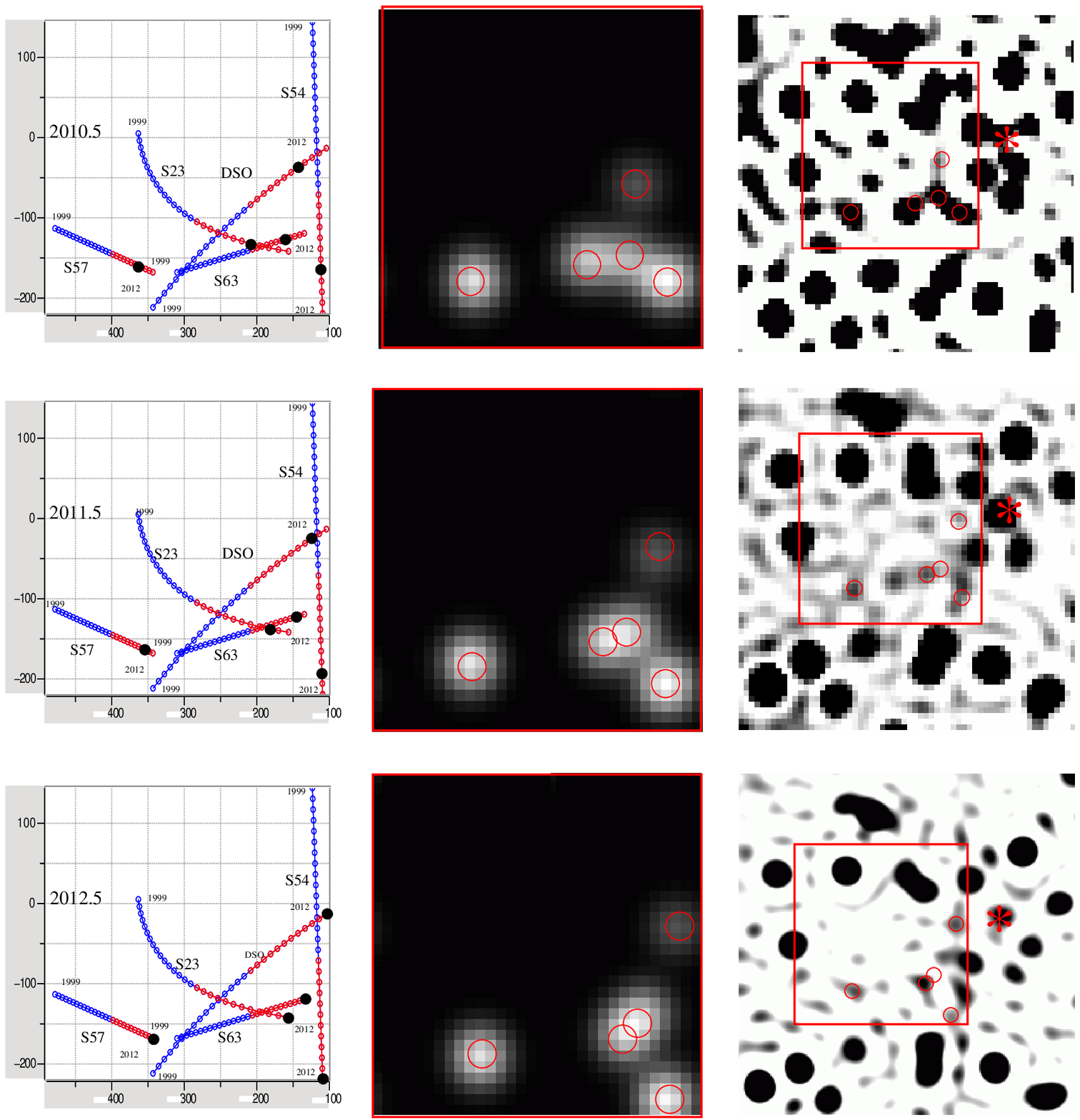}
\caption{\small
K$_s$-band identification of the DSO - continued.
}
\label{eckartfig-14-3}
\end{figure*}


\noindent
\begin{figure*}
\centering
\includegraphics[width=15cm,angle=-00]{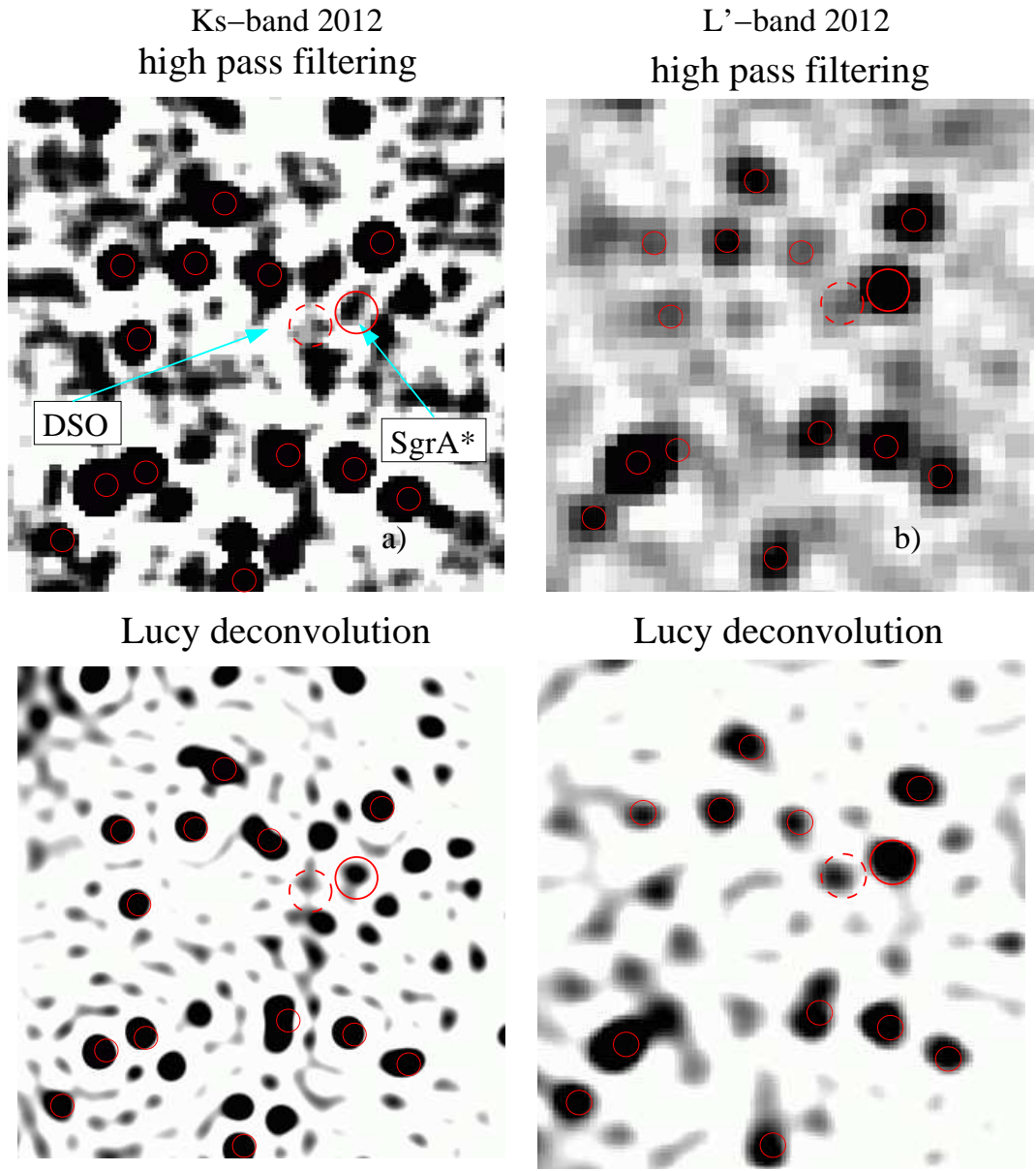}
\caption{\small Retrieving the DSO in 2012 using different methods.
Here we show a 
comparison of results from different data reduction methods and different
K$_s$- and $L'$-band datasets for 2012.
In both bands different methods result in a clear detection of a K$_s$- and $L'$-band 
counterpart of the DSO
(see Ott, Eckart \&  Genzel 1999 for a detailed comparison of deconvolution algorithms
in the GC field).
As the DSO is close to the confusion limit, the peak of the flux distribution within 
the diffraction limit results in small deviations in the apparent peak position.
}
\label{eckartfig-34}
\end{figure*}

\noindent
\begin{figure*}
\centering
\includegraphics[width=15cm,angle=-00]{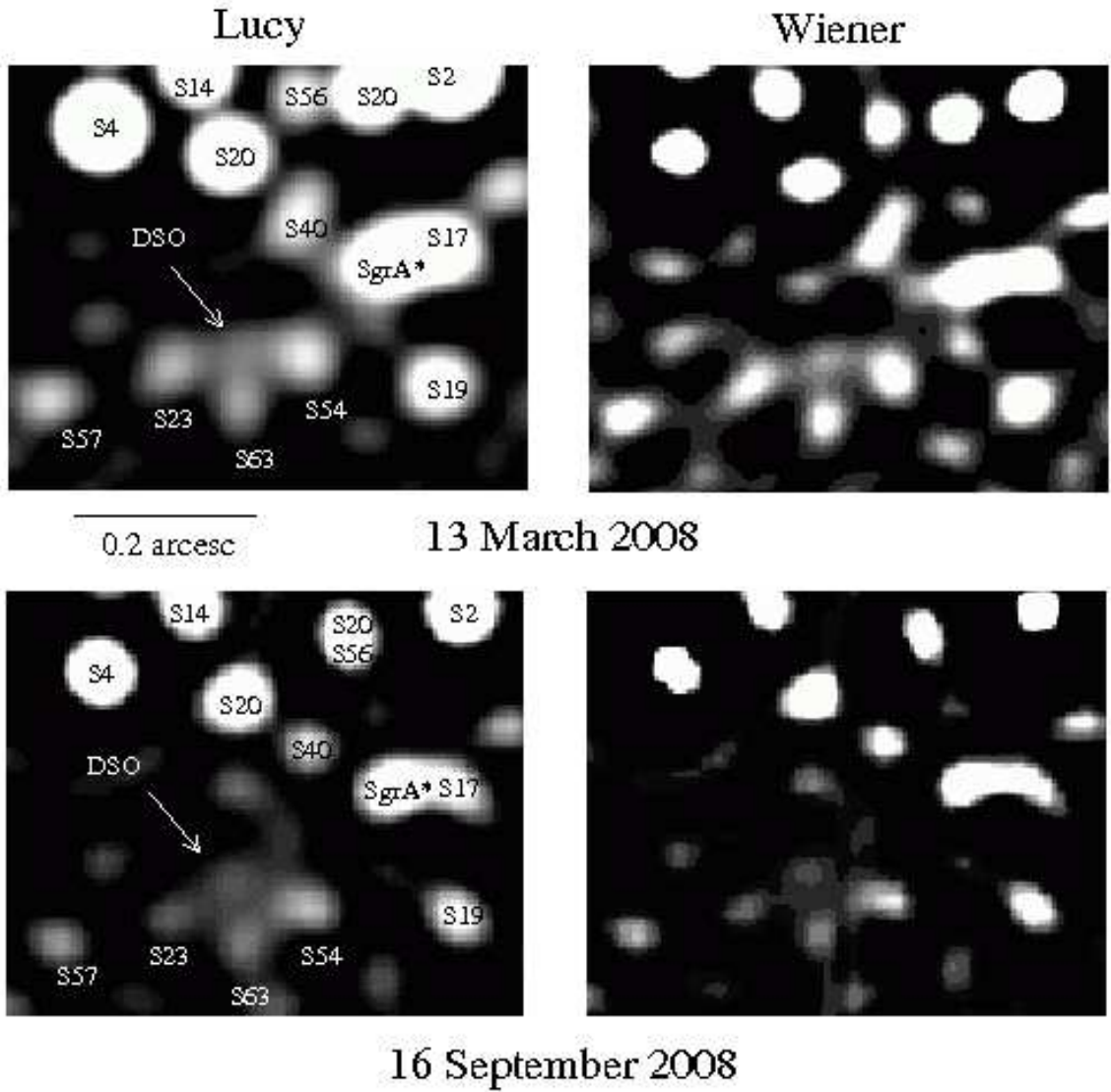}
\caption{\small Retrieving the DSO in 2008 using different methods.
Here we show a 
comparison of results from different data reduction methods and different
K$_s$-band datasets for 2008.
At both (selected as an example) epochs the different methods result in a 
clear detection of a K$_s$-band counterpart of the DSO
(see also Ott, Eckart \&  Genzel 1999).
}
\label{eckartfig-36}
\end{figure*}


\noindent
\begin{figure*}[ht!]
\centering
\includegraphics[width=15cm,angle=-00]{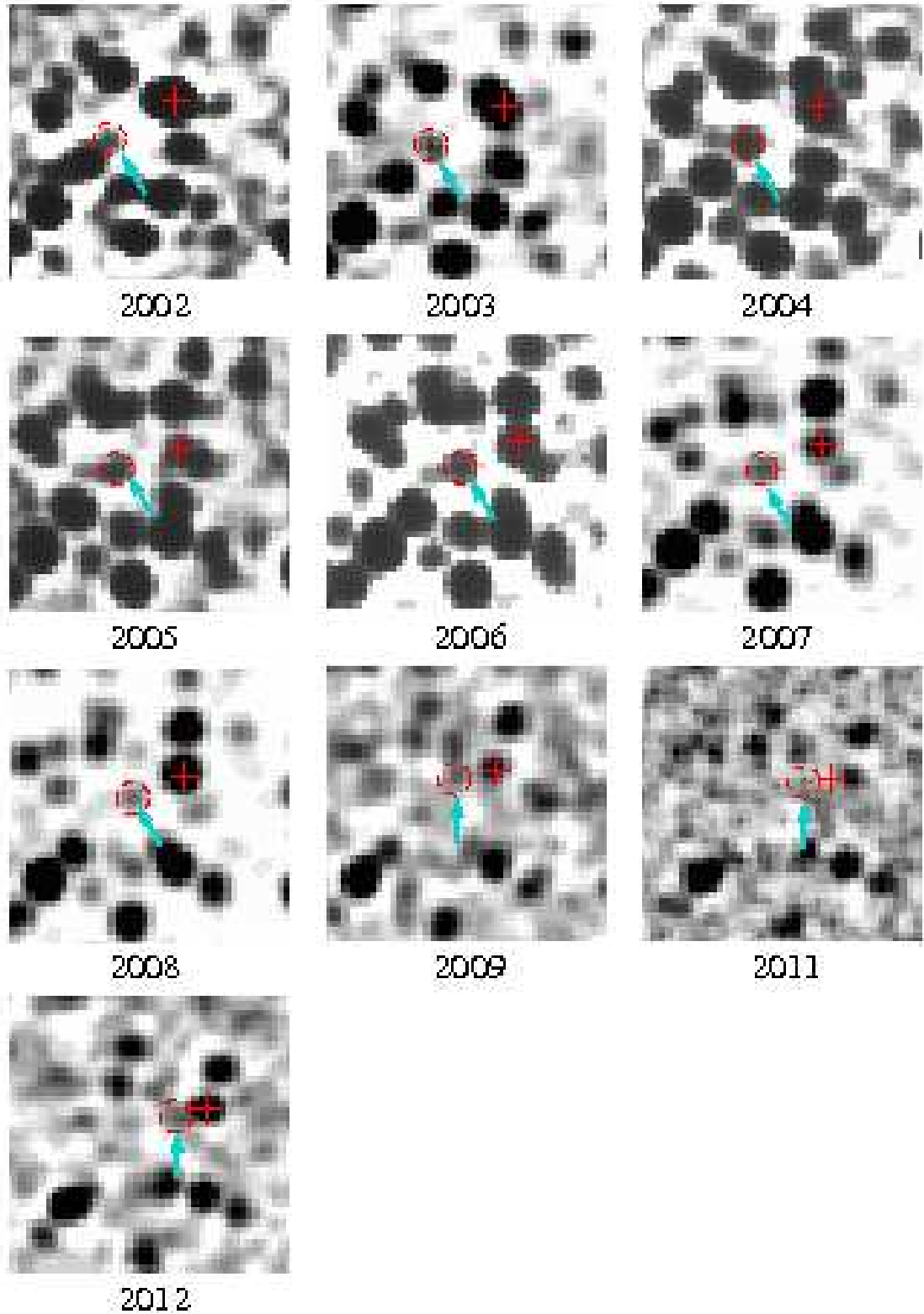}
\caption{\small $L'$-band identification of the DSO,
obtained via high-pass filtering. Here the source can always
clearly be identified. 
For the year 2010 we have no high-quality $L'$-band data at hand.
}
\label{eckartfig-21}
\end{figure*}

\noindent
\begin{figure*}[ht!]
\centering
\includegraphics[width=15cm,angle=-00]{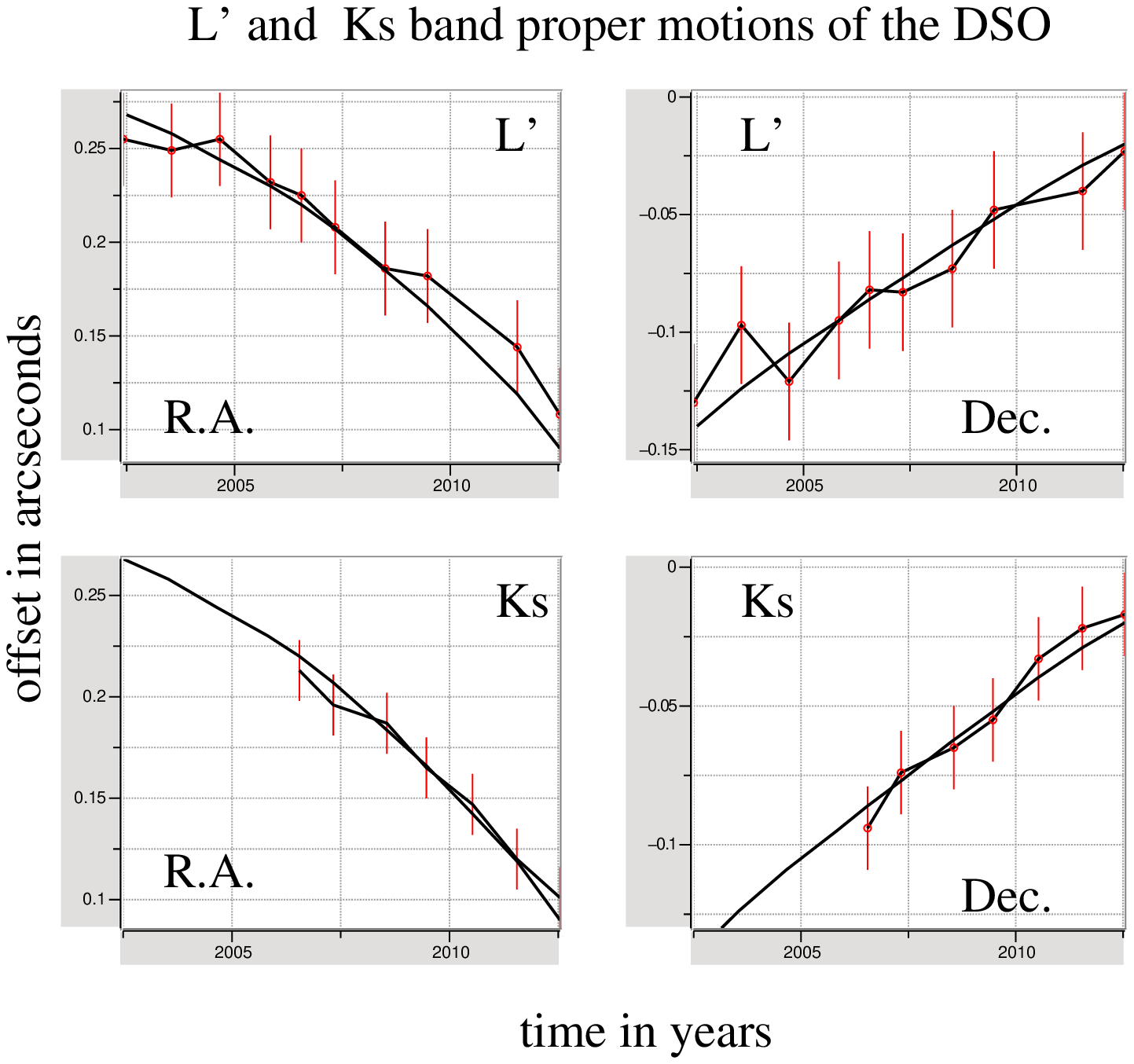}
\caption{\small  $L'$- and K$_s$-band coordinates of the identification of the DSO
shown separately for R.A. and Dec. values as a function of time.
The 1$\sigma$ deviations between the expected positions and the identifications
range between $\pm$13~mas and $\pm$25~mas (i.e., a quarter of the K$_s$- and L'-band beam).
}
\label{eckartfig-23}
\end{figure*}


\noindent
\begin{figure*}
\centering
\includegraphics[width=15cm,angle=-00]{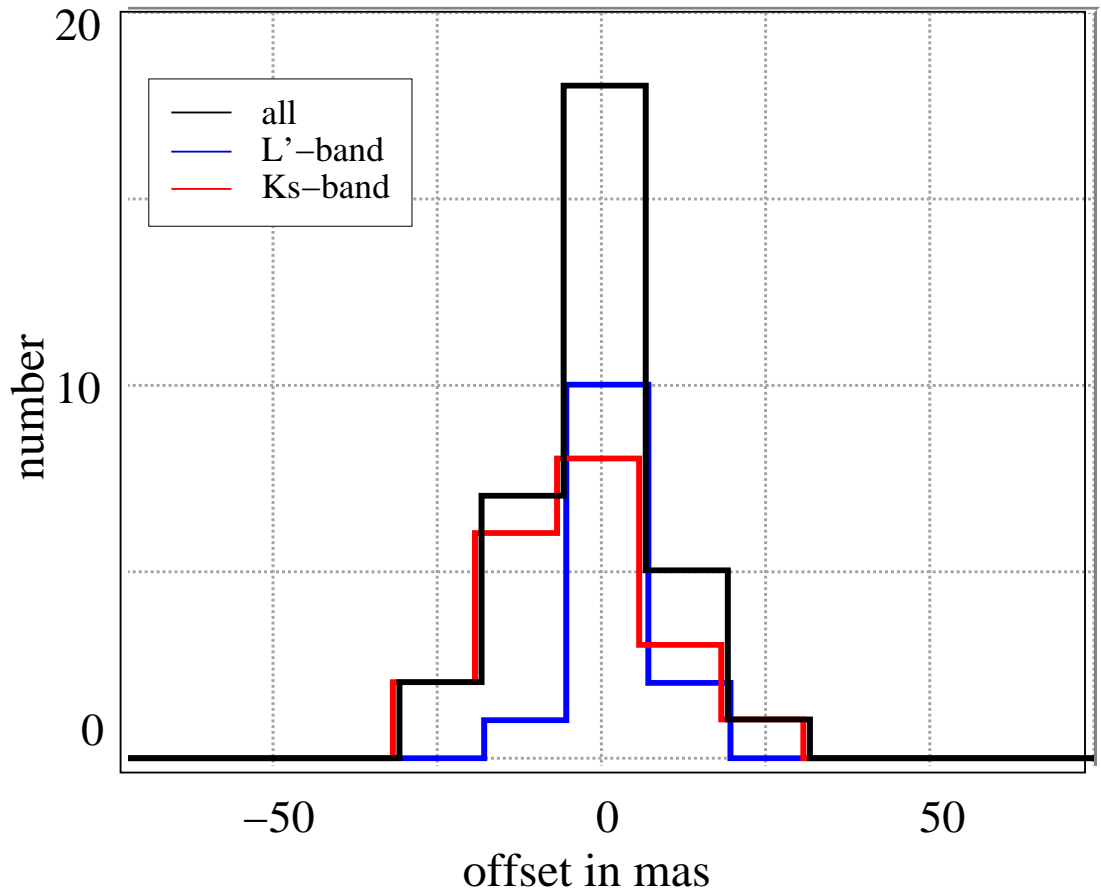}
\caption{\small Histograms of the deviations from the proposed orbit (Gillessen et al. 2012) 
for all bands. The 1$\sigma$ deviations between the expected positions and the identifications
in the different bands range between $\pm$13~mas and $\pm$25~mas.
}
\label{eckartfig-24}
\end{figure*}

\clearpage

\noindent
\begin{figure*}
\centering
\includegraphics[width=15cm,angle=-00]{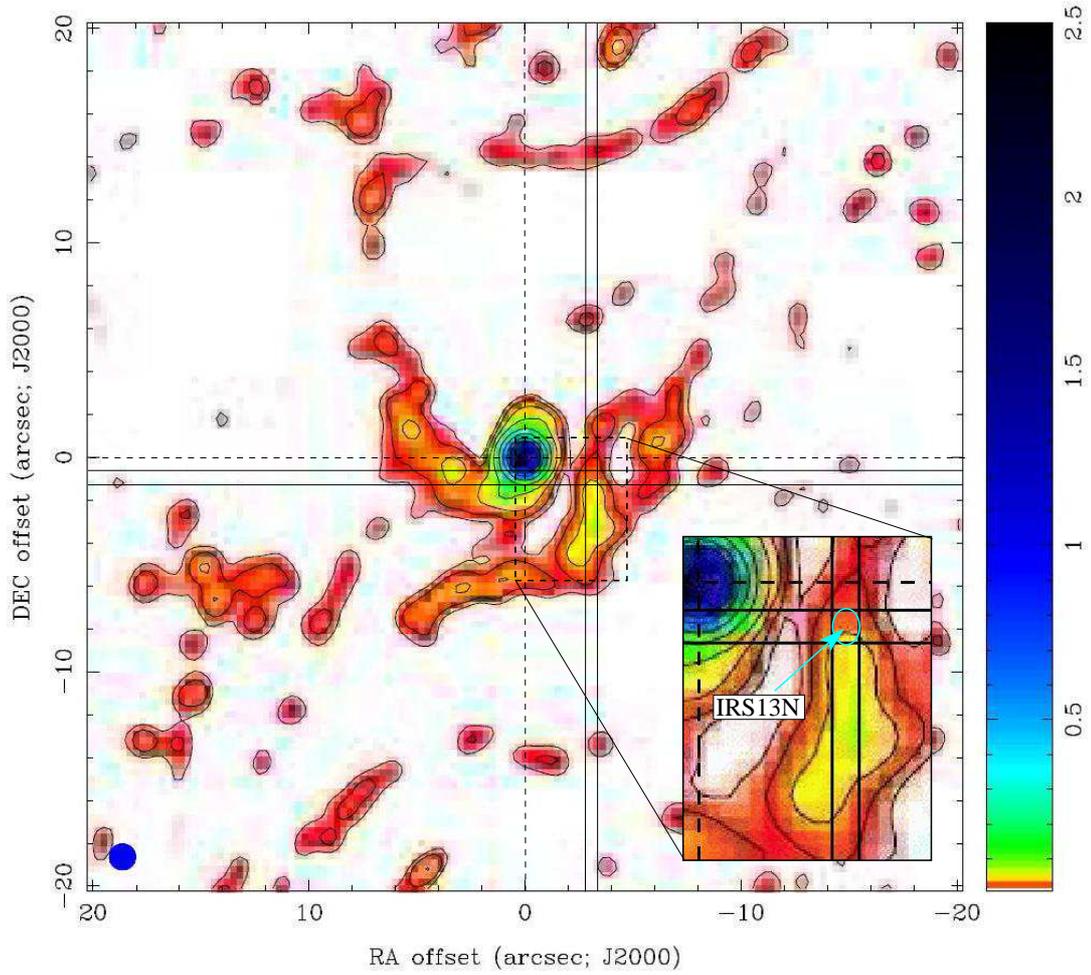}
\caption{\small Mini-spiral region imaged with CARMA in the 1.3\,mm CD array configuration 
with a synthesized circular beam size of 1.2$''$
presented by Kunneriath et al. (2011). 
We show the map to outline the
continuum emission from IRS13N.
Contour levels are 0.015, 0.02, 0.03, 0.04, 0.05, 0.1, 0.6, 0.9, 1.2, 1.5, and 2.5 Jy/beam.
Solid black lines and a thin cyan ellipse (in the inset) indicate in the map and the inset 
indicates the region in which the IRS13N sources are located.}
\label{eckartfig-16}
\end{figure*}

\clearpage

\noindent
\begin{figure*}
\centering
\includegraphics[width=15cm,angle=-00]{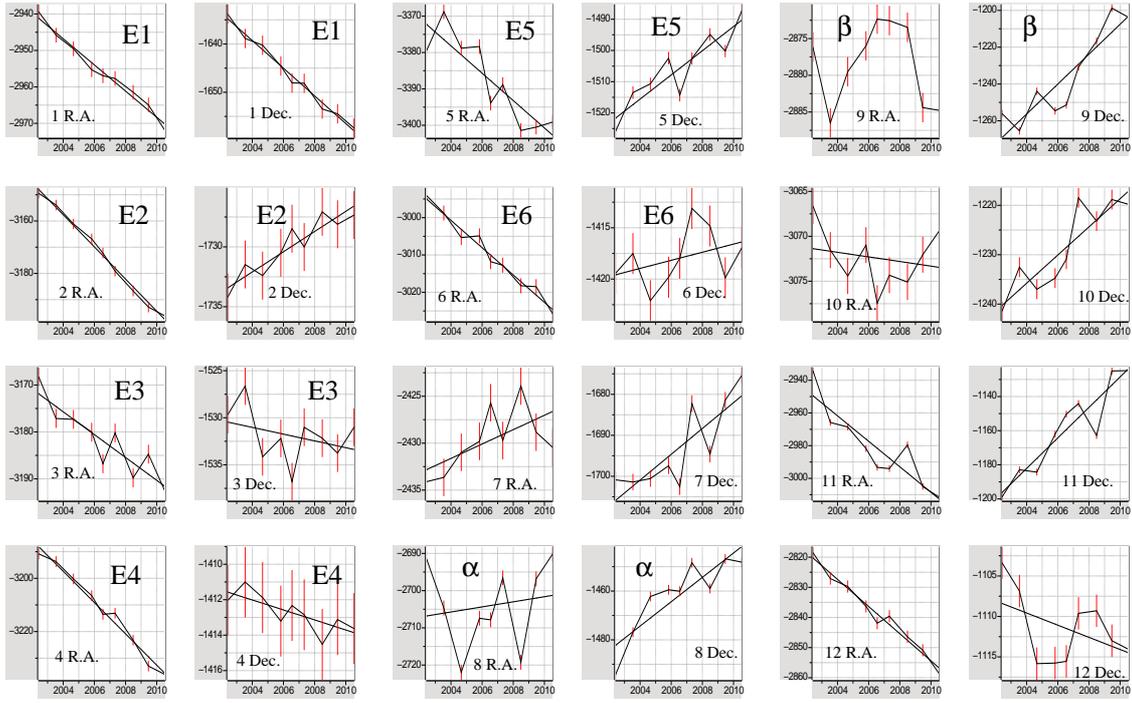}
\caption{\small 
In Figs.~\ref{eckartfig-25}, \ref{eckartfig-26}, and \ref{eckartfig-27}
we present all K$_s$-band proper motions of the 34 sources in and close 
to the IRS13E and IRS13N field.
Offsets in R.A. and Dec. in milliarcseconds are plotted against the time 
in years.
Source labels are given as indicated in Fig.~\ref{eckartfig-02}
and Tab.~\ref{Tab:13N}
}
\label{eckartfig-25}
\end{figure*}


\noindent
\begin{figure*}
\centering
\includegraphics[width=15cm,angle=-00]{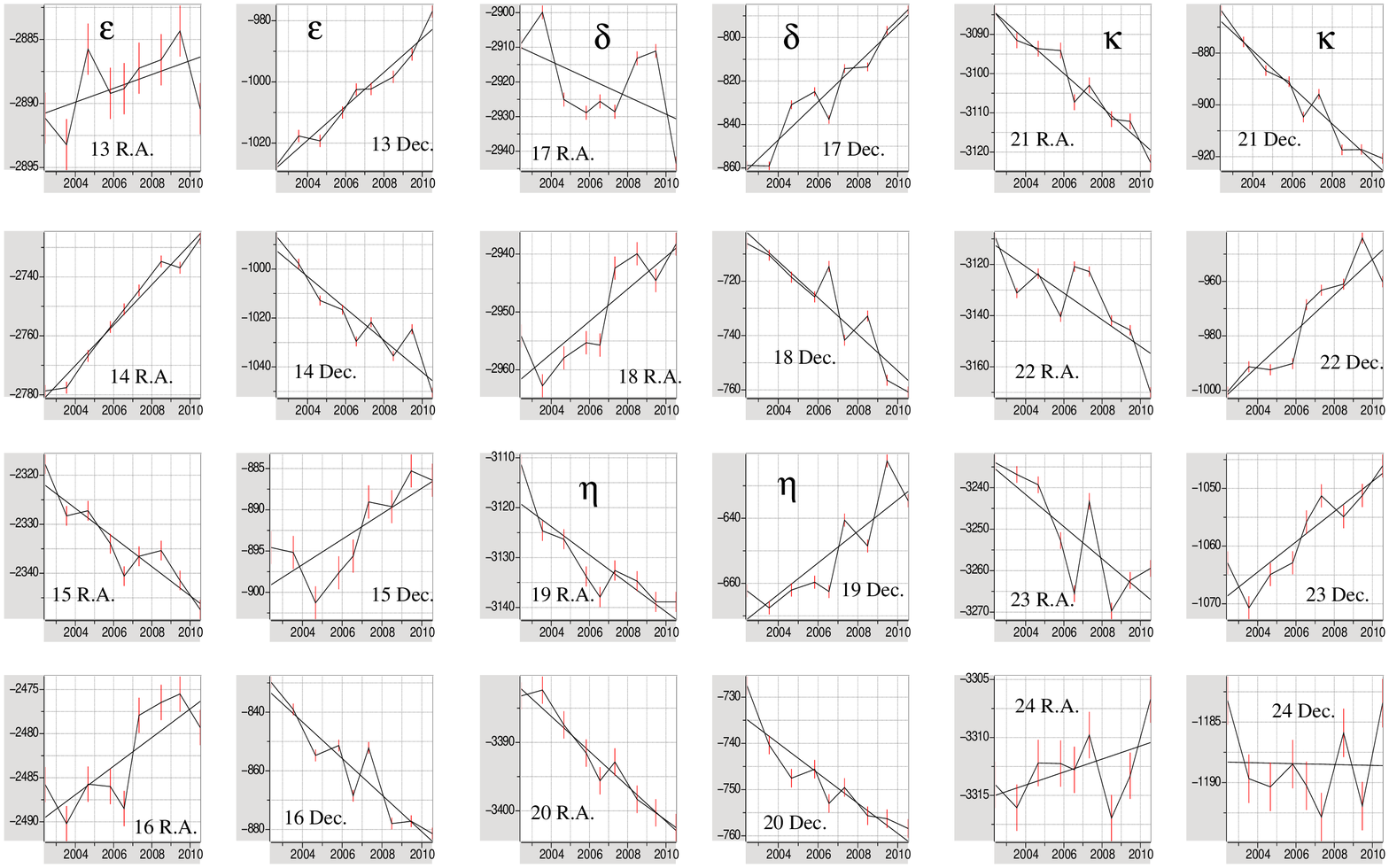}
\caption{\small Fig.~\ref{eckartfig-25} continued.
}
\label{eckartfig-26}
\end{figure*}


\noindent
\begin{figure*}
\centering
\includegraphics[width=15cm,angle=-00]{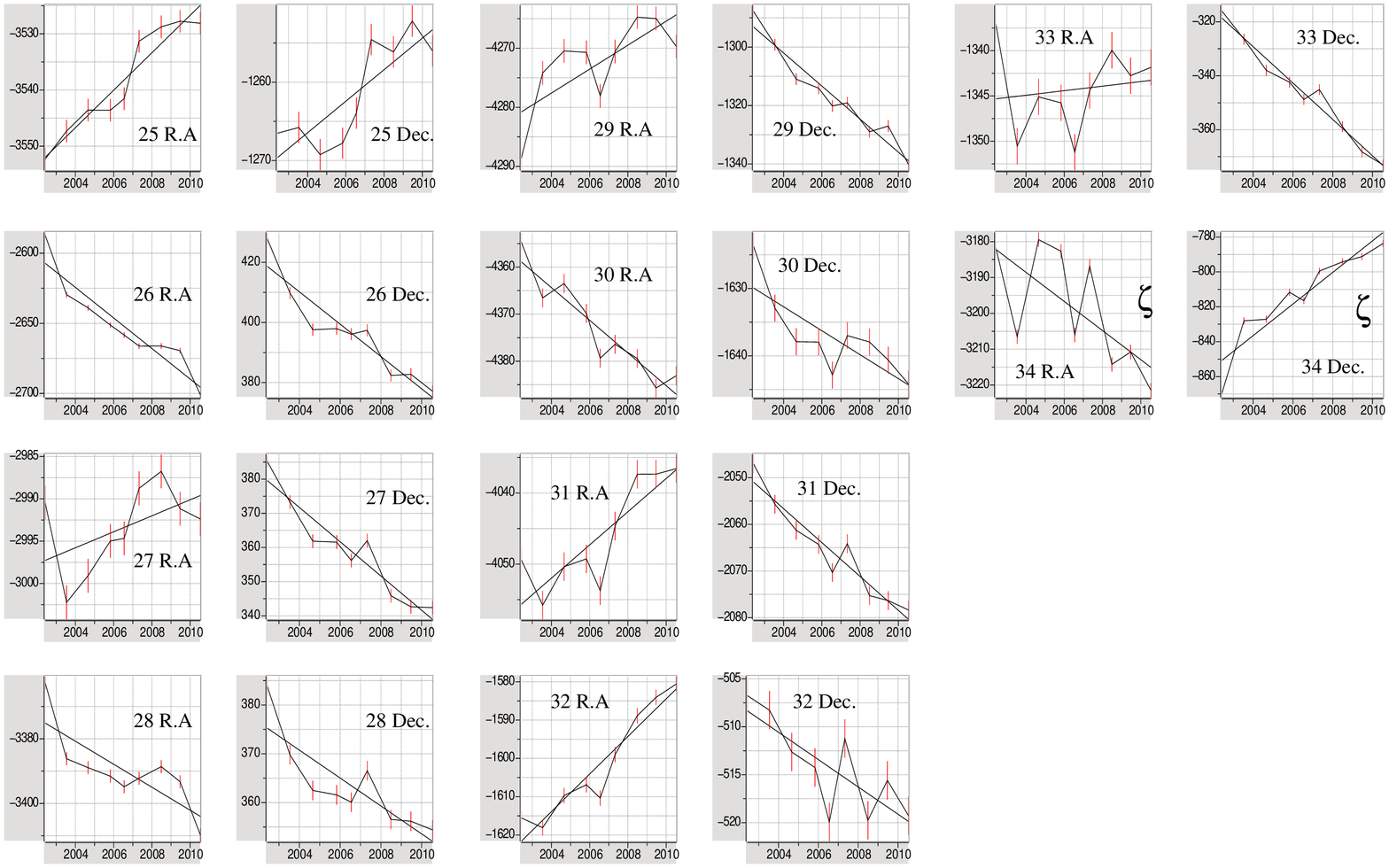}
\caption{\small Fig.~\ref{eckartfig-25} continued.
}
\label{eckartfig-27}
\end{figure*}

\clearpage

\noindent
\begin{figure*}
\centering
\includegraphics[width=15cm,angle=-00]{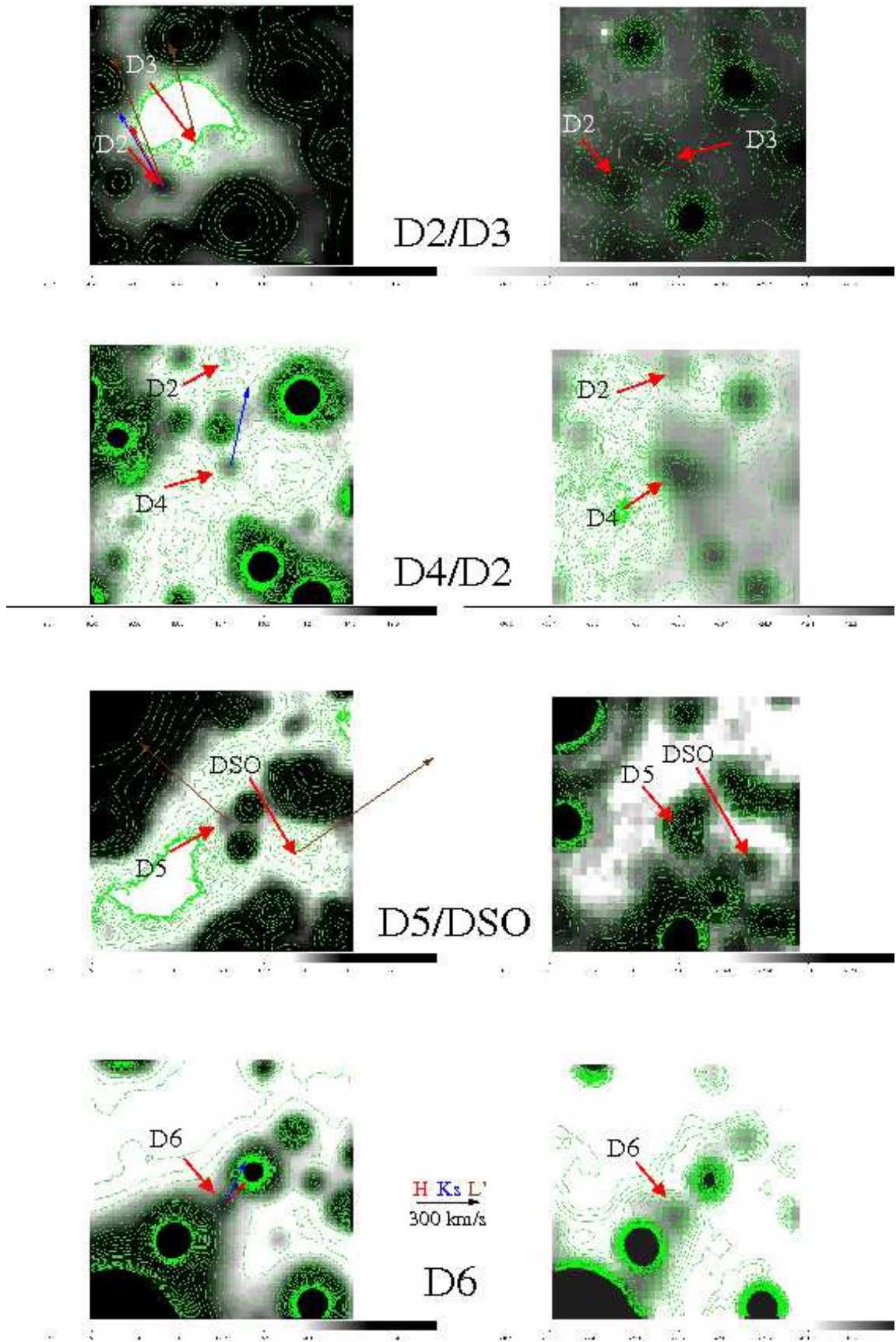}
\caption{\small Comparison of the K$_s$- (left) and L'-band (right) identifications.
L'-band excess sources in the field around SgrA*
as shown in the overview map in Fig.~\ref{eckartfig-01} and listed in Tab.~\ref{Tab:ExcessPM}.
The arrow color-coding for the observing bands and the proper motions are given
between the images in the lower part of the panel.
The 2004 map sizes are 
0.932''$\times$0.932'' (70x70 pixels) in the K$_s$-band 
and
0.945''$\times$0.945'' (35x35 pixels) in the L'-band. 
We have indicated the sources with arrows and labels. 
The gray-scaling is given in bars and the contours are 
logarithmically spaced and given for display purposes only.
The DSO is visible at the lowest contour levels in the appropriate plot
in Fig.~\ref{eckartfig-18}.
In Fig.~\ref{eckartfig-19} we marked the region in which D7 is located with
a dashed circle. Some contour excursions are visible, but no clear detection of a 
compact source is possible.
For source D4/X7 see Muzic et al. (2010).
}
\label{eckartfig-18}
\end{figure*}


\noindent
\begin{figure*}
\centering
\includegraphics[width=15cm,angle=-00]{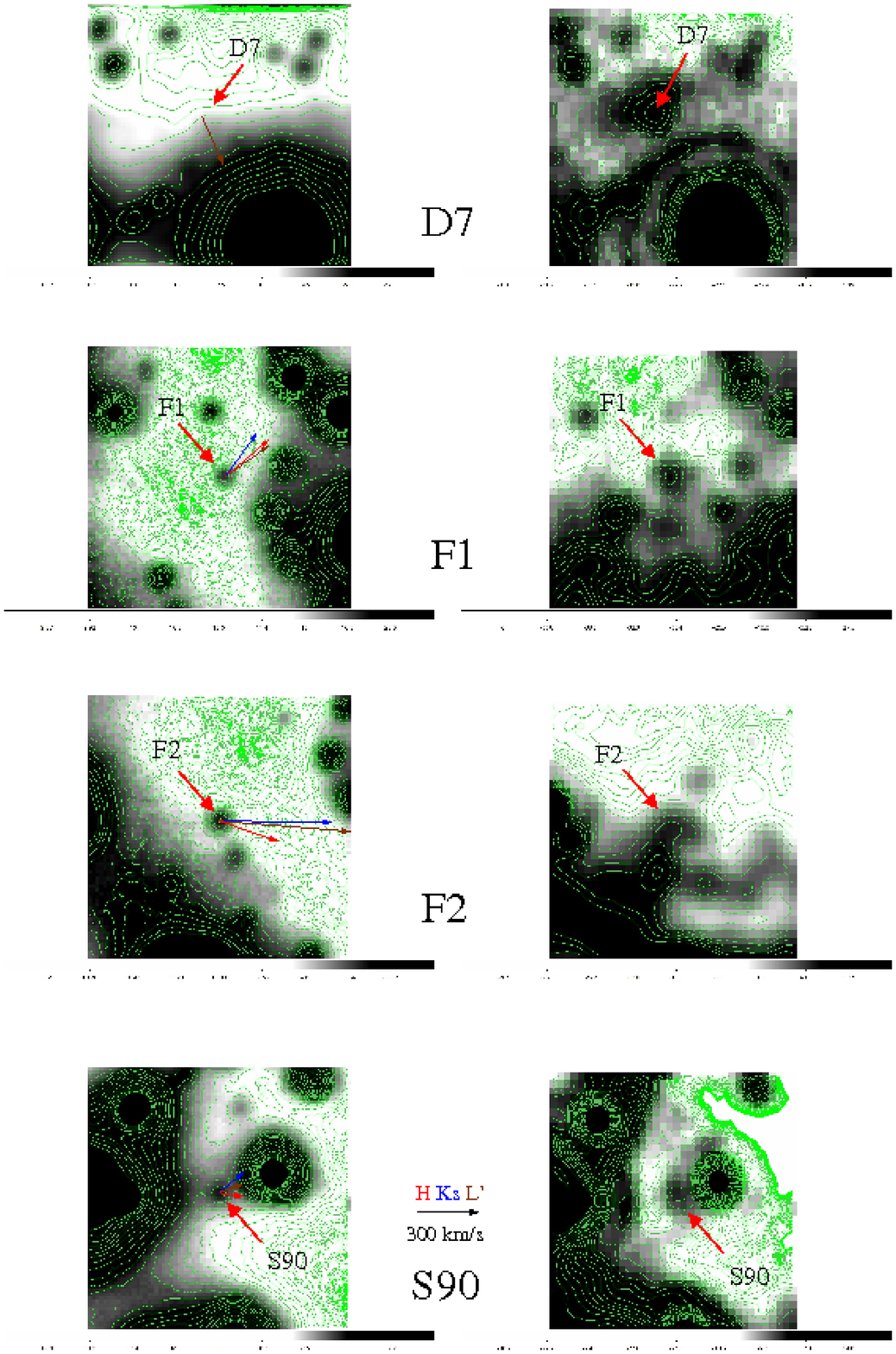}
\caption{\small Fig.~\ref{eckartfig-18} continued.
}
\label{eckartfig-19}
\end{figure*}


\noindent
\begin{figure*}
\centering
\includegraphics[width=15cm,angle=-00]{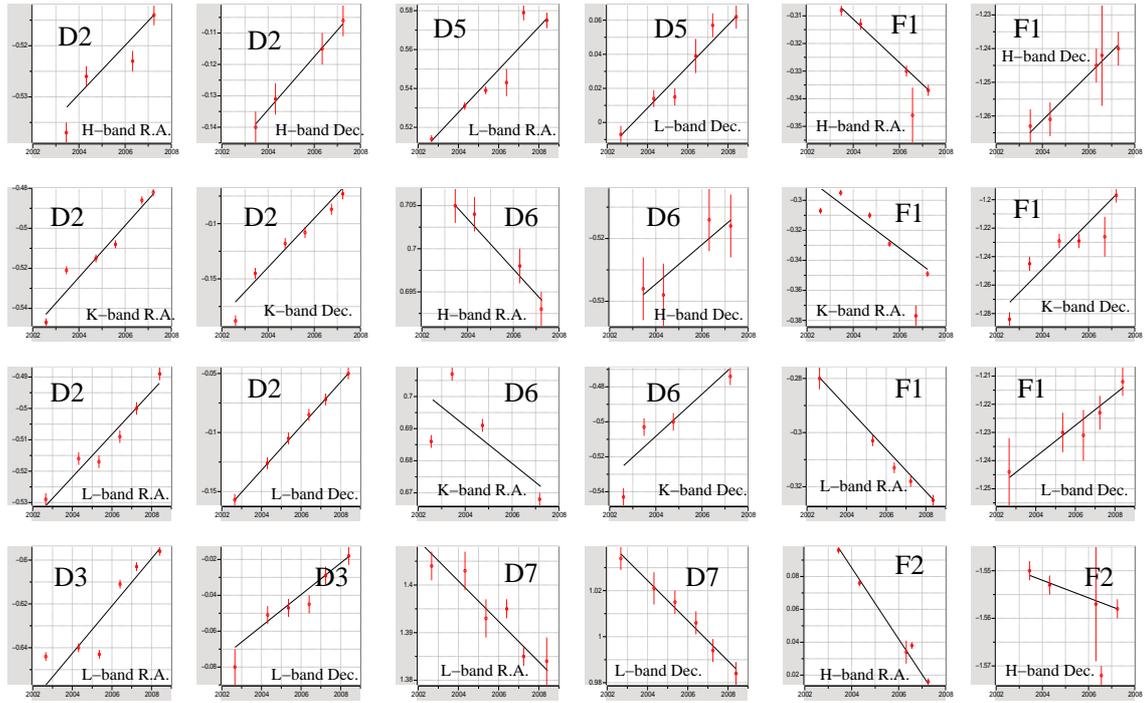}
\caption{\small Proper motions of infrared excess sources in the central
2'' close to SgrA*.
Offsets in R.A. and Dec. in milliarcseconds are plotted against the time 
in years.
Source labels and observing bands are given as indicated in Fig.~\ref{eckartfig-01}
and Tab.~\ref{Tab:ExcessPM}.
The source positions agree to within better than about $\pm$25~mas,
i.e., a quarter of the L'-band beam.
The motions obtained at different wavebands agree
within the uncertainties. For these sources for which S-star identifications are given,
the positions and motions agree on average to within about 30~mas and 150km/s.
}
\label{eckartfig-29}
\end{figure*}


\noindent
\begin{figure*}
\centering
\includegraphics[width=15cm,angle=-00]{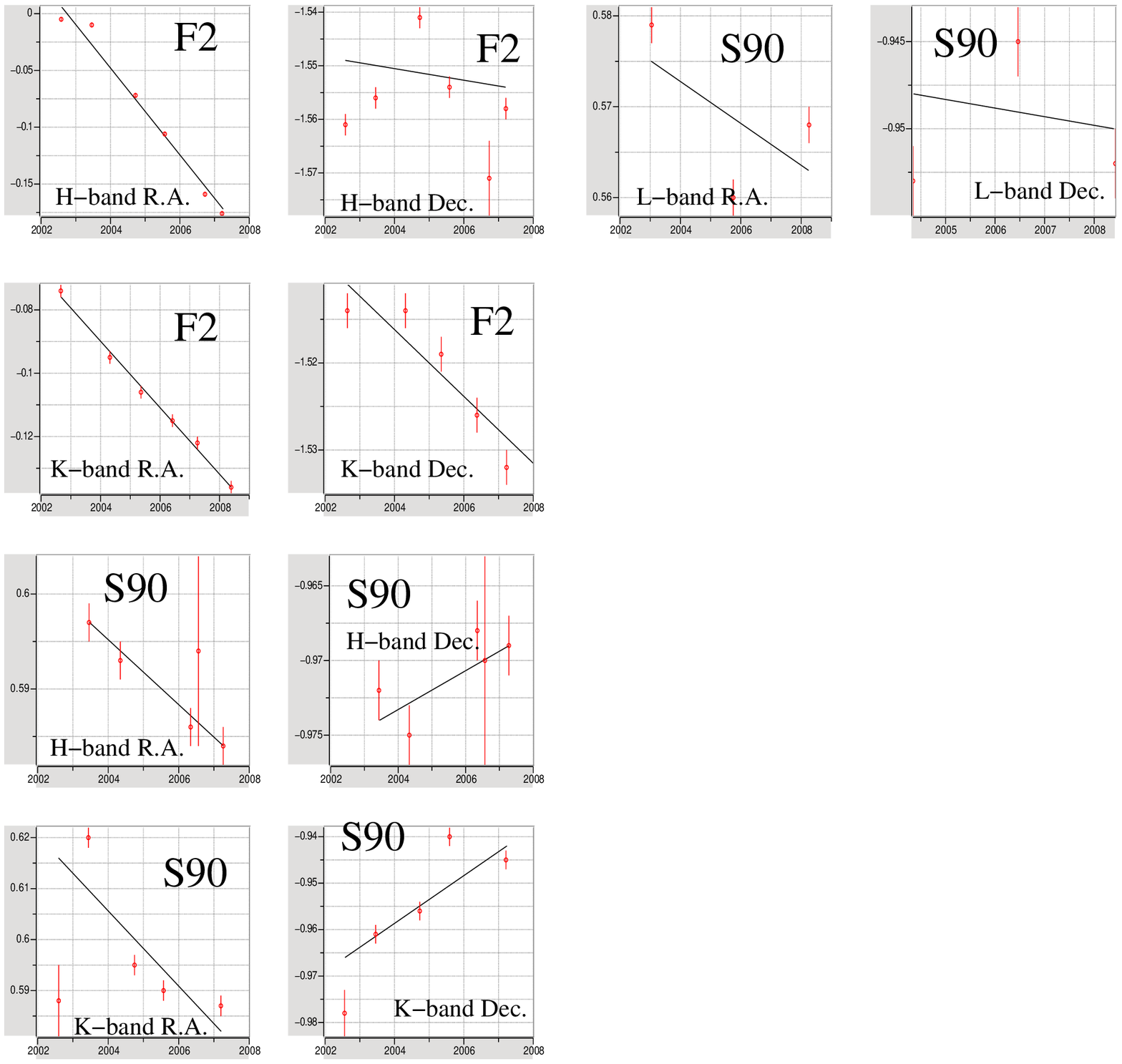}
\caption{\small Fig.~\ref{eckartfig-29} continued.
}
\label{eckartfig-30}
\end{figure*}

\clearpage

\begin{table*}[htb]
\begin{center}
\begin{tabular}{lrrrrrrrrrrrrrrrrrrrrrrrr}\hline 
names & 
comb. &
comb. &
L'  &
L'  &
K$_s$ &
K$_s$ &
H  &
H   \\ 
     & 
$\Delta$$\alpha$ &
$\Delta$$\delta$ &
$\Delta$$\alpha$ &
$\Delta$$\delta$ &
$\Delta$$\alpha$ &
$\Delta$$\delta$ &
$\Delta$$\alpha$ &
$\Delta$$\delta$  \\ \hline
D2, S43   & {\bf -0.515} &{\bf -0.114} & -0.510 & -0.099 & -0.510 & -0.120 & -0.525 & -0.123\\
D6, S79   & {\bf +0.699} &{\bf -0.520} &  0.708 & -0.532 &  0.688 & -0.505 &  0.700 & -0.523\\
S90       & {\bf +0.585} &{\bf -0.959} &  0.569 & -0.950 &  0.596 & -0.956 &  0.591 & -0.971\\
F1        & {\bf -0.321} &{\bf -1.238} & -0.308 & -1.228 & -0.328 & -1.235 & -0.327 & -1.250\\
F2        & {\bf -0.048} &{\bf -1.546} & -0.108 & -1.523 & -0.088 & -1.557 &  0.052 & -1.558\\
D3        & {\bf -0.623} &{\bf -0.045} & -0.623 & -0.045 &        &        &        &       \\
D5        & {\bf +0.547} &{\bf +0.030} &  0.547 &  0.030 &        &        &        &       \\
D4,S50,X7 & {\bf -0.540} &{\bf -0.520} & -0.540 & -0.520 &        &        &        &       \\
D7        & {\bf +1.338} &{\bf +1.003} &  1.394 &  1.009 &  1.384 &  1.002 &  1.236 &  0.998\\
\hline 
\end{tabular}
\end{center}
\caption{L'-, K$_s$, and H-band positions of infrared excess sources in the 
central 2'' field close to SgrA*. Typical uncertainties are $\pm$0.030''.
For D3 and D5 no reliable H- and K$_s$-band identification could be given.
D4 may be associated with S50 (Muzic et al. 2008).
The adopted combined relative positions are printed in boldface.
}
\label{Tab:Field1}
\end{table*}

\begin{table*}[htb]
\begin{center}
\begin{tabular}{lrrrrrrrrrrrrrrrrrrrrrrrr}\hline 
names & 
L'  &
K$_s$ &
H  &
comb. &
L'  &
K$_s$ &
H  &
comb. &
 \\ 
     & 
$v$$_{\alpha}$ &
$v$$_{\alpha}$ &
$v$$_{\alpha}$ &
$v$$_{\alpha}$ &
$v$$_{\delta}$ &
$v$$_{\delta}$ &
$v$$_{\delta}$ &
$v$$_{\delta}$ &
 \\ \hline
D2, S43   & 178 &  241 &  261 &{\bf +226 $\pm$ 44} & 317 & 416 &  709 &{\bf +480 $\pm$203} \\
D6, S79   &-114 & -109 &      &{\bf -112 $\pm$ 20} & 122 & 223 &      &{\bf +172 $\pm$ 50} \\
S90       &-133 & -140 & -108 &{\bf -127 $\pm$ 20} &  51 &  97 &  -15 &{\bf  +44 $\pm$ 60} \\
F1        &-297 & -217 & -281 &{\bf -265 $\pm$ 43} & 260 & 308 &  212 &{\bf +260 $\pm$ 50} \\
F2        &-811 & -702 & -393 &{\bf -635 $\pm$200} & -70 & -21 & -141 &{\bf  -77 $\pm$ 81} \\
D3        &-410 &      &      &{\bf +410 $\pm$212} & 331 &     &      &{\bf +331 $\pm$134} \\
D5        &-419 &      &      &{\bf +419 $\pm$150} & 468 &     &      &{\bf +468 $\pm$131} \\
D4,S50,X7 & -52 &      &      &{\bf  -52 $\pm$ 12} & 546 &     &      &{\bf +546 $\pm$ 20} \\
D7        &-132 &      &      &{\bf -132 $\pm$ 98} &-318 &     &      &{\bf -318 $\pm$ 70} \\
DSO       & 700 &  720 &  900 &{\bf -710 $\pm$170} & 550 & 550 &  400 &{\bf +550 $\pm$160} \\
\hline 
\end{tabular}
\end{center}
\caption{L'-, K$_s$, and H-band proper motion velocities 
of infrared excess sources in the 
central 2'' field close to SgrA*
including results from a linear fit to the DSO data.
The adopted combined velocities are printed in boldface.
For D3, D4 and D5 no reliable H- and K$_s$-band motion could be given.
}
\label{Tab:Field2}
\end{table*}

\begin{table*}[htb]
\begin{center}
\begin{tabular}{lrrrrrrrrrrrrrrrrrrrrrrrr}\hline 
name    & 
$\Delta$$\alpha$ &
$\Delta$$\delta$ &
$v$$_{\alpha}$ &
$v$$_{\delta}$ 
 \\ \hline
S43   & -0.488 & -0.126  & +215  & +310 \\
S79   &  0.647 & -0.532  &  +44  &  +52 \\
S90   &  0.533 &  0.969  &  +33  &  +20 \\
S50   & -0.507 &  0.519  & -114  & +380 \\
\hline 
\end{tabular}
\end{center}
\caption{H-band positions and proper motion velocities for the S-source presented by 
Gillessen et al. (2009) calculated for the epoch 2006.5 and assuming that
1~mas/year corresponds to 40 km/s at the distance of the Galactic Center.
Following the information given by Gillessen et al. (2009), the typical 
uncertainties are $\pm$1~mas for the positions and $\pm$6 km/s for the velocities.
For other S-sources used in the modeling presented in
Figs.~\ref{eckartfig-14-1}, \ref{eckartfig-14-2} and \ref{eckartfig-14-3}
see Tab.~\ref{Tab:DSOsim}.
}
\label{Tab:Field3}
\end{table*}

\end{appendix}

\end{document}